\documentclass[draftclsnofoot,onecolumn]{IEEEtran}

\usepackage{amsmath,epsfig,cite,amsfonts,amssymb,psfrag,subfig,color,setspace,paralist}
\usepackage[usenames,dvipsnames]{xcolor}
\usepackage{pgf}
\usepackage{tikz}
\usepackage{filecontents}
\usepackage{graphicx}
\usepgflibrary{shapes.geometric}
\usetikzlibrary{arrows,decorations.pathmorphing,decorations.footprints,fadings,calc,trees,mindmap,shadows,decorations.text,patterns,positioning,shapes,matrix,fit,chains}
\def\no{\nonumber}

\newtheorem{theorem}{Theorem}
\newtheorem{pro}{Proposition}
\newtheorem{col}{Corollary}
\newtheorem{lem}{Lemma}
\newtheorem{rem}{Remark}

\newtheorem{defi}{Definition}
\newtheorem{example}{Example}

\begin{document}

\title{Strong Secrecy in Pairwise Key Agreement over a Generalized Multiple Access Channel}

\author{\IEEEauthorblockN{Somayeh Salimi, Matthieu Bloch, Fr\'{e}d\'{e}ric Gabry, Mikael Skoglund, Panos Papadimitratos}}

\maketitle
\vspace{-1.cm}
\doublespacing
\begin{abstract}
This paper considers the problem of pairwise key agreement without public communication between three users connected through a generalized multiple access channel (MAC). While two users control the channel inputs, all three users observe noisy outputs from the channel and each pair of users wishes to agree on a secret key hidden from the remaining user. We first develop a ``pre-generated'' key-agreement scheme based on secrecy codes for the generalized MAC, in which the channel is only used to distribute pre-generated secret keys. We then extend this scheme to include an additional layer of rate-limited secret-key generation by treating the observed channel outputs as induced sources. We characterize inner and outer bounds on the strong secret-key capacity region for both schemes. For a special case of the ``pre-generated'' scheme, we obtain an exact characterization. We also illustrate with some binary examples that exploiting the generalized nature of the generalized MAC may lead to significantly larger key-agreement rates.
\end{abstract}
\let\thefootnote\relax\footnote{Somayeh Salimi, Mikael Skoglund and Panos Papadimitratos are with ACCESS Linnaeus Center, School of Electrical Engineering, KTH Royal Institute of Technology, Stockholm, Sweden, {emails: somayen@kth.se, skoglund@kth.se, papadim@kth.se}.

Matthieu Bloch is with School of Electrical and Computer Engineering, Georgia Institute of Technology, Atlanta GA, USA, {email: matthieu.bloch@ece.gatech.edu}.

Fr\'{e}d\'{e}ric Gabry is with the Mathematical and Algorithmic Sciences Lab at the Huawei France Research Center, Paris, France, {email: frederic.gabry@huawei.com}.
 }
\section{Introduction}
\label{intro}
\noindent 
Key management and key distribution in modern communication networks are becoming increasingly challenging because of the dynamic and heterogenous nature of the networks.
Among the proposed solutions to address these challenges, secret-key sharing at the physical layer offers a promising approach to complement traditional public-key and secret-key cryptographic techniques by obviating the need for pre-shared keys. The premise of secret-key sharing at the physical layer is to exploit the randomness in the medium as a resource to generate secret keys, which may then be exploited at the upper layers; one could, for instance, enhance confidentiality, authentication, and integrity of communications.

The canonical information-theoretic models of secret-key agreement have been introduced in~\cite{Ahlswede,Maurer}. These models consider a situation in which two legitimate terminals, observing the outputs of a noisy source or connected through a noisy channel, attempt to generate a secret key by discussion over a public channel in the presence of an eavesdropper. These models have since been extended in several directions to analyze situations involving more terminals, e.g., \cite{helper,multiterminal}, or requiring multiple pairs of keys to be generated, e.g., \cite{threeter}-\cite{salimi-pairwise}.

In this paper, we study a three-user model in which two users control the input of a generalized discrete memoryless multiple-access channel (GDMMAC) while all three users observe the outputs of the channel; each pair of users attempts to agree on a secret key concealed from the remaining user. This scenario models, for instance, a noisy environment with honest-but-curious users, in which two users communicate with a base station through the uplink and overhear each other's communications. Each user wishes to share a secret key with the base station hidden from the other user and simultaneously share a secret key with the other user hidden from the base station for their own private communications. Unlike the related work \cite{salimi-pairwise}, we do not assume that a public channel is available; this allows us to capture some limitations of realistic systems in which communications are inherently rate-limited. In that regard, results and techniques to study secret-key agreement with rate-limited public communication \cite{helper} are particularly relevant to the present work. Our key-sharing scenario encompasses most previous works on secret-key agreement between three users; for instance, if we just consider key sharing between the two users controlling the channel inputs, our setting reduces to the problem of key sharing or secret message(s) transmission over two-way channels as in \cite{twoway1}-\cite{tw-bloch}. If we ignore key sharing between the two users controlling the channel inputs and only consider key sharing between each of these users and the third user simultaneously, our setting reduces to the problem of key sharing or secret message(s) transmission over multiple access channel as in \cite{mac-poor}.

Specifically, we investigate the performance of two distinct pairwise key sharing schemes. We first study a ``pre-generated'' key sharing scheme, in which each of the two active users randomly generates keys that are then encoded for secure transmission over the channel. This scheme does not exploit the generalized nature of the MAC, and merely relies on the use of wiretap codebooks combined with rate splitting. We derive inner and outer bounds on the secret-key capacity region (Theorem~\ref{th1} and Theorem~\ref{th2}), and we identify a special case in which the bounds match (Corollary~\ref{cor:corollary_1}). We then study a ``generalized scheme'' in which the active terminals also exploit the observations from their noisy channel outputs to generate a key. This scheme extends the ``pre-generated'' key sharing scheme by combining rate-limited secret key generation with wiretap codebooks and rate splitting. We again establish inner and outer bounds on the secret-key capacity region (Theorem~\ref{th3} and Theorem~\ref{th4}). We illustrate the performance of both schemes with examples of binary channels that are amenable to numerical calculations (Section III-C and Section IV-B). In particular, these examples show the potential performance gains brought by the extraction of secret keys from channel output observations. Another contribution of this work is to establish strong secrecy results, by leveraging and combining coding techniques for channel resolvability \cite{ch.resolv,Hayashi2006} and channel intrinsic randomness  \cite{ch.intrins1,ch.intrins2,ch.intrins4}. It should be noted that ``pre-generated'' key sharing scheme was partially investigated in \cite{ISWCS2013} with weak secrecy constraints.

The remainder of the paper is organized as follows. Section~\ref{genralmodel} formally introduces the  general model of pairwise key agreement. Section \ref{firstscheme} analyzes the ``pre-generated'' key-sharing scheme and characterizes its performance. The secret-key capacity region is fully characterized for a special case, and results are illustrated with two examples of binary generalized MAC. Section \ref{secondscheme} studies the performance of the generalized key-sharing scheme, which is again illustrated with examples. The technical details of most proofs are relegated in the appendices to streamline presentation.



\section{Pairwise Secret Key Agreement Model}
\label{genralmodel}
\begin{figure}
\centering
\includegraphics[width=9cm]{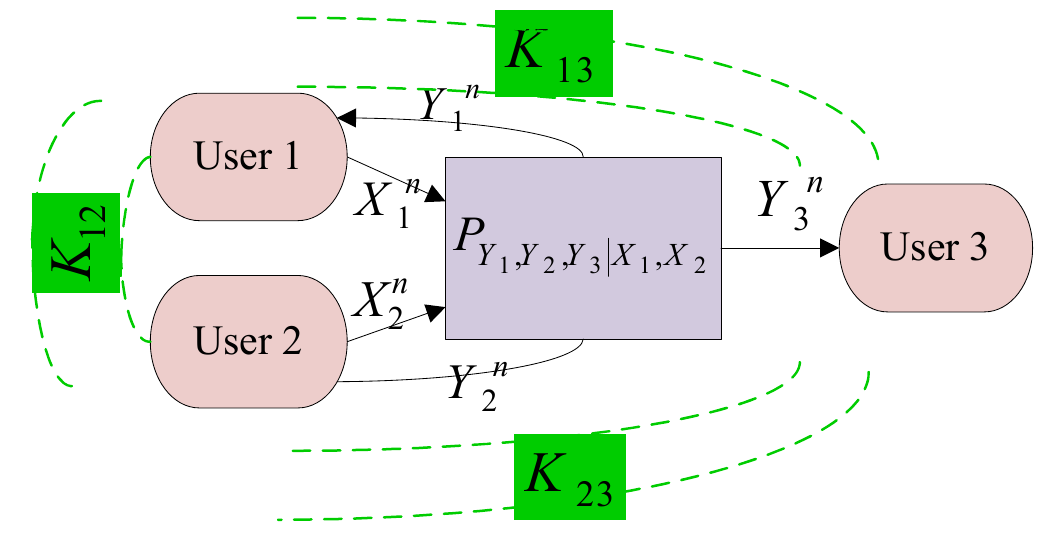}
\vspace{-.3cm}
\caption{\footnotesize{Pairwise key sharing over GDMMAC}}
\vspace{-.5cm}
\label{fig:generic_model}
\end{figure}

In this section, we introduce our model for pairwise secret key agreement. As illustrated in Fig.~\ref{fig:generic_model}, we consider a memoryless generalized MAC $(\mathcal{X}_1,\mathcal{X}_2,P_{Y_1Y_2Y_3|X_1X_2},\mathcal{Y}_1,\mathcal{Y}_2,\mathcal{Y}_3)$. User 1 and User 2 control the inputs $X_1$ and $X_2$, respectively, while Users 1, 2, and 3 observe the outputs $Y_1$, $Y_2$, $Y_3$. respectively. The objective is for each pair of users to share a secret key while keeping it concealed from the remaining user. Formally, a code for pairwise key agreement consists of the following.
\begin{itemize}
\item Two randomization sequences $\mathcal{V}_j$ for $j\in\{1,2\}$, used as sources of local randomness for User 1 and User 2.
\item Key sets $\mathcal{K}_{jl}=\{1,\cdots,2^{nR_{jl}}\}$ for $j<l\in\{1,2,3\}$, in which the pairwise keys take values.
\item Two sequences of encoding functions $f^{(i)}_j:\mathcal{V}_j\times\mathcal{Y}_j^{i-1}\rightarrow\mathcal{X}_j$ for $j\in\{1,2\}$ and $i\in\{1,\cdots,n\}$ allowing User $j$, $j\in\{1,2\}$, to generate channel input $X_j^{i}$ at time instant $i$, as a function of a randomization sequence $V_j$ and past observed channel outputs $Y_j^{i-1}$.
\item Key generation functions $g_{l}$, $l\in\{1,2,3\}$ available at User $l$
  \begin{align}
    g_{1}&:\mathcal{V}_1\times\mathcal{Y}_1^n\rightarrow\mathcal{K}_{12}\times\mathcal{K}_{13} \\
    g_{2}&:\mathcal{V}_2\times\mathcal{Y}_2^n\rightarrow\mathcal{K}_{12}\times\mathcal{K}_{23} \\
    g_{3}&:\mathcal{Y}_3^n\rightarrow\mathcal{K}_{13}\times\mathcal{K}_{23}.
  \end{align}
that allow Users $j$ and $l$ to generate $K_{jl}$ and $\hat{K}_{jl}$, respectively, as the shared key between them for $j<l\in\{1,2,3\}$.
\end{itemize}

\begin{defi}
\label{def1}
A secret-key rate triple $(R_{12},R_{13},R_{23})$ is achievable if for every $\varepsilon>0$ and all sufficiently large $n$, we have:
\begin{align}
&{\forall j<l\in\{1,2,3\}\quad \frac{1}{n}H(K_{jl})=\frac{1}{n}\log|\mathcal{K}_{jl}|\geq R_{jl}-\epsilon} \label{eq1}\\
&{\forall j<l\in\{1,2,3\}\quad\Pr\{K_{jl}\neq\hat{K}_{jl}\}<\epsilon} \label{eq2}\\
&{I(K_{12};Y_3^n)<\epsilon, \ \text{and} \  \forall j\neq \bar{j}\in\{1,2\}\quad I(K_{j3};X_{\bar{j}}^n,Y_{\bar{j}}^n)<\epsilon}  \label{eq3}
 \end{align}
The set of all the achievable secret-key rate triples $(R_{12},R_{13},R_{23})$ is the secret-key capacity region.
\end{defi}
\noindent\normalsize Equation \eqref{eq1} means that the rates $R_{12},R_{13}$ and $R_{23}$ are the rates of nearly uniform secret keys between Users 1 and 2, Users 1 and 3, and Users 2 and 3, respectively.
Equation \eqref{eq2} means that each user can correctly estimate the related keys. Equation \eqref{eq3} means that each user effectively has no information about the other users' secret keys.
The key rates in Definition \ref{def1} are strongly secure since only the total leakage information is bounded by $\varepsilon$.  All the above keys take values in some finite sets.


\section{The Pre-Generated Keys Scheme of Pairwise Secret Key Agreement over GDMMAC}
\label{firstscheme}
In this section, we specialize the generic scheme of Section~\ref{genralmodel} to a simpler key-sharing scheme, in which the keys are ``pre-generated''. Specifically, the idea is to have User 1 and User 2 generate keys with their local randomness and then secretly transmit them to the other users without using the generalized feedback.

\subsection{Principle of Pre-Generated Keys Scheme}
\label{prem1}
\begin{figure}
\centering
\includegraphics[width=9cm]{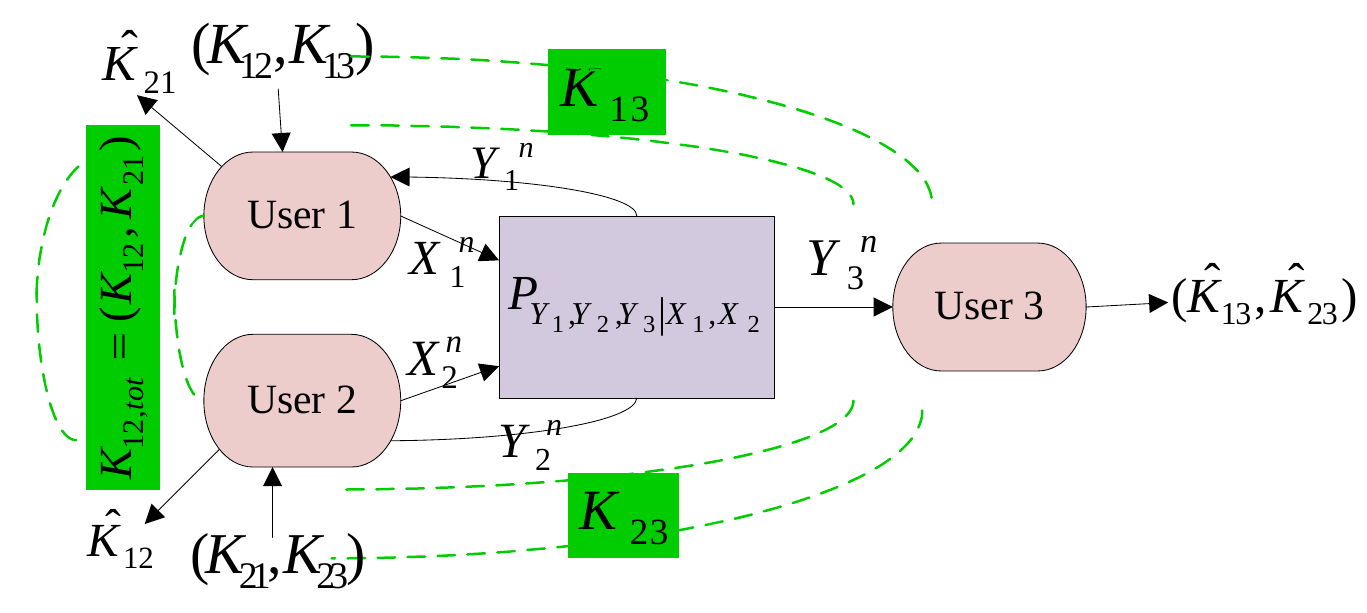}
\vspace{-.3cm}
\caption{\footnotesize{Pairwise key sharing over GDMMAC using the pre-generated keys scheme}}
\vspace{-.5cm}
\label{fig_sim}
\end{figure}

As illustrated in Fig.~\ref{fig_sim}, the pre-generated key sharing scheme consists of the following simplifications in the generic scheme.
\begin{itemize}
\item User 1 uses part of its common randomness to generate local keys $K_{12}\in\mathcal{K}_{12}$ and $K_{13}\in\mathcal{K}_{13}$ to share with Users 2 and 3, respectively. The remaining part is denoted $\widetilde{\mathcal{V}}_1$.
\item User 2 uses part of its common randomness to generate local keys $K_{21}\in\mathcal{K}_{21}$ and $K_{23}\in\mathcal{K}_{23}$ to share with Users 1 and 3, respectively. The remaining part is denoted $\widetilde{\mathcal{V}}_2$.
\item User 1's deterministic encoding and decoding functions are
  \begin{align*}
    f_1:\widetilde{\mathcal{V}}_1\times \mathcal{K}_{12}\times \mathcal{K}_{13}\rightarrow\mathcal{X}_1^n\quad\text{and}\quad
    g_1:\mathcal{Y}_1^n\rightarrow \mathcal{K}_{21}.
  \end{align*}
\item User 2's deterministic encoding and decoding functions are
  \begin{align*}
    f_2:\widetilde{\mathcal{V}}_2\times \mathcal{K}_{21}\times \mathcal{K}_{23}\rightarrow\mathcal{X}_3^n\quad\text{and}\quad
    g_2:\mathcal{Y}_2^n\rightarrow \mathcal{K}_{12}.
  \end{align*}
\item User 3's deterministic decoding function is
 \begin{align*}
    g_{3}:\mathcal{Y}_3^n\rightarrow\mathcal{K}_{13}\times\mathcal{K}_{23}
  \end{align*}
\end{itemize}
At the end of a transmission, the key pair $K_{12,tot}=(K_{12},K_{21})$ is shared between User l and User 2, $K_{13}$ is shared between User 1 and User 3, and $K_{23}$ is shared between User 2 and User 3. Note that the shared key between Users 1 and 2 ($K_{12,tot}$) consists of two secret keys and hence, rate $R_{12}$ defined in (\ref{eq1}) is the total key rate. This specialization reduces the generic problem to the simpler problem of secret key distribution, and our analysis of this scheme only relies on the use of \emph{wiretap codes}.

\vspace{-.3cm}

\subsection{Main Results}
\label{main1}
Our main results for the pre-generated key sharing scheme consist of inner and outer bounds on the secret-key capacity region. We first define the following rates:
\begin{align}\no
&{{\bf r}_{{\bf 12}} =[I(S_{12} ;X_{2},Y_{2})-I(S_{12} ;Y_{3} ,S_{13},S_{23})]^{+} ,}\no\displaybreak[0]
\\&{{\bf r}_{{\bf 21}} =[I(S_{21} ;X_{1},Y_{1})-I(S_{21} ;Y_{3},S_{13},S_{23})]^{+} ,}\no\displaybreak[0]
\\&{{\bf I}_{{\bf 12}} =I(S_{12};S_{21} |Y_{3} ,S_{13} ,S_{23} )},\no\displaybreak[0]
\\&{{\bf r}_{{\bf 13}} =[I(S_{13} ;Y_{3}|S_{23})-I(S_{13} ;X_{2},Y_{2},S_{12}|S_{23})]^{+} ,}\no\displaybreak[0]
\\&{{\bf r}_{{\bf 23}} =[I(S_{23} ;Y_{3}|S_{13})-I(S_{23} ;X_{1},Y_{1},S_{21} |S_{13})]^{+} ,}\no\displaybreak[0]
 \\&{{\bf I}_{{\bf 3}} =I(S_{13} ;S_{23} |Y_{3} )}\label{rate-define1}
\end{align}
in which $[x]^{+}=\max(x,0)$.

\begin{theorem}
\label{th1}
In the pre-generated keys scheme, all rate triples in the closure of the convex hull of the set of rate triples $(R_{12},R_{13},R_{23})$ that satisfy the following conditions are achievable:
\begin{align}
&{R_{12} \ge 0,R_{13} \ge 0,R_{23} \ge 0,}\no
\\& {R_{12} \le {\bf r}_{{\bf 12}} {\bf +r}_{{\bf 21}} -{\bf I}_{{\bf 12}} ,}\no
\\&{R_{13} \le {\bf r}_{\bf 13},} \no
\\&{R_{23} \le {\bf r}_{\bf 23},}\no
\\&{R_{13} +R_{23}\le {\bf r}_{\bf 13} {\bf +r}_{\bf 23}-{\bf I}_{\bf 3},}\no
\end{align}
for random variables taking values in finite sets and with joint distribution factorizing as:
\begin{align*}
  p(\!s_{12},\!s_{13},s_{21},\!s_{23} ,\!x_{1} ,\!x_{2} ,\!y_{1} ,\!y_{2} ,\!y_{3}\!)\!
  =\!p(s_{12}\!)p(s_{13}\!)p(s_{21}\!)p(s_{23}\!)p(\!x_{1}|s_{12},\!s_{13}\!)p(\!x_{2}|s_{21},\!s_{23}\!)p(\!y_{1},\!y_{2},\!y_{3}|x_{1} ,\!x_{2}\!).
\end{align*}
\end{theorem}

The proof of Theorem~\ref{th1} is actually a special case of the proof of Theorem~\ref{th3}, which may be found in Appendix~\ref{App1}. We only provide here a high-level description of the scheme achieving the rate region, which is essentially a combination of wiretap codebooks and rate splitting. The channel model is split into two broadcast channels with confidential messages; one from User 1 to Users 2 and 3 and the other from User 2 to Users 1 and 3, where in each broadcast channel, the receivers are eavesdroppers of each other's key. The rates $\bf {r_{12}}$ and $\bf {r_{13}}$ correspond to the well known rates of secure communication between Users 1 and 2 and Users 1 and 3, respectively. Similarly, $\bf {r_{21}}$ and $\bf {r_{23}}$ are the rates of secure communication between Users 2 and 1 and, Users 2 and 3, respectively. The bound on the total key rate between Users 1 and 2 is the sum of the bounds ${\bf r}_{{\bf 12}}$ and ${\bf r}_{{\bf 21}}$ minus a penalty term ${\bf I}_{{\bf 12}}$, which results from the required independence of the transmitted keys. Similarly, the sum rate of the keys to User 3 is the sum rate ${\bf r}_{\bf 13} {\bf +r}_{\bf 23}$ penalized by ${\bf I}_{{\bf 3}}$.

\begin {rem}
If we ignore key sharing between Users 1 and 2, our result reduces to the secrecy rate region of the generalized multiple access channel with confidential messages~\cite[Corollary 1]{mac-poor} by substituting $S_{12}=S_{21}=\emptyset$ in Theorem \ref{th1}. Similarly, if we ignore key sharing between Users 1 and 3 as well as Users 2 and 3, our result reduces to secret-key sharing in the two-way channel with an external eavesdropper~\cite[Corollary 1]{tw-bloch} by substituting $S_{13}=S_{23}=\emptyset$ in Theorem \ref{th1}.
\end {rem}

\begin{theorem}\label{th2}
In the pre-generated keys scheme, if a
key rate triple is achievable, then it belongs to the set of all rate triples $(R_{12},R_{13},R_{23})$ that satisfy:
\begin{align*}
  0\leq R_{12} &\le I(X_{1};\!Y_{2} |X_{2},\!Y_{3} )\!+\!I(X_{2};\!Y_{1} |X_{1},\!Y_{3} )\!+\!I(Y_{1} ;\!Y_{2} |X_{1},X_{2},\!Y_{3} )\!+\!I(X_{1};Y_{3}|X_{2},U)-I(X_{1};Y_{3}|U),\\
  0\leq R_{13} &\le I(X_{1} ;Y_{3}|X_{2},Y_{2} ),\\
  0\leq R_{23} &\le I(X_{2} ;Y_{3}|X_{1},Y_{1} ),
\end{align*}
for random variables
$U,X_1,X_2,Y_1,Y_2,Y_3$, all taking values in finite sets, such that $U-(X_1,X_2)-(Y_1,Y_2,Y_{3})$ forms a Markov chain.
\end{theorem}
\begin{IEEEproof}
See Appendix~\ref{App3}.
\end{IEEEproof}


\subsection{Special Case and Examples}
\label{spec1}
We now investigate a special case of the model in which the inner and outer bounds in Theorems~\ref{th1} and~\ref{th2} match, hence providing a complete characterization of the secret-key capacity region.
\begin{col}
\label{cor:corollary_1}
When the GDMMAC inputs and outputs form Markov chains as $ X_{1}- (X_{2},Y_{2} ) - Y_{3} - Y_{1}$ and $ X_{1} - Y_{3} - X_{2}$, the secret-key capacity region is
\begin{equation}
\left\{(R_1,R_2):
  \begin{array}{l}
    0\le R_{12} \le I(X_{1};Y_{2}|X_{2}Y_{3}),\\
    R_{13}=0,\\
    0\leq R_{23}\le I(X_{2};Y_{3}|X_{1},Y_{1})
  \end{array}
\right\}
\end{equation}
\end{col}
\begin{IEEEproof}
The achievability can be inferred from Theorem~\ref{th1} by substituting $S_{12}\!=\!X_{1}, S_{13}\!=\!\emptyset, S_{21}\!=\!\emptyset,S_{23}\!=\!X_{2}$ and using the Markov chains in the statement of Corollary~\ref{cor:corollary_1}. The converse follows from Theorem~\ref{th2} since
\begin{multline*}
  I(X_{1}; Y_{2} |X_{2}, Y_{3} ) + I(X_{2}; Y_{1} |X_{1}, Y_{3} ) + I(Y_{1} ; Y_{2} |X_{1},X_{2}, Y_{3} ) + I(X_{1}; Y_{3}|X_{2},U)-I(X_{1};Y_{3}|U)\\
  \begin{split}
    &\mathop{=}\limits^{{\rm(a)}} I(X_{1}; Y_{2} |X_{2}, Y_{3}    ) + I(X_{1}; Y_{3}|X_{2},U)-I(X_{1};Y_{3}|U), \\
    &\mathop{=}\limits^{{\rm(b)}} I(X_{1}; Y_{2} |X_{2}, Y_{3}    ) + H(X_{1}|X_{2},U)-H(X_{1}|U, Y_{3})-I(X_{1};Y_{3}|U), \\
    &\leq I(X_{1};Y_{2}|X_{2}, Y_{3}),
\end{split}
\end{multline*}
and
\begin{align*}
  I(X_{1} ;Y_{3}|X_{2},Y_{2} )\mathop{=}\limits^{{\rm(c)}}0 .
\end{align*}
In the above equations (a) and (c) follow from the Markov chain $\!X_{1}-\!(X_{2},Y_{2}\!)\!-\!Y_{3}\!-\!Y_{1}$ while (b) follows from the Markov chain $\!X_{1}\!-\!Y_{3}\!-\!X_{2}$.
\end{IEEEproof}

We now introduce an example in which the Markov chains of Corollary~\ref{cor:corollary_1} hold. The following lemma turns out to be useful.

\begin{lem}
\label{lm:lemma_1}
In the pre-generated keys scheme, if two GDMMACs have the same marginal channel transition probability distributions $p(y_{1}|x_{1},x_{2})$, $p(y_{2}|x_{1},x_{2})$ and $p(y_{3}|x_{1},x_{2})$, then they have the same secret-key capacity region.
\end{lem}
Since in the pre-generated keys scheme, the channel outputs are not involved in the encoding, Lemma~\ref{lm:lemma_1} can be proved using the same approach as in~\cite[Lemma 1]{mac-poor}.

\begin{figure}
\centering
\includegraphics[width=9cm]{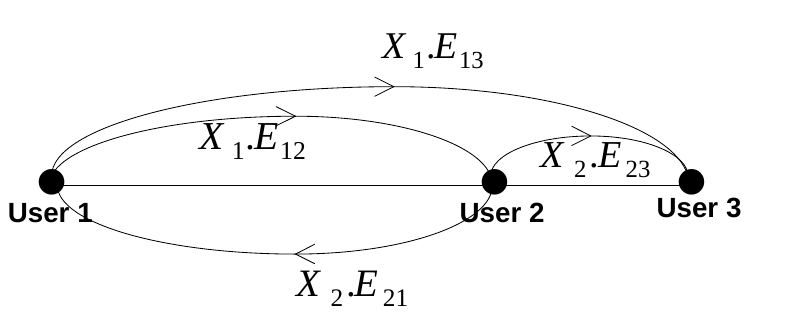}
\vspace{-.3cm}
\caption{\footnotesize{Erasure Example 1}
}
\vspace{-.5cm}
\label{fig_sim11}
\end{figure}

\begin{example}
  \label{ex:example_1}
  Consider a GDMMAC with inputs alphabet $\{ -1,1\}$ where erased versions of the inputs are received by the users according to Fig. \ref{fig_sim11} as:
\begin{align*}
Y_{1} &=X_{2} \times E_{21},\quad Y_{2} =X_{1} \times E_{12},\quad Y_{3} =(X_{1} \times E_{13},X_{2} \times E_{23}),
\end{align*}
in which $E_{ij}$ takes values in $\{ 0,1\}$ with distribution $\Pr(E_{ij}=0)=p_{ij}$. Operation $\times$ has the usual meaning of multiplication and the random variables $(X_{1},X_{2},E_{12},E_{21},E_{13},E_{23})$ are independent of each other. In this example, we assume that the channels between Users 1 and 2 are symmetric and hence $p_{12}=p_{21}$ and furthermore $p_{13}\geq p_{12}\geq p_{23}$. Since $p_{13}\geq p_{12}$, User 1 can only share secret key with User 2 and hence, in Theorem 1, we set $S_{13}\!=\emptyset$ and $S_{12}\!=\!X_{1}$ where $X_{1}$ is uniformly distributed over $\{ -1,1\}$. On the other hand, since $p_{12}\geq p_{23}$, User 2 dedicates the whole channel input to share a secret key with User 3 by substituting $S_{21}\!=\emptyset$ and $S_{23}\!=\!X_{2}$ where $X_{2}$ is uniformly distributed over $\{ -1,1\}$. By substituting the auxiliary random variables in Theorem~\ref{th1} as described above, we obtain the following achievable secret-key rate region:
\begin{align}
  \left\{(R_1,R_2):
  \begin{array}{l}
    0\leq R_{12} \le p_{13}- p_{12},\\
    R_{13} =0,\\
    0\leq R_{23} \le p_{12}- p_{23}.
  \end{array}
\right\}\label{ex1}
\end{align}
We now use Lemma~\ref{lm:lemma_1} to show that the rate region in (\ref{ex1}) is the capacity region. Consider a new GDMMAC with channel outputs $y_{1},y'_{2},y'_{3}$ at Users 1, 2 and 3, respectively, where
\begin{align*}
  Y_{1} =X_{2} \times E_{21},\quad Y'_{2} =X_{1} \times E'_{12},\quad Y'_{3} =(X_{1} \times E_{13},X_{2} \times E'_{23}),
\end{align*}
in which $E_{21}$ and $E_{13}$ are the erasure random variables in Example~\ref{ex:example_1} and $E'_{12}$ and $E'_{23}$ are erasure random variables with erasure probabilities $p_{12}$ and $p_{23}$, respectively. $E_{21}$ and $E_{13}$ are correlated with $E'_{12}$ and $E'_{23}$ according to Fig. \ref{fig_sim12}. It can be shown that the following relationships hold between the new channel outputs
\begin{align*}
  Y_{1}=X_{2} \times E'_{23}\times E_{x},\quad Y'_{2} =X_{1} \times E'_{12},\quad Y'_{3} =(Y'_{2}\times E_{y},X_{2} \times E'_{23}).
\end{align*}
\begin{figure}
\centering
\includegraphics[width=12cm]{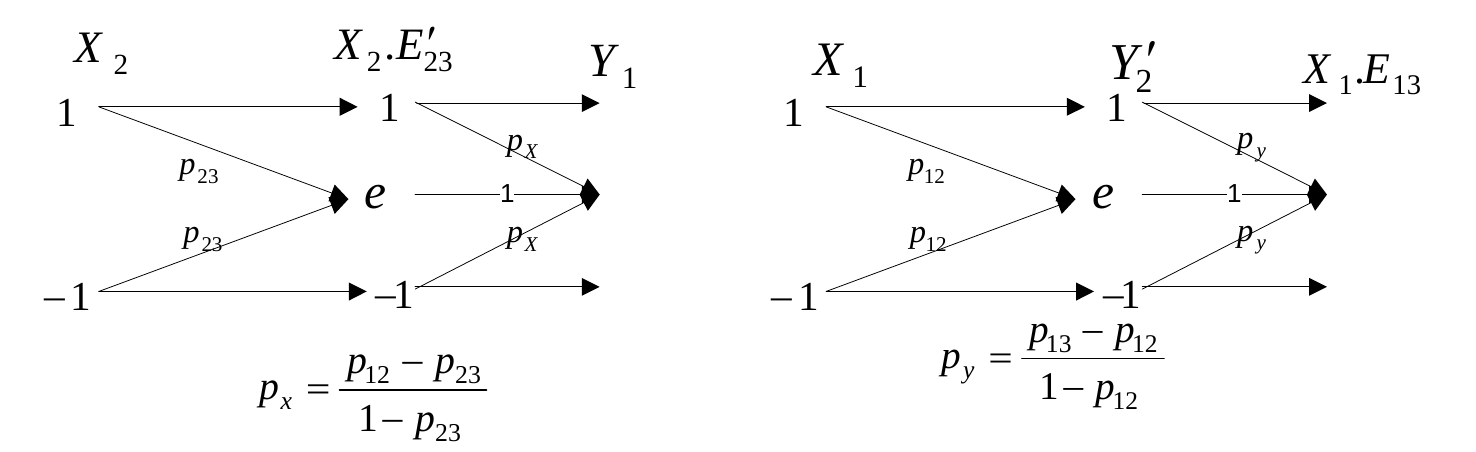}
\vspace{-.3cm}
\caption{\footnotesize{New erasure channel outputs relationships in Example~\ref{ex:example_1}}
}
\vspace{-.5cm}
\label{fig_sim12}
\end{figure}
The new channel outputs satisfy the Markov chains of Corollary~\ref{cor:corollary_1} (replacing $Y_{2}$ and $Y_{3}$ with $Y'_{2}$ and $Y'_{3}$). Therefore, the secret-key capacity region of the new channel is
\begin{align}
  \left\{(R_1,R_2):
  \begin{array}{l}
    0\leq R_{12} \le p_{13}- p_{12},\\
    R_{13} =0,\\
    0\leq R_{23} \le p_{12}- p_{23}.
  \end{array}
  \right\}\label{ex1}
\end{align}
Since $Y'_{2}$ and $Y'_{3}$ have the same marginal distributions as $y_{2}$ and $y_{3}$, respectively, according to Lemma~\ref{lm:lemma_1}, the original channel has the same secret-key capacity region as the new channel and hence, the secret-key capacity region in (\ref{ex1}) is the capacity region.
\end{example}

\begin{example}
  \label{ex:example_2}
  Consider a binary GDMMAC where the relationships between the channel inputs and outputs are according to:
\begin{align}
{Y_{i} =X_{1}+X_{2}+Z_{i}{\rm \; \; \; \; \; \; \ \ }i=1,2,3,}\no
\end{align}
in which $Z_{i}$ is a binary random variable with distribution $\Pr(Z_{i}=1)=p_{i}$. Operation $+$ is the binary summation and the random variables $(X_{1},X_{2},Z_{1},Z_{2},Z_{3})$ are independent of each other. We assume that $0\leq p_{2}\leq p_{3}\leq p_{1}\le0.5$. The other cases can be similarly considered. At first glance, it may seem that because of $p_{2}\leq p_{3}$, the best strategy for User 1 is to set $S_{12}\!=\!X_{1},S_{13}\!=\emptyset$ where $\!X_{1}$ is uniformly distributed over $\{ 0,1\}$. This is the best strategy for User 1 to maximize his secret key rate $R_{12}$, but it would result in $R_{23}=0$ since $X_{1}$ is uniformly distributed over $\{ 0,1\}$ and hence $I(S_{23};Y_{3})=0$. Based on this argument, we assume that $X_{1}$ is a binary random variable with parameter $\alpha$ where $0\leq \alpha\leq0.5$. On the other hand, if User 2 sets $S_{21}\!=\emptyset,S_{23}\!=\!X_{2}$ with $\!X_{2}$ uniformly distributed over $\{ 0,1\}$, the maximum rate of $R_{23}$ will be achievable, however it decreases $R_{12}$ since User 3 decodes $X_{2}$ and the leakage rate $I(S_{12};Y_{3},S_{23})$ in the expression of $R_{12}$ will increase. Hence, we assume $\!S_{23}$ be a binary random variable connected to $X_{2}$ by another binary symmetric channel with parameter $\beta$, as in Fig. \ref{fig.ex1} where $0\leq \beta\leq0.5$.
\begin{figure}
\centering
\includegraphics[width=5cm]{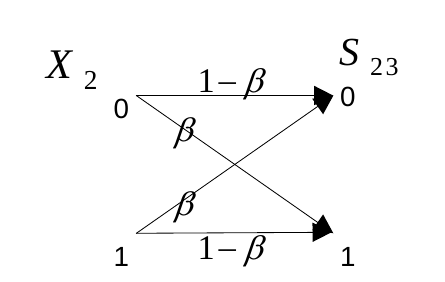}
\vspace{-.3cm}
\caption{\footnotesize{Auxiliary random variable $S_{23}$ in Example~\ref{ex:example_2}}
}
\vspace{-.5cm}
\label{fig.ex1}
\end{figure}
By substituting the auxiliary random variables in Theorem 1 as described, the following rate region is achievable:
\begin{align}
&{R_{12} \le h(\alpha\ast p_{2})+h(\beta\ast p_{3})-h(\alpha\ast\beta\ast p_{3})-h(p_{2}),}\no\\
&{R_{13} =0,}\no\\
&{R_{23} \le h(\beta\ast p_{1})-h(\beta\ast \alpha\ast p_{3})}\label{innerex2}
\end{align}
for all values of $0\leq \alpha,\beta \leq0.5$, where
\begin{align}
&{\alpha\ast\beta\triangleq \alpha(1-\beta)+\beta(1-\alpha)}\label{stardef}\\
&{h(p)\triangleq -p\log p-(1-p)\log(1-p)}\label{entdef}
\end{align}
\begin{figure}
\centering
\includegraphics[width=9cm]{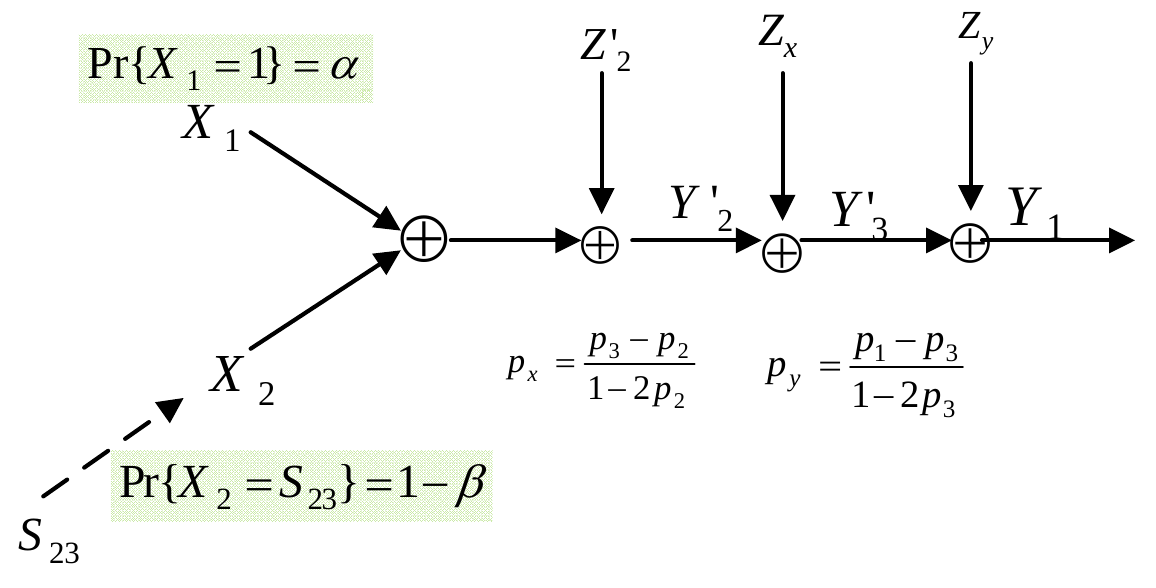}
\vspace{-.3cm}
\caption{\footnotesize{New channel outputs relationships in Example~\ref{ex:example_2}}
}
\vspace{-.5cm}
\label{fig_sim21}
\end{figure}
\indent To derive the outer bound on the secret-key capacity region, we use Lemma~\ref{lm:lemma_1} as in Example~\ref{ex:example_1} to introduce stochastic degradedness. Since $0\leq p_{2}\leq p_{3}\leq p_{1}\le0.5$, we define a new channel with the same marginal distributions as in Example~\ref{ex:example_2} where:
\begin{align}
&{Y'_{2}=X_{1}+X_{2}+Z'_{2},}\no\\
&{Y'_{3}=X_{1}+X_{2}+Z'_{2}+Z_{x},}\no\\
&{Y_{1}=X_{1}+X_{2}+Z'_{2}+Z_{x}+Z_{y}=X_{1}+X_{2}+Z_{1},}\no
\end{align}
\noindent in which $Z'_{2}$ has the same distribution as $Z_{2}$ and $Z_{x}$ and $Z_{y}$ are binary random variables with distributions $\Pr(Z_{x}=1)=p_{x},\Pr(Z_{y} =1)=p_{y}$ where $p_{x}$ and $p_{y}$ are defined as in Fig. \ref{fig_sim21}. The random variables $(Z'_{2},Z_{x},Z_{y})$ are independent of each other.

The Markov chain $(X_{1},X_{2})-Y'_{2}-Y'_{3}-Y_{1}$ holds between the new channel inputs and outputs and the following region is an outer bound on the secret-key capacity region of the new channel:
\begin{equation} R_{12}\leq1-h(p_{2}),{\rm \; \; \; \; \; \; \; \; \; \; }R_{13}=0,{\rm \; \; \; \; \; \; \; \; \; \; }R_{23} \le h( p_{1})-h(p_{3})\label{outerex2}
\end{equation}
\noindent where the upper bounds on $R_{13}$ and $R_{23}$ are directly inferred from Theorem~\ref{th2} by considering the Markov chain $(X_{1},X_{2})-Y'_{2}-Y'_{3}-Y_{1}$. To deduce the upper bound on $R_{12}$, using Theorem~\ref{th2}, we have:
\noindent \begin{align}
R_{12}&\le I(X_{1};\!Y'_{2} |X_{2},\!Y'_{3} )\!+\!I(X_{2};\!Y_{1} |X_{1},\!Y'_{3} )\!+\!I(Y_{1} ;\!Y'_{2} |X_{1},\!X_{2},\!Y'_{3} )\!+\!I(X_{1};\!Y'_{3}|X_{2},\!U)\!-\!I(X_{1};\!Y'_{3}|U)\no
\\&{\mathop{=}\limits^{{\rm(a)}}I(X_{1};\!Y'_{2} |X_{2},\!Y'_{3} )\!+\!I(X_{1};Y'_{3}|X_{2},U)-I(X_{1};Y'_{3}|U)}\no
\\&{\le I(X_{1};\!Y'_{2} |X_{2},\!Y'_{3} )\!+\!I(X_{1};Y'_{3}|X_{2},U)}\no
\\&{\mathop{=}\limits^{{\rm(b)}} I(X_{1};\!Y'_{2} |X_{2},\!Y'_{3} )\!+\!H(Y'_{3}|X_{2},U)\!-\!H(Y'_{3}|X_{1},X_{2})}\no
\\&{\le I(X_{1};\!Y'_{2} |X_{2},\!Y'_{3} )\!+I(X_{1};Y'_{3}|X_{2})}\no
\\&{=I(X_{1};\!Y'_{2},\!Y'_{3}|X_{2})}\no
\\&{\mathop{=}\limits^{{\rm(c)}}I(X_{1};\!Y'_{2}|X_{2})}\no
\end{align}
where (a) and (c) are deduced from the Markov chain $(X_{1},X_{2})-Y'_{2}-Y'_{3}-Y_{1}$ and (b) from the distribution of $U$. Since the new channel and the original channel have the same marginal distributions, then the outer bound in (\ref{outerex2}) holds for the original channel.

The rate region $(R_{23},R_{12})$ in (\ref{innerex2}) along with the outer bound in (\ref{outerex2}) are shown in Fig. \ref{fig.ex2} for different values of $p_{1},p_{2}$ and $p_{3}$. In addition, the rate region in (\ref{innerex2}) is compared with the rate region obtained from the time sharing between Users 1 and 2.
\begin{figure}
\centering
\includegraphics[width=9cm]{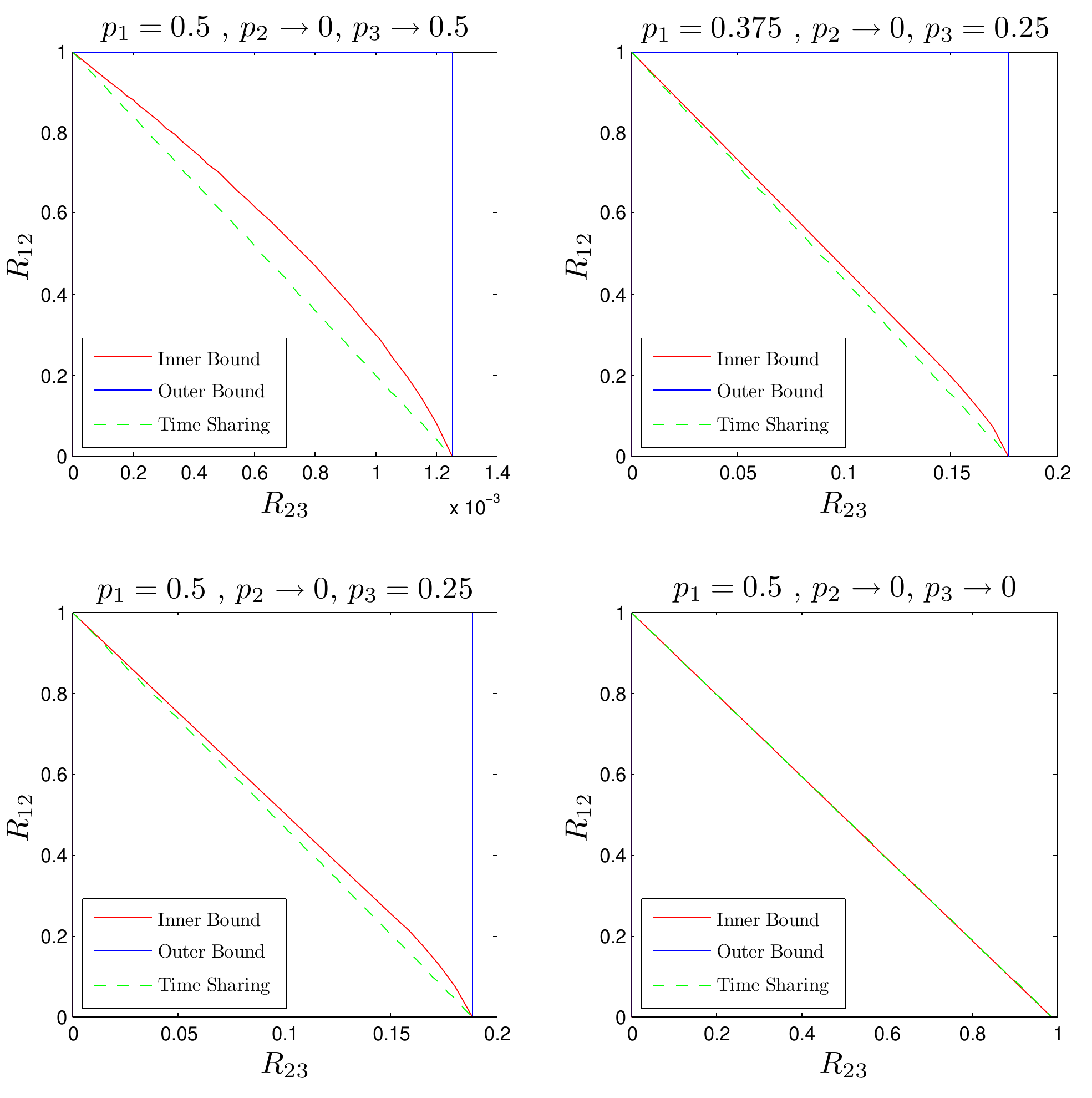}
\caption{\footnotesize{Comparison of the bounds for different values of $p_{1},p_{2}$ and $p_{3}$ in Example 2}
}
\vspace{-.6cm}
\label{fig.ex2}
\end{figure}
\end{example}


\section{The Generalized Scheme of Pairwise Secret Key Agreement over GDMMAC}
\label{secondscheme}
We now analyze the ``generalized key sharing'' scheme, in which the channel outputs allowed by the generalized feedback are used as induced sources for key generation. In contrast to the pre-generated keys scheme in Section~\ref{firstscheme}, the channel outputs at Users 1 and 2 are used as inputs to the encoders and hence, the channel inputs are stochastic functions of not only the pre-generated keys but also the previous channel outputs. The encoding and decoding functions refer to those introduced in Section~\ref{genralmodel}. In the generalized scheme, key sharing is achieved in two stages so that the final key $K_{ij}$ shared between User $i$ and User $j$ for $i<j\in \{1,2,3\}$ is composed of two sub-keys; a randomly pre-generated key randomly produced by User $i$ and another key generated as a stochastic function of the received channel output $Y_i^n$ at user $i$. This procedure is performed over multiple blocks of $n$ channel uses, and the detailed achievable scheme is given in the following subsection.


\subsection{Main Results}
\label{main2}
We first define the following rates:
\begin{align}\no
&{{\bf r}_{{\bf 12,p}} =[I(S_{12} ;X_{2},Y_{2})-I(S_{12};Y_{3} ,S_{13},S_{23},T_{13},T_{23})]^{+} ,}\no
\\&{{\bf r}_{{\bf 21,p}} =[I(S_{21} ;X_{1},Y_{1})-I(S_{21} ;Y_{3} ,S_{13},S_{23},T_{13},T_{23})]^{+} ,}\no
\\&{{\bf I}_{{\bf 12,p}} =I(S_{12};S_{21}|Y_{3} ,S_{13},S_{23},T_{13},T_{23}),}\no
\\&{{\bf r}_{{\bf 13,p}} =[I(S_{13} ;Y_{3}|S_{23})-I(S_{13} ;X_{2},Y_{2},S_{12},T_{12}|S_{23})]^{+} ,}\no
\\&{{\bf r}_{{\bf 23,p}} =[I(S_{23} ;Y_{3}|S_{13})-I(S_{23} ;X_{1},Y_{1},S_{21},T_{21}|S_{13})]^{+} ,}\no
\\&{{\bf I}_{{\bf 3,p}} =I(S_{13} ;S_{23} |Y_{3} )}\no
\\&{{\bf r}_{{\bf 12,s}} =[I(T_{12} ;X_{2},Y_{2}|S_{12},S_{21})-I(T_{12};Y_{3},S_{13},S_{23},T_{13},T_{23}|S_{12},S_{21})]^{+} ,}\no
\\&{{\bf r}_{{\bf 21,s}} =[I(T_{21} ;X_{1},Y_{1}|S_{12},S_{21})-I(T_{21};Y_{3},S_{13},S_{23},T_{13},T_{23}|S_{12},S_{21})]^{+} ,}\no
\\&{{\bf I}_{{\bf 12,s}} =I(T_{12};T_{21}|Y_{3},S_{13},S_{23},T_{13},T_{23},S_{12},S_{21}),}\no
\\&{{\bf r}_{{\bf 13,s}} =[I(T_{13} ;Y_{3}|S_{13},S_{23},T_{23})-I(T_{13} ;X_{2},Y_{2},S_{12},T_{12}|S_{13},S_{23},T_{23})]^{+} ,}\no
\\&{{\bf r}_{{\bf 23,s}} =[I(T_{23} ;Y_{3}|S_{13},S_{23},T_{13})-I(T_{23} ;X_{1},Y_{1},S_{21},T_{21}|S_{13},S_{23},T_{13})]^{+} ,}\no
 \\&{{\bf I}_{{\bf 3,s}} =I(T_{13} ;T_{23} |Y_{3},S_{13},S_{23})}\label{rate-define2}
\end{align}

\begin{theorem}
\label{th3}
In the generalized scheme, all rate triples in the closure of the convex hull of the set of rate triples $(R_{12},R_{13},R_{23})$ that satisfy the following conditions are achievable:

\vspace{-.55cm}
\begin{align}
&{R_{12} \ge 0,R_{13} \ge 0,R_{23} \ge 0,}\no
\\& {R_{12} \le [{\bf r}_{{\bf 12,p}}+{\bf r}_{{\bf 21,p}}-{\bf I}_{{\bf 12,p}}]^{+}+[{\bf r}_{{\bf 12,s}}+{\bf r}_{{\bf 21,s}}-{\bf I}_{{\bf 12,s}}]^{+},}\no
\\&{R_{13} \le {\bf r}_{\bf 13,p}+{\bf r}_{\bf 13,s},} \no
\\&{R_{23} \le {\bf r}_{\bf 23,p}+{\bf r}_{\bf 23,s},}\no
\\&{R_{13} +R_{23}\le [{\bf r}_{\bf 13,p} {\bf +r}_{\bf 23,p}-{\bf I}_{\bf 3,p}]^{+}+[{\bf r}_{\bf 13,s} {\bf +r}_{\bf 23,s}-{\bf I}_{\bf 3,s}]^{+},}\no
\end{align}
\vspace{.01cm}
\noindent for random variables taking values in finite sets and with joint distribution factorizing as:
\vspace{-.4cm}
\begin{align}
&{p(\!s_{12},\!s_{13},s_{21},\!s_{23},t_{12},\!t_{13},t_{21},\!t_{23},\!x_{1},\!x_{2},\!y_{1} ,\!y_{2} ,\!y_{3})\!=\!p(\!s_{12})p(\!s_{13})p(\!s_{21})p(\!s_{23})p(\!x_{1}|s_{12},\!s_{13})p(\!x_{2}|s_{21},\!s_{23})}\no
\\&{p(\!y_{1},\!y_{2},\!y_{3}|x_{1} ,\!x_{2}\!)p(\!t_{12}|x_{1},\!y_{1},\!s_{12})p(t_{13}|x_{1},\!y_{1},s_{13})p(\!t_{21}|x_{2},\!y_{2},\!s_{21})p(t_{23}|x_{2},\!y_{2},s_{23})}\label{distri} \end{align}
and subject to the constraints:
\begin{align}
&{I(T_{12};X_{1},Y_{1}|X_{2},Y_{2},S_{12},S_{21}) \le I(S_{12};X_{2},Y_{2}),}\no
\\&{I(T_{13};X_{1},Y_{1}|Y_{3},S_{13},S_{23},T_{23}) \le I(S_{13};Y_{3}|S_{23}),}\no
\\&{I(T_{21};X_{2},Y_{2}|X_{1},Y_{1},S_{12},S_{21}) \le I(S_{21};X_{1},Y_{1}),}\no
\\&{I(T_{23};X_{2},Y_{2}|Y_{3},S_{13},S_{23},T_{13}) \le I(S_{23};Y_{3}|S_{13}),}\no
\\&{I(T_{13},T_{23};X_{1},Y_{1},X_{2},Y_{2}|Y_{3},S_{13},S_{23})\le I(S_{13},S_{23};Y_{3})}\label{constraints}
\end{align}
\end{theorem}

The proof of Theorem~\ref{th3} is given in Appendix~\ref{App1}. Note that each individual rate bound consists of two parts: a primary rate (denoted by subscript ``p'') and a secondary rate (denoted by subscript ``s''). This split reflect the two-step key generation process behind the achievability proof. The primary rates are associated with the pre-generated keys randomly generated and sent by Users 1 and 2 through the channel, as in Theorem \ref{th1}. The secondary rates are generated by Users 1 and 2 after receiving the channel outputs, which are exploited as induced sources to generate additional keys. Intuitively, the form of the bound for $R_{12}$ originates from the two-way channel between Users 1 and 2 in which User 3 acts as an external eavesdropper. Similarly, the form of the bounds for $R_{13}$ and $R_{23}$ stems from the generalized MAC from Users 1 and 2 to User 3, in which Users 1 and 2 eavesdrop each other. For the secondary keys, a combination of secret sharing codebooks and superposition coding is used at Users 1 and 2 as well as random binning based on Probability Mass Function (PMF) approximation arguments \cite{yassaee}. Finally, note that the constraints in \eqref{constraints} reflect the absence of public channel, sot that all the required information to reconstruct the secondary keys should be sent through noisy channels.

This procedure is performed over multiple blocks. At each block, by $n$ uses of the channel, each of Users 1 and 2 encodes pre-generated keys (primary keys) plus the secondary keys induced from the channel outputs received at the end of the previous block. In Theorem \ref{th3}, $S_{ij}$ and $T_{ij}$ are the auxiliary random variables relevant to the primary and secondary keys $K_{ij,p}$ and $K_{ij,s}$, respectively, which are generated by User $i$ to be shared with User $j$ where $i\in \{1,2\}$, $j\in \{1,2,3\}$ and $i\neq j$.

\begin{figure}
\centering
\includegraphics[width=15cm]{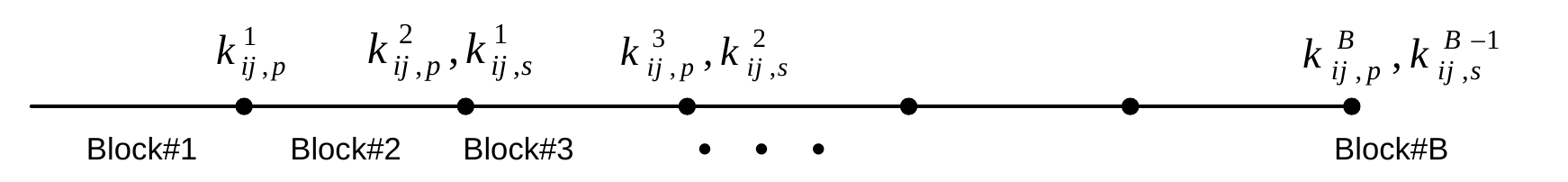}
\caption{\footnotesize{Key sharing scheme associated with Theorem \ref{th3}}
}
\vspace{-.5cm}
\label{fig_sim61}
\end{figure}

A simple illustration of the encoding is shown in Fig. \ref{fig_sim61}.
The primary key $k_{ij,p}^{1}$ is randomly generated to be shared between Users $i$ and $j$ in the first block of $n$ channel uses. At the end of the first block, by receiving the corresponding outputs, the secondary key of the first block $k_{ij,s}^{1}$ is generated as a stochastic function of the received outputs to be shared between Users $i$ and $j$. The required information of the secondary key is sent through $n$ channel uses in the second block along with the required information of the second block primary key, i.e., $k_{ij,p}^{2}$ which is generated randomly and independently of the first block secondary key. The constraints in \eqref{constraints} reflect this fact. In this scheme, the secondary key $k_{ij,s}^{1}$ is decoded and shared between Users $i$ and $j$ at the end of the second block in addition to the primary key of the second block $k_{ij,p}^{2}$. The same procedure is performed $B$ blocks. Assuming $r_{ij,p}$ and $r_{ij,s}$ be the rates of the primary and secondary keys, respectively, the total rate of the shared key between Users $i$ and $j$ at the end of block $B$ is:
$$\bar{R}_{ij}=\frac{nBr_{ij,p}+n(B-1)r_{ij,s}}{nB}$$
which is approximately equal to $r_{ij,p}+r_{ij,s}$ if $B$ is large enough.

\begin {rem}
The auxiliary random variables $S_{12}$ and $S_{13}$ (resp. $S_{21}$ and $S_{23}$) could be made dependent in~\eqref{distri}. If this were the case, additional constraints should be added to \eqref{constraints} according to Marton's bound for the broadcast channel from User 1 to Users 2 and 3 (resp. from User 2 to Users 1 and 3). For the sake of simplicity, we assume that $S_{12}$ and $S_{13}$ (resp. $S_{21}$ and $S_{23}$) are independent.
\end {rem}

\begin {rem}
If we cancel the secondary keys generation by setting $T_{12}=T_{13}=T_{21}=T_{23}=\emptyset$ in Theorem \ref{th3}, then the rate region  reduces to the one in Theorem \ref{th1}.
\end {rem}

\begin{theorem}
\label{th4}
In the generalized scheme, if a
key rate triple is achievable, then it belongs
to the set of all rate triples $(R_{12},R_{13},R_{23})$ that satisfy:
\[\begin{array}{l} {0\leq R_{12} \le I(X_{1};\!Y_{2}|\!X_{2},\!Y_{3} )\!+\!I(X_{2};\!Y_{1}|\!X_{1},\!Y_{3} )\!+\!I(Y_{1} ;\!Y_{2}|\!X_{1},\!X_{2},\!Y_{3} )\!+\!I(X_{1};\!Y_{3}|\!X_{2},\!U)\!-\!I(X_{1};\!Y_{3}|U),}
\\ {0\leq R_{13} \le I(X_{1},Y_{1} ;Y_{3}|X_{2},Y_{2} ) ,}
\\ {0\leq R_{23} \le I(X_{2},Y_{2} ;Y_{3}|X_{1},Y_{1} ),} \end{array}\]
for random variables $U,X_1,X_2,Y_1,Y_2,Y_3$, all taking values in finite sets, such that $U-(X_1,X_2)-(Y_1,Y_2,Y_{3})$ forms a Markov chain.
\end{theorem}
\begin{IEEEproof}
See Appendix~\ref{App5}.
\end{IEEEproof}

\subsection{A Binary Example}
\label{spec2}
%


In this section, we discuss a binary example to illustrate the benefits of the generalized scheme. To clarify the effect of involving the channel outputs in the pairwise key sharing, we modify Example~\ref{ex:example_2} in such a way that the binary noises received over the channel are correlated and hence, given the channel inputs, the channel outputs can be considered as correlated sources.

%
%
%

\begin{example}
Consider a binary GDMMAC where the relationships between the channel inputs and outputs are according to:

\begin{align}
&{Y_{1} =X_{1}+X_{2}+Z_{1}+Z_{2}+Z_{3},}\no\\
&{Y_{2} =X_{1}+X_{2}+Z_{2},}\no\\
&{Y_{3} =X_{1}+X_{2}+Z_{2}+Z_{3},}\no
\end{align}

\noindent in which $Z_{1}$, $Z_{2}$ and $Z_{3}$ are binary random variables with distributions $\Pr(Z_{i} =1)=p_{i}$ where $0<p_{i}\le0.5$ for $i=1,2,3$. Operation $+$ is binary summation and the random variables $(X_{1},X_{2},Z_{1},Z_{2},Z_{3})$ are independent of each other. In this example, there is a physical degradedness in the channel and the Markov chain $(X_{1},X_{2})-Y_{2}-Y_{3}-Y_{1}$ holds between the channel inputs and outputs. Since the received noises over the channel are not independent, the channel outputs have the role of correlated sources to produce the secondary keys. In particular, because of the Markov chain $Y_{2}-Y_{3}-Y_{1}$ between the channel outputs, Users 1 and 2 can, respectively, share secondary keys $K_{13,s}$ and $K_{23,s}$ with User 3 but they cannot share any secondary key with each other, i.e., $K_{12,s}=K_{21,s}=\emptyset$. Since the required information of secondary key $K_{13,s}$ should be sent through the channel form User 1 to User 3, we can not set $S_{13}=\emptyset$ as in Example 2. We change the auxiliary random variables of the pre-generated keys scheme in Example 2 such that $X_{1}=S_{12}+S_{13}$ where $\Pr(S_{12}=1)=\alpha$ and $\Pr(S_{13}=1)=\alpha'$. The auxiliary random variables $S_{21}$ and $S_{23}$ are substituted the same as in Example 2, i.e., $S_{21}\!=\emptyset$ and $\Pr\{S_{23}\!=\!X_{2}\}=1-\beta,$ where $X_{2}=Ber(0.5)$.

For the secondary key generation, Users 1 and 2, respectively, consider auxiliary random variables $T_{13}$ and $T_{23}$ to share secret keys with User 3 and set $T_{12}=T_{21}=\emptyset$ since they can not share a secondary key between themselves due to Markov chain $Y_{2}-Y_{3}-Y_{1}$. According to the constraints in (\ref{constraints}), noisy versions of the channel outputs at Users 1 and 2 are considered as auxiliary random variables of the secondary keys, i.e., $T_{13}=Y_{1}+Z'_{1},T_{23}=Y_{2}+Z'_{2}$ where $Z'_{1}$ and $Z'_{2}$ are independently binary noises such that $\Pr(Z'_{1}=1)=\alpha''$ and $\Pr(Z'_{2}=1)=\beta'$.
By substituting the auxiliary random variables in Theorem \ref{th3} as described, the following key rate region is achievable:
\begin{align}
&{R_{12} \le [h_{x}-h_{y}-h(\alpha'\ast p_{2})+h(\alpha\ast\alpha'\ast p_{2})]^{+}},\no
\\&{R_{13} \le h(p_{1}\ast p_{3}\ast\alpha'')-h(p_{1}\ast\alpha'')},\no
\\&{R_{23} \le [h(\beta\ast p_{1}\ast p_{2}\ast p_{3})-h(\alpha\ast\beta\ast p_{2}\ast p_{3})]^{+}+}\no
\\& {\ \ \ \ \ \ \ \ [h_{z}-h_{y}+h(\alpha\ast\beta\ast p_{2}\ast p_{3})-h(\beta\ast p_{1}\ast p_{2}\ast p_{3})]^{+}}\label{binaryex32}
\end{align}
subject to:
\begin{align}
&{h(\alpha''\ast p_{1})-h(\alpha'')\le h(\alpha\ast\alpha'\ast\beta\ast p_{2}\ast p_{3})-h(\alpha\ast\beta\ast p_{2}\ast p_{3})},\no
\\&{h(\alpha''\ast p_{1})-h(\alpha'')+h_{y}-h(\beta')\le 1},\no
\end{align}
where $h_{x}$,$h_{y}$ and $h_{z}$ are defined as:
\begin{align}
&{h_{x}=f(\beta\ast p_{2},p_{3},\beta'),}\no
\\&{h_{y}=f(\alpha\ast\beta\ast p_{2},p_{3},\beta'),}\no
\\&{h_{z}=f(\beta\ast p_{2},p_{1}\ast p_{3},\beta'),}\no
\\&{f(a,b,c)=-(abc+\overline{a}\overline{b}\overline{c})\log (abc+\overline{a}\overline{b}\overline{c})}\no
\\&{-(a\overline{b}c+\overline{a}b\overline{c})\log (a\overline{b}c+\overline{a}b\overline{c})}\no
\\&{-(ab\overline{c}+\overline{a}\overline{b}c)\log(ab\overline{c}+\overline{a}\overline{b}c)}\no
\\&{-(a\overline{b}\overline{c}+\overline{a}bc)\log(a\overline{b}\overline{c}+\overline{a}bc)},\no
\end{align}
for all values of $0\leq \alpha,\alpha',\alpha'',\beta,\beta', \leq0.5$.

In the above equations $\overline{p}=1-p$ and, operations $*$ and $h(.)$ are defined the same as in \eqref{stardef} and \eqref{entdef}.

\indent If we substitute $S_{13}=T_{13}=T_{23}=\emptyset$ in Example 3 or equivalently $\alpha''=\beta'=0.5$ and $\alpha'=0$ in rate region (\ref{binaryex32}), then the rate region of the pre-generated keys scheme is deduced as:
\begin{align}
&{R_{12} \le [h(\beta\ast p_{2}\ast p_{3})-h(\alpha\beta\ast p_{2}\ast p_{3})-h(\alpha'\ast p_{2})+h(\alpha\ast\alpha'\ast p_{2})]^{+}},\no
\\&{R_{13}=0},\no
\\&{R_{23} \le [h(\beta\ast p_{1}\ast p_{2}\ast p_{3})-h(\alpha\ast\beta\ast p_{2}\ast p_{3})]^{+}}.\label{binaryex31}
\end{align}

The following region is an outer bound of the secret-key capacity region for the generalized scheme in Example 3:
\begin{align}
&{R_{12} \le 1- h(p_{2})},\no
\\&{R_{13} \le h(p_{1}\ast p_{3})-h(p_{1})},\no
\\&{R_{23} \le h(p_{1}\ast p_{3})-h(p_{3})}\label{binaryex3out}
\end{align}
which is directly deduced from Theorem \ref{th4} since we have:
\begin{align*}
  R_{12} &\le I(X_{1}; Y_{2} |X_{2}, Y_{3} ) + I(X_{2}; Y_{1} |X_{1}, Y_{3} ) + I(Y_{1} ; Y_{2} |X_{1}, X_{2}, Y_{3} ) + I(X_{1}; Y_{3} |X_{2}, U) - I(X_{1}; Y_{3}|U),\\
   &\mathop{=}\limits^{(a)}I(X_{1}; Y_{2} |X_{2}, Y_{3} ) + I(X_{1};Y_{3}|X_{2},U)-I(X_{1};Y_{3}|U)
  \\ &\le I(X_{1}; Y_{2} |X_{2}, Y_{3} ) + I(X_{1};Y_{3}|X_{2},U),
  \\ &= I(X_{1}; Y_{2} |X_{2}, Y_{3} ) + H(Y_{3}|X_{2},U)-H(Y_{3}|X_{1},X_{2},U)
  \\ &\le I(X_{1}; Y_{2} |X_{2}, Y_{3} ) + H(Y_{3}|X_{2})-H(Y_{3}|X_{1},X_{2},U),
  \\ &\mathop{=}\limits^{(b)}I(X_{1}; Y_{2} |X_{2}, Y_{3} ) + H(Y_{3}|X_{2})-H(Y_{3}|X_{1},X_{2}),
  \\ &=I(X_{1}; Y_{2} |X_{2}, Y_{3} ) + I(X_{1};Y_{3}|X_{2})=I(X_{1}; Y_{2}, Y_{3}|X_{2}),
  \\ &\mathop{=}\limits^{(c)}I(X_{1}; Y_{2}|X_{2})\le 1-h(p_{2}),
\end{align*}
\begin{align*}
 R_{23} \le I(X_{2},Y_{2} ;Y_{3}|X_{1},Y_{1}) &=H(Y_{3}|X_{1},Y_{1})-H(Y_{3}|X_{1},Y_{1},X_{2},Y_{2})
\\ &=H(X_{2}+Z_{2}+Z_{3}|X_{2}+Z_{1}+Z_{2}+Z_{3})-H(Z_{2}+Z_{3}|Z_{2},Z_{1}+Z_{2}+Z_{3})
\\ &=H(X_{2}+Z_{2}+Z_{3}|X_{2}+Z_{1}+Z_{2}+Z_{3})-H(Z_{3}|Z_{1}+Z_{3})
\\ &\le h(p_{1})-H(Z_{3}|Z_{1}+Z_{3})=h(p_{1}\ast p_{3})-h(p_{3}).
\end{align*}
and
\begin{align*}
  R_{13} \le I(X_{1},Y_{1} ;Y_{3}|X_{2},Y_{2} ) \mathop{=}\limits^{(d)}I(Y_{1} ;Y_{3}|Y_{2})=h(p_{1}\ast p_{3})-h(p_{1}).
\end{align*}

In the above equations, (a), (c) and (d) are inferred from the Markov chain $(X_{1},X_{2})-Y_{2}-Y_{3}-Y_{1}$ and (b) is deduced from the distribution of $U$.

The key rate regions of the pre-generated keys and the generalized schemes in (\ref{binaryex31}) and (\ref{binaryex32}) along with the outer bound in (\ref{binaryex3out}) are compared in Fig. \ref{twoschemes} for different values of the  noises. In order to clarify the regions, we projected each 3-D region into three 2-D regions. It is seen that the generalized scheme strictly has a better performance compared to the pre-generated keys scheme. By using the pre-generated keys scheme, we can only achieve two non-zero secret-key rates while using the generalized scheme, we attain all three non-zero rates and obtain rate regions which are significantly larger compared to the former scheme.

\begin{figure}[htp]

\subfloat[$p_{1}=0.09,p_{2}=0.1,p_{3}=0.07$]{%
\includegraphics[width=16cm]{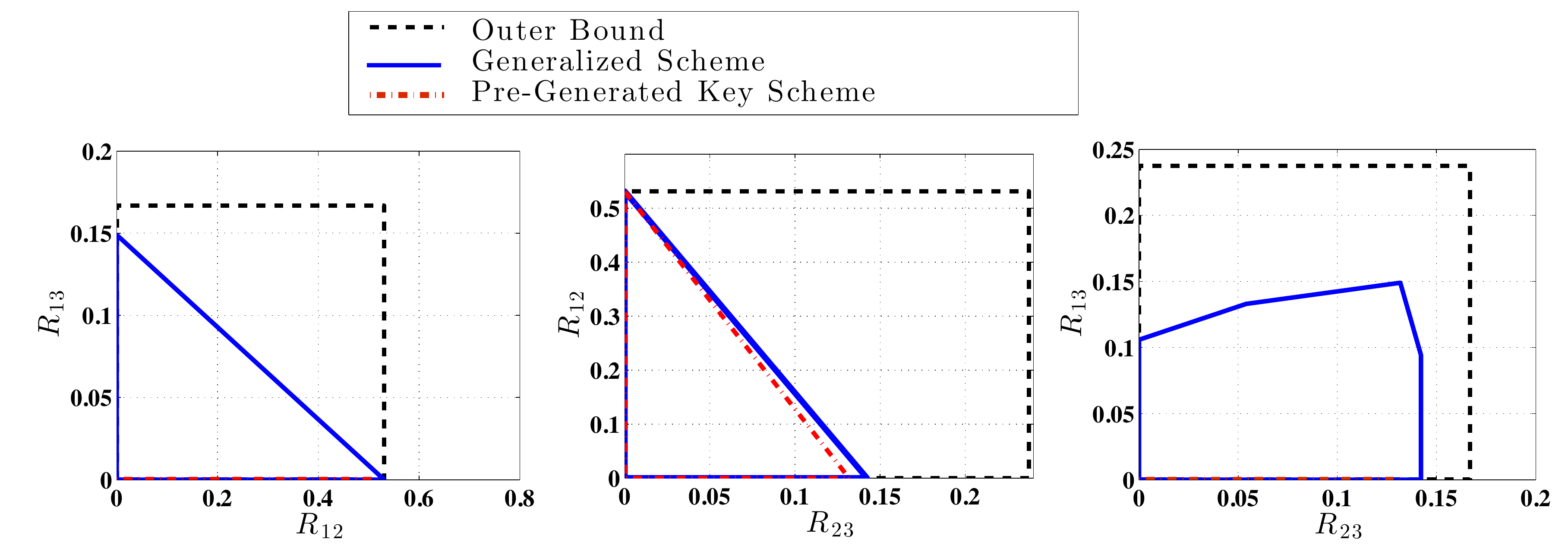}%
}

\subfloat[$p_{1}=0.01,p_{2}=0.02,p_{3}=0.01$]{%
\includegraphics[width=16cm]{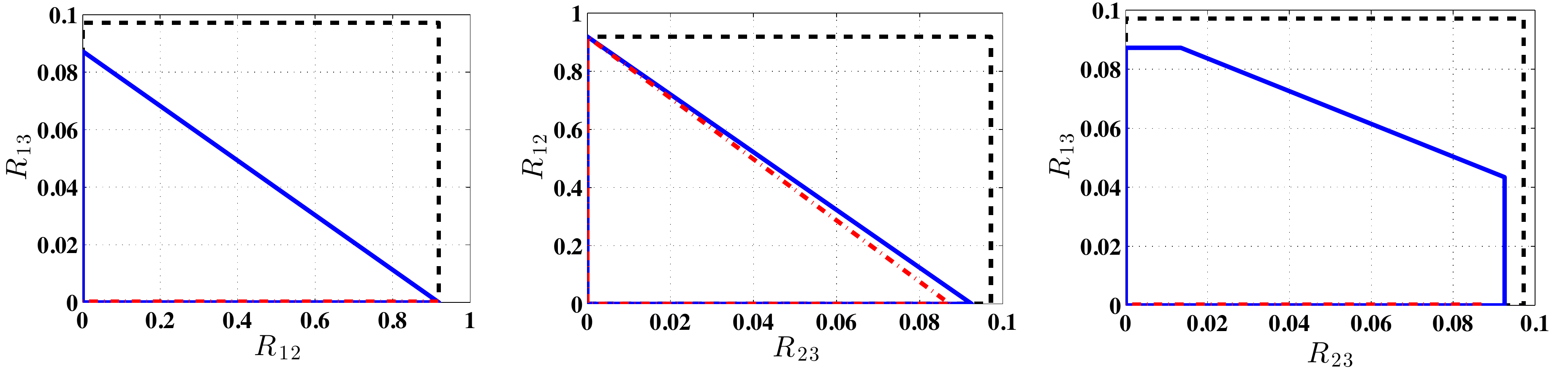}%
}

\subfloat[$p_{1}=0.03,p_{2}=0.05,p_{3}=0.02$]{%
\includegraphics[width=16cm]{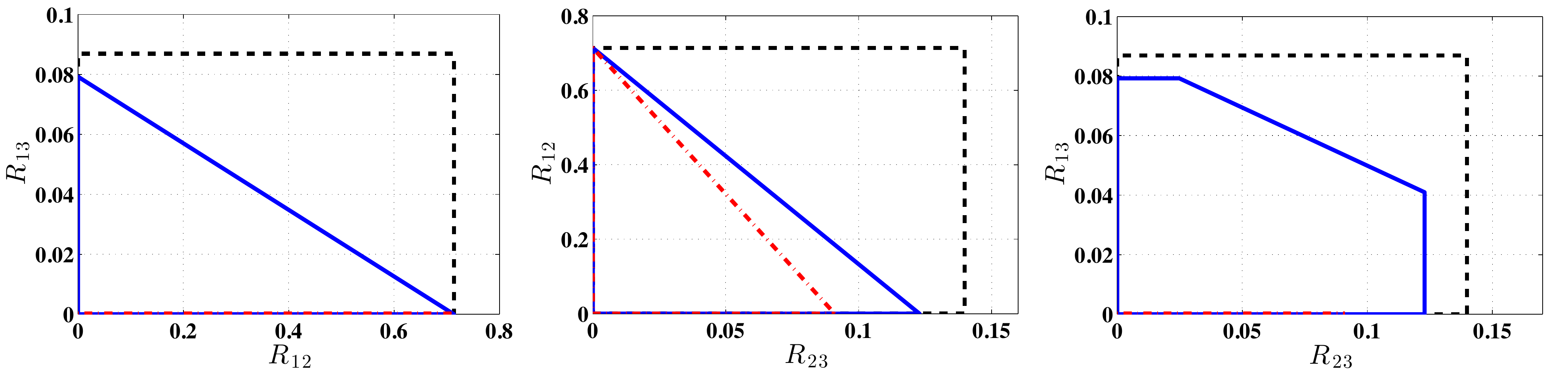}%
}
\caption{Comparison of the two schemes and the outer bound for Binary Example 3}
\label{twoschemes}
\end{figure}
\end{example}

\appendices
\section{Proof of the inner bound in Theorem \ref{th3} (including proof of Theorem \ref{th1})}
\label{App1}
We fix the distribution of all random variables involved in the coding scheme as defined Theorem \ref{th3}. Key sharing is performed over $B$ blocks, each comprised of $n$ uses of the channel. In what follows, we describe the code construction, the associated encoding and decoding, and the security analysis, in the $i$-th block. The boldface random variable $\textbf{X}$ (resp. $\textbf{x}$) denotes $n$ repetitions of random variable $X$, i.e., $X^{n}$ (resp. its realization $x$). $\textbf{X}^{i}$ (resp. $\textbf{x}^{i}$) denotes $X_{(i-1)n+1}^{in}$ (resp. $x_{(i-1)n+1}^{in}$), i.e, $n$ successive repetitions of random variable $X$ associated with block $i$ (resp. its successive realizations). $\textbf{X}^{i:j}$ denotes $X_{(i-1)n+1}^{jn}$ and correspondingly, $\textbf{x}^{i:j}$ denotes $n(j-i+1)$ realizations of $X$ from block $i$ to $j$.

\subsection*{Code Construction}
User $j$ independently generates $2^{n(\!r_{jl,p}\!+\!r'_{jl,p}\!)}$ codewords $\textbf{s}_{jl}$, for $j<l\in\{1,2,3\}$ according to the i.i.d. distribution $\prod_{i=1}^{n}p(s_{jl,i})$ which are labeled as:
\[\begin{array}{l}
\hspace{-.2cm}{\textbf{s}_{jl}(k_{jl,p},\!k'_{jl,p})\!,\!k_{jl,p}\!\in \!\mathcal K_{jl,p}\!=\!\{1,\!...,\!2^{nr_{jl,p}}\}\!,\!k'_{jl,p}\!\in\!\mathcal K'_{jl,p}\!=\!\{1,\!...,\!2^{nr'_{jl,p}}\}.}
\end{array}\]
Each sequence $\textbf{s}_{jl}$ can be determined if the indices $(k_{jl,p},k'_{jl,p})$ are known. The index $k_{jl,p}$ represents the pre-generated key to be shared between Users $j$ and $l$ while $k'_{jl,p}$ is a randomization index drawn from the local randomness.

Moreover, User $j$ chooses sequences $\textbf{t}_{jl}$ according to i.i.d. distribution $\prod_{i=1}^{n}p(t_{jl,i})$ and, it randomly and independently bins all sequences $\textbf{t}_{jl}$ as follows:
\begin{itemize}
  \item index $\phi_{jl}(\textbf{t}_{jl})$ is uniformly generated over $[1,\!2^{nr_{jl}}]$. We set $k_{jl,s}=\phi(\textbf{t}_{jl})$ as the secondary key to be shared between Users $j$ and $l$.
  \item index $\psi_{jl}(\textbf{t}_{jl})$ is uniformly generated over $[1,\!2^{nr'_{jl}}]$. We set $k'_{jl,s}=\psi(\textbf{t}_{jl})$ as the index sent to User $l$ such that it can reconstruct $\textbf{T}_{jl}$.
  \item index $\theta_{jl}(\textbf{t}_{jl})$ is uniformly generated over $[1,\!2^{nr''_{jl}}]$. We set $k''_{jl,s}=\theta(\textbf{t}_{jl})$ as the index also used by User $l$ to reconstruct $\textbf{t}_{jl}$. We assume that a specific index $k''_{jl,s}$ is publicly shared between Users $j$ and $l$ ahead of time.
\end{itemize}
Note that, unlike traditional source models for key generation, there is no public channel over which to transmit the index $k'_{jl,s}$. To transmit the index over the noisy channel, we will have to explicitly define how to encode it into a codeword $\textbf{s}_{jl}$. To this end, for $j\in \{1,2\}$, $l\in \{1,2,3\}$, $j\neq l$, we define functions
\[\begin{array}{l}
{{f_{jl}:{\tilde{\mathcal S}_{jl}}\to{\rm{\mathcal K}}'_{jl,s}},}\end{array}\]
where ${\tilde{\mathcal S}_{jl}}$ is the set of $2^{n(r_{jl,p}\!+\!r'_{jl,p})}$ indices pairs $(\!k_{jl,p} ,k'_{jl,p}\!)$. Each $f_{jl}$ is a random partitioning of ${\tilde{\mathcal S}_{jl}}$ into $2^{nr'_{jl,s}}$ equal-sized parts. Elements of part $i$ are labeled as $(\!\tilde{{\mathcal S}}_{jl}\!)_{i}$. We implicitly assume that the following condition holds
\begin{align}
{r'_{jl,s}<(r_{jl,p}+r'_{jl,p})}\label{necessary cond}
\end{align}
Later on, It will be seen that this assumption holds because of the rate constraints (\ref{constraints}) in Theorem \ref{th3}.

\subsection*{Encoding in block $i$}
At the beginning of block $i$, we assume that $k_{jl,s}^{''i-1}$ is chosen uniformly at random from its corresponding set and is publicly shared between Users $j$ and $l$. We will later show that under specific conditions, the users can agree on specific values of $k_{jl,s}^{''i-1}$ ahead of time so that this assumption can be dropped. Sequences $\textbf{y}_{1}^{1:i-1},\textbf{x}_{1}^{1:i-1},\textbf{s}_{12}^{1:i-1},\textbf{s}_{13}^{1:i-1},\textbf{t}_{12}^{1:i-2},\textbf{t}_{13}^{1:i-2}$ are available at User 1, who can also decode sequences $\textbf{s}_{21}^{1:i-1}\textbf{t}_{21}^{1:i-2}$.

User 1 then generates sequence $\textbf{t}_{12}^{i-1}$ according to distribution $P_{\textbf{T}_{12}|\textbf{Y}_{1}\textbf{X}_{1}\textbf{S}_{12}K''_{12,s}}$, and sequence $\textbf{t}_{13}^{i-1}$ according to distribution $P_{\textbf{T}_{13}|\textbf{Y}_{1}\textbf{X}_{1}\textbf{S}_{13}K''_{13,s}}$. Similarly, User 2 generates sequences $\textbf{t}_{21}^{i-1}$ and $\textbf{t}_{23}^{i-1}$. Subsequently, User $j$ computes the secondary key of block $i-1$ for sharing with User $l$ as $k_{jl,s}^{i-1}=\phi_{jl}(\textbf{t}_{jl}^{i-1})$ and the index to be sent to User $l$ to reconstruct $\textbf{t}_{jl}^{i-1}$ as $k_{jl,s}^{'i-1}=\psi_{jl}(\textbf{t}_{jl}^{i-1})$. As stated earlier $k_{jl,s}^{''i-1}=\theta_{jl}(\textbf{t}_{jl}^{i-1})$ is already shared between Users $j$ and $l$.

User $j$ encodes $k_{jl,s}^{'i-1}$ in such a way that he finds the respective part $(\tilde{{\mathcal S}}_{jl})_{k_{jl,s}^{'i-1}}$ according to partitioning function $f_{jl}$, and then it randomly chooses a pair $(k_{jl,p},k'_{jl,p})$ from $(\tilde{{\mathcal S}}_{jl})_{k_{jl,s}^{'i-1}}$.
For the selected $(k_{jl,p},k'_{jl,p})$, User $j$ picks up sequence $\textbf{s}_{jl}^{i}(k_{jl,p},k'_{jl,p})$.
Then, User $j$ selects the respective index $k_{jl,p}$ of $\textbf{s}_{jl}^{i}$ as the primary key of block $i$ for sharing with User $l$, i.e., $k_{jl,p}^{i}$. In this way, Users 1 and 2 choose $(\textbf{s}_{12}^{i},\textbf{s}_{13}^{i})$ and $(\textbf{s}_{21}^{i},\textbf{s}_{23}^{i})$, respectively. The channel inputs of block $i$ are sent over the GDMMAC according to the distributions $p(\!x_{1}|s_{12},s_{13})$ and $p(\!x_{2}|s_{21},s_{23})$ by Users 1 and 2, respectively, through $n$ channel uses.

\subsection*{Decoding of block $i$}
We let $\mathcal{T}_{\epsilon'}^{n}(\!P_{X}\!)$ denote the set of $\epsilon'-$strongly typical sequences $x^{n}$ with respect to distribution $p(\!x)$. At the end of block $i$, Users 1, 2 and 3 receive $\textbf{y}_{1}^{i},\textbf{y}_{2}^{i}$ and $\textbf{y}_{3}^{i}$ through the channel, respectively. With access to $(\textbf{x}_{1}^{i},\textbf{y}_{1}^{i})$, User 1 declares an error unless there exists a unique $\textbf{s}_{21}^{i}(k_{21,p}^{i},k_{21,p}^{'i})$ such that $(\textbf{x}_{1}^{i},\textbf{y}_{1}^{i},\textbf{s}_{21}^{i})\!\!\in\!\!\mathcal{T}_{\epsilon_{1}}^{n}(\!P_{X_{1},Y_{1},S_{21}}\!)$. User 2 acts in the symmetric way and decodes $\textbf{s}_{12}^{i}$. With access to $\textbf{y}_{3}^{i}$, User 3 declares an error unless there exists a unique pair $(\textbf{s}_{13}^{i}(k_{13,p}^{i},k_{13,p}^{'i}),\textbf{s}_{23}^{i}(k_{23,p}^{i},k_{23,p}^{'i}))$ such that $(\textbf{y}_{3}^{i},\textbf{s}_{13}^{i},\textbf{s}_{23}^{i})\!\!\in\!\!\mathcal{T}_{\epsilon_{1}}^{n}(\!P_{Y_{3},S_{13},S_{23}}\!)$.

After decoding the primary keys of block $i$, users decode the secondary keys of block $i-1$. In particular, using function $f_{21}$, User 1 finds mapping $({\tilde{\mathcal S}_{21})}_{q_{21}}$ of the pair indices $(k_{21,p}^{i},k_{21,p}^{'i})$ related to the decoded $\textbf{s}_{21}^{i}$ and sets $k_{21,s}^{'i-1}=q_{21}$. Furthermore, as we assumed at the beginning of the encoding step, index $k_{21,s}^{''i-1}$ is available at User 1. With access to indices $k_{21,s}^{'i-1}$ and $k_{21,s}^{''i-1}$, and the sequences $(\textbf{x}_{1}^{i-1},\textbf{y}_{1}^{i-1},\textbf{s}_{12}^{i-1},\textbf{s}_{21}^{i-1})$, User 1 decodes sequence $\textbf{t}_{21}^{i-1}$ if $(\textbf{t}_{21}^{i-1}(k_{21,s}^{i-1},k_{21,s}^{'i-1},k_{21,s}^{''i-1}),\textbf{x}_{1}^{i-1},\textbf{y}_{1}^{i-1},\textbf{s}_{12}^{i-1},\textbf{s}_{21}^{i-1})\in \mathcal{T}_{\epsilon_{2}}^{n}(P_{T_{21},X_{1},Y_{1}|S_{12},S_{21}}),$ when such $\textbf{t}_{21}^{i-1}$ exists and is unique. Otherwise, it declares an error. User 2 exploits mapping $f_{12}$ to find $k_{12,s}^{'i-1}$ and decodes $\textbf{t}_{12}^{i-1}$ in the symmetric way. User 3 uses mappings $f_{13}$ and $f_{23}$ to find $k_{13,s}^{'i-1}$ and $k_{23,s}^{'i-1}$. With access to the indices $k_{13,s}^{'i-1},k_{23,s}^{'i-1}$, and the shared indices $k_{13,s}^{''i-1},k_{23,s}^{''i-1}$ and the sequences $(\textbf{y}_{3}^{i-1},\textbf{s}_{13}^{i-1},\textbf{s}_{23}^{i-1})$, User 3 decodes sequence pair $(\textbf{t}_{13}^{i-1},\textbf{t}_{23}^{i-1})$ if $(\textbf{t}_{13}^{i-1}(k_{13,s}^{i-1},k_{13,s}^{'i-1},k_{13,s}^{''i-1}),\textbf{t}_{23}^{i-1}(k_{23,s}^{i-1},k_{23,s}^{'i-1},k_{23,s}^{''i-1}),\textbf{y}_{3}^{i-1},\textbf{s}_{13}^{i-1},\textbf{s}_{23}^{i-1})\in \mathcal{T}_{\epsilon_{2}}^{n}(P_{T_{13},T_{23},Y_{3}|S_{13},S_{23}}),$ when such pair $(\textbf{t}_{13}^{i-1},\textbf{t}_{23}^{i-1})$ exists and is unique. Otherwise, it declares an error.

\subsection*{Reliability analysis}
\label{sec:reliability-analysis}
If we set:
\begin{align}
&{r_{12,p} +r'_{12,p}<I(S_{12};X_{2},Y_{2}) }\label{dec1}
\\&{r_{21,p} +r'_{21,p}<I(S_{21};X_{1},Y_{1})}\label{dec2}
\\&{r_{13,p} +r'_{13,p}<I(S_{13};Y_{3}|S_{23})}\label{dec3}
\\&{r_{23,p} +r'_{23,p}<I(S_{23};Y_{3}|S_{13})}\label{dec4}
\\&{r_{13,p} +r'_{13,p}+r_{23,p} +r'_{23,p}<I(S_{13},S_{23};Y_{3})}\label{dec5}
\end{align}
and if we choose $\epsilon_{1}=\frac{\epsilon}{16B}$, it can be shown with standard arguments that the average of the primary decoding error probabilities $P_{ej,p}^{(n)}$ at User $j$ are bounded by
\begin{align}
&{E(P_{e1,p}^{(n)})\le 2\epsilon _{1}=\frac{\epsilon}{8B}}\label{n.12}
\\&{E(P_{e2,p}^{(n)})\le 2\epsilon _{1}=\frac{\epsilon}{8B}}\label{n.13}
\\&{E(P_{e3,p}^{(n)})\le 4\epsilon _{1}=\frac{\epsilon}{4B}}\label{n.14}
\end{align}
for $n$ sufficiently large. In \eqref{n.12}-\eqref{n.14}, expectation is taken over the randomly generated code and all the binning functions where the error probabilities are conditioned to them. Notice that \eqref{dec1}-\eqref{dec2} reflect the point-to-point nature of the channel between Users 1 and 2, while \eqref{dec3}-\eqref{dec5} reflect the multiple access nature of the channel to User 3.

The analysis of the error probability for the secondary keys requires slightly more care. Note that the induced probability distribution by the encoding scheme is:
\begin{align}
&{\tilde{P}_{K_{12,s}K'_{12,s}K''_{12,s}K_{13,s}K'_{13,s}K''_{13,s}K_{21,s}K'_{21,s}K''_{21,s}K_{23,s}K'_{23,s}K''_{23,s}\mathbf{T}_{12}\mathbf{T}_{13}\mathbf{T}_{21}\mathbf{T}_{23}\mathbf{X}_{1}\mathbf{X}_{2}\mathbf{Y}_{1}\mathbf{Y}_{2}\mathbf{Y}_{3}\mathbf{S}_{12}\mathbf{S}_{13}\mathbf{S}_{21}\mathbf{S}_{23}}=}\no
\\&{P_{\!K_{\!12,s}|\mathbf{T}_{\!12}}P_{\!K'_{\!12,s}|\mathbf{T}_{\!12}}P_{\!K''_{\!12,s}}P_{\!K_{\!13,s}|\mathbf{T}_{\!13}}P_{\!K'_{\!13,s}|\mathbf{T}_{\!13}}P_{\!K''_{\!13,s}}P_{\!K_{\!21,s}|\mathbf{T}_{\!21}}P_{\!K'_{\!21,s}|\mathbf{T}_{\!21}}P_{\!K''_{\!21,s}}P_{\!K_{\!23,s}|\mathbf{T}_{\!23}}P_{\!K'_{\!23,s}|\mathbf{T}_{\!23}}P_{\!K''_{\!23,s}}\!\times}\no
\\&{P_{\mathbf{T}_{12}|\mathbf{X}_{1}\mathbf{Y}_{1}\mathbf{S}_{12}K''_{12,s}}P_{\mathbf{T}_{13}|\mathbf{X}_{1}\mathbf{Y}_{1}\mathbf{S}_{13}K''_{13,s}}P_{\mathbf{T}_{21}|\mathbf{X}_{2}\mathbf{Y}_{2}\mathbf{S}_{21}K''_{21,s}}P_{\mathbf{T}_{23}|\mathbf{X}_{2}\mathbf{Y}_{2}\mathbf{S}_{23}K''_{23,s}}P_{\mathbf{X}_{1}\mathbf{X}_{2}\mathbf{Y}_{1}\mathbf{Y}_{2}\mathbf{Y}_{3}\mathbf{S}_{12}\mathbf{S}_{13}\mathbf{S}_{21}\mathbf{S}_{23}}}\label{dist.contruc3}
\end{align}
We will show that this distribution is nearly indistinguishable from the distribution
\begin{align}
&{P_{K_{12,s}K'_{12,s}K''_{12,s}K_{13,s}K'_{13,s}K''_{13,s}K_{21,s}K'_{21,s}K''_{21,s}K_{23,s}K'_{23,s}K''_{23,s}\mathbf{T}_{12}\mathbf{T}_{13}\mathbf{T}_{21}\mathbf{T}_{23}\mathbf{X}_{1}\mathbf{X}_{2}\mathbf{Y}_{1}\mathbf{Y}_{2}\mathbf{Y}_{3}\mathbf{S}_{12}\mathbf{S}_{13}\mathbf{S}_{21}\mathbf{S}_{23}}=}\no
\\&{P_{\!K_{\!12,s}|\mathbf{T}_{\!12}}P_{\!K'_{\!12,s}|\mathbf{T}_{\!12}}P_{\!K''_{\!12,s}|\mathbf{T}_{\!12}}P_{\!K_{\!13,s}|\mathbf{T}_{\!13}}P_{\!K'_{\!13,s}|\mathbf{T}_{\!13}}P_{\!K''_{\!13,s}|\mathbf{T}_{\!13}}P_{\!K_{\!21,s}|\mathbf{T}_{\!21}}P_{\!K'_{\!21,s}|\mathbf{T}_{\!21}}P_{\!K''_{21,s}|\mathbf{T}_{\!21}}P_{K_{\!23,s}|\mathbf{T}_{\!23}}\times}\no
\\&{P_{K'_{23,s}|\mathbf{T}_{23}}P_{K''_{23,s}|\mathbf{T}_{23}}P_{\mathbf{T}_{12}|\mathbf{X}_{1}\mathbf{Y}_{1}\mathbf{S}_{12}}P_{\mathbf{T}_{13}|\mathbf{X}_{1}\mathbf{Y}_{1}\mathbf{S}_{13}}P_{\mathbf{T}_{21}|\mathbf{X}_{2}\mathbf{Y}_{2}\mathbf{S}_{21}}P_{\mathbf{T}_{23}|\mathbf{X}_{2}\mathbf{Y}_{2}\mathbf{S}_{23}}P_{\mathbf{X}_{1}\mathbf{X}_{2}\mathbf{Y}_{1}\mathbf{Y}_{2}\mathbf{Y}_{3}\mathbf{S}_{12}\mathbf{S}_{13}\mathbf{S}_{21}\mathbf{S}_{23}}}\label{dist.contruc1}
\end{align}
To this aim, we use the fact that index $k''_{jl,s}$ is generated at random by User $j$ independently of the other available sequences for $j=1,2$. In particular, $K''_{12,s}$ is independent of $(\mathbf{Y}_{1},\mathbf{X}_{1},\mathbf{S}_{12})$ and $K''_{13,s}$ is independent of $(\mathbf{Y}_{1},\mathbf{X}_{1},\mathbf{S}_{13})$. Symmetrically, $K''_{21,s}$ is independent of $(\mathbf{Y}_{2},\mathbf{X}_{2},\mathbf{S}_{21})$ and $K''_{23,s}$ is independent of $(\mathbf{Y}_{2},\mathbf{X}_{2},\mathbf{S}_{23})$. It follows from standard results~\cite{yassaee} that if
\begin{align}
&{r''_{12,s}<H(T_{12}|Y_{1},X_{1},S_{12})}\label{c12.cond}
\\&{r''_{13,s}<H(T_{13}|Y_{1},X_{1},S_{13})}\label{c13.cond}
\\&{r''_{21,s}<H(T_{21}|Y_{2},X_{2},S_{21})}\label{c21.cond}
\\&{r''_{23,s}<H(T_{23}|Y_{2},X_{2},S_{23})}\label{c23.cond}
\end{align}
then there exist $\alpha_{12}>0,\alpha_{13}>0,\alpha_{21}>0$ and $\alpha_{23}>0$ such that:
\begin{align}
&{E(D(P_{K''_{12,s}\mathbf{Y}_{1}\mathbf{X}_{1}\mathbf{S}_{12}}\parallel{q_{K''_{12,s}}P_{\mathbf{Y}_{1}\mathbf{X}_{1}\mathbf{S}_{12}}}))\leq 2^{-n\alpha_{12}}}\label{n.1}
\\&{E(D(P_{K''_{13,s}\mathbf{Y}_{1}\mathbf{X}_{1}\mathbf{S}_{13}}\parallel{q_{K''_{13,s}}P_{\mathbf{Y}_{1}\mathbf{X}_{1}\mathbf{S}_{13}}}))\leq 2^{-n\alpha_{13}}}\label{n.2}
\\&{E(D(P_{K''_{21,s}\mathbf{Y}_{2}\mathbf{X}_{2}\mathbf{S}_{21}}\parallel{q_{K''_{21,s}}P_{\mathbf{Y}_{2}\mathbf{X}_{2}\mathbf{S}_{21}}}))\leq 2^{-n\alpha_{21}}}\label{n.3}
\\&{E(D(P_{K''_{23,s}\mathbf{Y}_{2}\mathbf{X}_{2}\mathbf{S}_{23}}\parallel{q_{K''_{23,s}}P_{\mathbf{Y}_{2}\mathbf{X}_{2}\mathbf{S}_{23}}}))\leq 2^{-n\alpha_{23}}}\label{n.4}
\end{align}
in which $q$ represents uniform distribution over the respective set. We now set:
\begin{align}
&{\mathbf{O}_{12}=(\mathbf{X}_{1},\mathbf{Y}_{1},\mathbf{S}_{12})}\label{n.5}
\\&{\mathbf{O}_{13}=(\mathbf{X}_{1},\mathbf{Y}_{1},\mathbf{S}_{13})}\label{n.6}
\\&{\mathbf{O}_{21}=(\mathbf{X}_{2},\mathbf{Y}_{2},\mathbf{S}_{21})}\label{n.7}
\\&{\mathbf{O}_{23}=(\mathbf{X}_{2},\mathbf{Y}_{2},\mathbf{S}_{23})}\label{n.8}
\\&{\mathbf{O}_{3}=\mathbf{Y}_{3}}
\end{align}
Then, we have
{\small
\begin{align}
&{D(P_{\mathbf{T}_{12}\mathbf{T}_{13}\mathbf{T}_{21}\mathbf{T}_{23}\mathbf{O}_{12}\mathbf{O}_{13}\mathbf{O}_{21}\mathbf{O}_{23}\mathbf{O}_{3}}\parallel{\tilde{P}_{\mathbf{T}_{12}\mathbf{T}_{13}\mathbf{T}_{21}\mathbf{T}_{23}\mathbf{O}_{12}\mathbf{O}_{13}\mathbf{O}_{21}\mathbf{O}_{23}\mathbf{O}_{3}}})}\no
\\&{\mathop{\leq}\limits^{(a)}D(P_{K''_{12,s}K''_{13,s}K''_{21,s}K''_{23,s}\mathbf{T}_{12}\mathbf{T}_{13}\mathbf{T}_{21}\mathbf{T}_{23}\mathbf{O}_{12}\mathbf{O}_{13}\mathbf{O}_{21}\mathbf{O}_{23}\mathbf{O}_{3}}\parallel{\tilde{P}_{K''_{12,s}K''_{13,s}K''_{21,s}K''_{23,s}\mathbf{T}_{12}\mathbf{T}_{13}\mathbf{T}_{21}\mathbf{T}_{23}\mathbf{O}_{12}\mathbf{O}_{13}\mathbf{O}_{21}\mathbf{O}_{23}\mathbf{O}_{3}}})}\no
\\&{\mathop{=}\limits^{(b)}D(\!P_{K''_{12,s}K''_{13,s}\mathbf{T}_{12}\mathbf{T}_{13}\mathbf{O}_{12}\mathbf{O}_{13}}\!\!\parallel\!\!{\tilde{P}_{K''_{12,s}K''_{13,s}\mathbf{T}_{12}\mathbf{T}_{13}\mathbf{O}_{12}\mathbf{O}_{13}}}\!)\!+\!D(P_{\!K''_{21,s}K''_{23,s}\mathbf{T}_{21}\mathbf{T}_{23}\mathbf{O}_{21}\mathbf{O}_{23}}\!\!\parallel\!\!{\tilde{P}_{K''_{21,s}K''_{23,s}\mathbf{T}_{21}\mathbf{T}_{23}\mathbf{O}_{21}\mathbf{O}_{23}}}\!)}\no
\\&{=\sum_{k''_{12,s}k''_{13,s}\mathbf{t}_{12}\mathbf{t}_{13}\mathbf{o}_{12}\mathbf{o}_{13}}\!\!\!\!\!\!\!\!\!\!\!\!\!\!\!\!\!\!\!\!P(k''_{12,s},k''_{13,s},\mathbf{t}_{12},\mathbf{t}_{13},\mathbf{o}_{12},\mathbf{o}_{13})\log_2\frac{P(k''_{12,s},k''_{13,s},\mathbf{t}_{12},\mathbf{t}_{13},\mathbf{o}_{12},\mathbf{o}_{13})}{\tilde{P}(k''_{12,s},k''_{13,s},\mathbf{t}_{12},\mathbf{t}_{13},\mathbf{o}_{12},\mathbf{o}_{13})}}\no
\\&{+\sum_{k''_{21,s}k''_{23,s}\mathbf{t}_{21}\mathbf{t}_{23}\mathbf{o}_{21}\mathbf{o}_{23}}\!\!\!\!\!\!\!\!\!\!\!\!\!\!\!\!\!\!\!\!P(k''_{21,s},k''_{23,s},\mathbf{t}_{21},\mathbf{t}_{23},\mathbf{o}_{21},\mathbf{o}_{23})\log_2\frac{P(k''_{21,s},k''_{23,s},\mathbf{t}_{21},\mathbf{t}_{23},\mathbf{o}_{21},\mathbf{o}_{23})}{\tilde{P}(k''_{21,s},k''_{23,s},\mathbf{t}_{21},\mathbf{t}_{23},\mathbf{o}_{21},\mathbf{o}_{23})}}\no
\\&{=\sum_{k''_{12,s}k''_{13,s}\mathbf{t}_{12}\mathbf{t}_{13}\mathbf{o}_{12}\mathbf{o}_{13}}\!\!\!\!\!\!\!\!\!\!\!\!\!\!\!\!\!\!\!\!P(k''_{12,s},k''_{13,s},\mathbf{t}_{12},\mathbf{t}_{13},\mathbf{o}_{12},\mathbf{o}_{13})\log_2\frac{P(\mathbf{t}_{12},\mathbf{t}_{13}|k''_{12,s},k''_{13,s},\mathbf{o}_{12},\mathbf{o}_{13})P(k''_{12,s},k''_{13,s},\mathbf{o}_{12},\mathbf{o}_{13})}{P(\mathbf{t}_{12},\mathbf{t}_{13}|k''_{12,s},k''_{13,s},\mathbf{o}_{12},\mathbf{o}_{13})q_{k''_{12,s}}q_{k''_{13,s}}p(\mathbf{o}_{12},\mathbf{o}_{13})}}\no
\\&{+\sum_{k''_{21,s}k''_{23,s}\mathbf{t}_{21}\mathbf{t}_{23}\mathbf{o}_{21}\mathbf{o}_{23}}\!\!\!\!\!\!\!\!\!\!\!\!\!\!\!\!\!\!\!\!P(k''_{21,s},k''_{23,s},\mathbf{t}_{21},\mathbf{t}_{23},\mathbf{o}_{21},\mathbf{o}_{23})\log_2\frac{P(\mathbf{t}_{21},\mathbf{t}_{23}|k''_{21,s},k''_{23,s},\mathbf{o}_{21},\mathbf{o}_{23})P(k''_{21,s},k''_{23,s},\mathbf{o}_{21},\mathbf{o}_{23})}{P(\mathbf{t}_{21},\mathbf{t}_{23}|k''_{21,s},k''_{23,s},\mathbf{o}_{21},\mathbf{o}_{23})q_{k''_{21,s}}q_{k''_{23,s}}p(\mathbf{o}_{21},\mathbf{o}_{23})}}\no
\\&{\mathop{=}\limits^{(c)}\sum_{k''_{12,s}k''_{13,s}\mathbf{t}_{12}\mathbf{t}_{13}\mathbf{o}_{12}\mathbf{o}_{13}}P(k''_{12,s},k''_{13,s},\mathbf{t}_{12},\mathbf{t}_{13},\mathbf{o}_{12},\mathbf{o}_{13})\log_2\frac{P(k''_{12,s}|\mathbf{o}_{12})P(k''_{13,s}|\mathbf{o}_{13})p(\mathbf{o}_{12},\mathbf{o}_{13})}{q_{k''_{12,s}}q_{k''_{13,s}}p(\mathbf{o}_{12},\mathbf{o}_{13})}}\no
\\&{+\sum_{k''_{21,s}k''_{23,s}\mathbf{t}_{21}\mathbf{t}_{23}\mathbf{o}_{21}\mathbf{o}_{23}}P(k''_{21,s},k''_{23,s},\mathbf{t}_{21},\mathbf{t}_{23},\mathbf{o}_{21},\mathbf{o}_{23})\log_2\frac{P(k''_{21,s}|\mathbf{o}_{21})P(k''_{23,s}|\mathbf{o}_{23})p(\mathbf{o}_{21},\mathbf{o}_{23})}{q_{k''_{21,s}}q_{k''_{23,s}}p(\mathbf{o}_{21},\mathbf{o}_{23})}}\no
\\&{=\sum_{k''_{12,s}k''_{13,s}\mathbf{t}_{12}\mathbf{t}_{13}\mathbf{o}_{12}\mathbf{o}_{13}}P(k''_{12,s},k''_{13,s},\mathbf{t}_{12},\mathbf{t}_{13},\mathbf{o}_{12},\mathbf{o}_{13})\log_2\frac{P(k''_{12,s}|\mathbf{o}_{12})P(k''_{13,s}|\mathbf{o}_{13})}{q_{k''_{12,s}}q_{k''_{13,s}}}}\no
\\&{+\sum_{k''_{21,s}k''_{23,s}\mathbf{t}_{21}\mathbf{t}_{23}\mathbf{o}_{21}\mathbf{o}_{23}}P(k''_{21,s},k''_{23,s},\mathbf{t}_{21},\mathbf{t}_{23},\mathbf{o}_{21},\mathbf{o}_{23})\log_2\frac{P(k''_{21,s}|\mathbf{o}_{21})P(k''_{23,s}|\mathbf{o}_{23})}{q_{k''_{21,s}}q_{k''_{23,s}}}}\no
\\&{=\sum_{k''_{12,s}k''_{13,s}\mathbf{t}_{12}\mathbf{t}_{13}\mathbf{o}_{12}\mathbf{o}_{13}}P(k''_{12,s},k''_{13,s},\mathbf{t}_{12},\mathbf{t}_{13},\mathbf{o}_{12},\mathbf{o}_{13})(\log_2\frac{P(k''_{12,s}|\mathbf{o}_{12})}{q_{k''_{12,s}}}+\log_2\frac{P(k''_{13,s}|\mathbf{o}_{13})}{q_{k''_{13,s}}})}\no
\\&{+\sum_{k''_{21,s}k''_{23,s}\mathbf{t}_{21}\mathbf{t}_{23}\mathbf{o}_{21}\mathbf{o}_{23}}P(k''_{21,s},k''_{23,s},\mathbf{t}_{21},\mathbf{t}_{23},\mathbf{o}_{21},\mathbf{o}_{23})(\log_2\frac{P(k''_{21,s}|\mathbf{o}_{21})}{q_{k''_{21,s}}}+\log_2\frac{P(k''_{23,s}|\mathbf{o}_{23})}{q_{k''_{23,s}}})}\no
\\&{=D(P_{K''_{12,s}\mathbf{O}_{12}}\!\!\parallel\!\!{q_{K''_{12,s}}P_{\mathbf{O}_{12}}})\!+\!D(P_{K''_{13,s}\mathbf{O}_{13}}\!\!\parallel\!\!{q_{K''_{13,s}}P_{\mathbf{O}_{13}}})\!+\!D(P_{K''_{21,s}\mathbf{O}_{21}}\!\!\parallel\!\!{q_{K''_{21,s}}P_{\mathbf{O}_{21}}})\!+\!D(P_{K''_{23,s}\mathbf{O}_{23}}\!\!\parallel\!\!{q_{K''_{23,s}}P_{\mathbf{O}_{23}}})}\no
\\&{\mathop{\leq}\limits^{(d)}2^{-n\alpha_{12}}+2^{-n\alpha_{13}}+2^{-n\alpha_{21}}+2^{-n\alpha_{23}}}\label{dif.ent1}
\end{align}}
\normalsize{
\noindent where in the above equations, (a) follows from the relative entropy properties \cite{CoverBok}, (b) follows from the auxiliary random variables distributions in \eqref{dist.contruc3} and \eqref{dist.contruc1}, (c) follows from distribution $P$ in \eqref{dist.contruc1} that results in the following Markov chains:}
\begin{align}
&{K''_{12,s}-\mathbf{T}_{12}-\mathbf{O}_{12}-\mathbf{O}_{13}-\mathbf{T}_{13}-K''_{13,s}}\no
\\&{K''_{21,s}-\mathbf{T}_{21}-\mathbf{O}_{21}-\mathbf{O}_{23}-\mathbf{T}_{23}-K''_{23,s}}\no
\end{align}
\noindent and (d) is deduced from \eqref{n.1}-\eqref{n.4}.

Hence, the distance between the two probabilities $P$ and $\tilde{P}$ is arbitrarily small. Consequently the decoding error probability of the secondary keys can be analyzed for the much simpler distribution ${P}$.

It follows from standard results of Slepian-Wolf coding~\cite{SW} that if
\begin{align}
&{\!r'_{12,s}\!+\!r''_{12,s}\!>H(T_{12}|X_{2},Y_{2},S_{12},S_{21})}\label{dec6}
\\&{\!r'_{21,s}\!+\!r''_{21,s}\!>H(T_{21}|X_{1},Y_{1},S_{12},S_{21})}\label{dec7}
\\&{\!r'_{13,s}\!+\!r''_{13,s}\!>H(T_{13}|Y_{3},S_{13},S_{23},T_{23})}\label{dec9}
\\&{\!r'_{23,s}\!+\!r''_{23,s}\!>H(T_{23}|Y_{3},S_{13},S_{23},T_{13})}\label{dec10}
\\&{\!r'_{13,s}\!+\!r''_{13,s}\!+\!r'_{23,s}\!+\!r''_{23,s}\!>H(T_{13},T_{23}|Y_{3},S_{13},S_{23})}\label{dec11}
\end{align}
and if we set $\epsilon_{2}=\frac{\epsilon}{16B}$, then
\begin{align}
&{E(P_{e1,s}^{(n)})\le 2\epsilon _{2}=\frac{\epsilon}{8B}}\label{n.15}
\\&{E(P_{e2,s}^{(n)})\le 2\epsilon _{2}=\frac{\epsilon}{8B}}\label{n.16}
\\&{E(P_{e3,s}^{(n)})\le 4\epsilon _{2}=\frac{\epsilon}{4B}}\label{n.17}
\end{align}
for sufficiently large $n$.

Finally, we show that we can select a specific $k_{21,s}^{''*}$ ahead of time so that need not be transmitted. In fact by assuming $C$ as the random variable representing randomly generated code and all binning functions, we have:
\small{
\begin{align}
&{E(P_{e1,s}^{(n)})= E_{C}(\Pr\{\mathbf{T}_{21}\neq g_{1}(\mathbf{O}_{12},K'_{21,s},K''_{21,s})|C\})=}\no
\\&{E_{C}(\!\!\!\!\!\!\!\!\sum_{k''_{21,s}k'_{21,s}\mathbf{o}_{12}\mathbf{o}_{21}\mathbf{t}_{21}}\!\!\!\!\!\!\!\!\!\!\!\!\!\!\!\!\frac{1}{2^{nr''_{21,s}}}P_{\!K'_{21,s}|\mathbf{T}_{21}}(\!k'_{21,s}|\mathbf{t}_{21}\!)\!P_{\mathbf{T}_{21}|\mathbf{O}_{21}K''_{21,s}}(\!\mathbf{t}_{21}|\mathbf{o}_{21},k''_{21,s}\!)\!P_{\!\mathbf{O}_{12}\mathbf{O}_{21}}\!(\!\mathbf{o}_{12},\mathbf{o}_{21}\!)\textbf{1}[\mathbf{t}_{21}\!\neq\! g_{1}(\mathbf{o}_{12},k'_{21,s},k''_{21,s})|C])}\no
\\&{=E_{K''_{21,s}C}(\Pr\{\mathbf{T}_{21}\neq g_{1}(\mathbf{O}_{12},K'_{21,s},K''_{21,s})\}|K''_{21,s},C)\leq \frac{\epsilon}{8B}}
\end{align}}
\normalsize{
Hence, there exists $k_{21,s}^{''*}$ such that:}
\begin{equation}
E(P_{e1,s}^{(n)})=E_{C}(\Pr\{\mathbf{T}_{21}\neq g_{1}(\mathbf{O}_{12},K'_{21,s},k_{21,s}^{''*})|C\})< \frac{3\epsilon}{8B}\label{n.9}
\end{equation}
Similarly, there exist $k_{12,s}^{''*},k_{13,s}^{''*}$ and $k_{23,s}^{''*}$ such that:
\begin{align}
&{E(P_{e2,s}^{(n)})=E_{C}(\Pr\{\mathbf{T}_{12}\neq g_{2}(\mathbf{O}_{21},K'_{12,s},k_{12,s}^{''*})|C\})< \frac{3\epsilon}{8B}}\label{n.10}
\\&{E(P_{e3,s}^{(n)})=E_{C}(\Pr\{(\mathbf{T}_{13},\mathbf{T}_{23})\neq g_{3}(\mathbf{O}_{3},K'_{13,s},k_{13,s}^{''*},K'_{23},k_{23,s}^{''*})|C\})<\frac{3\epsilon}{4B}}\label{n.11}
\end{align}

Repeating the above encoding and decoding procedures over $B$ blocks, the total decoding error probability at User $j\in\{1,2,3\}$ is bounded as
\[\begin{array}{l} {E(P_{j}^{(nB)}) \le B(E(P_{ej,p}^{(n)})+E(P_{ej,s}^{(n)}))\mathop{<}\limits^{(a)}\epsilon,} \end{array}\]
where (a) is deduced from \eqref{n.12}-\eqref{n.14} and \eqref{n.9}-\eqref{n.11}.

\begin{rem}
combining \eqref{dec6}-\eqref{dec11} and \eqref{c12.cond}-\eqref{c23.cond}, we obtain:
\begin{align}
&{\!r'_{12,s}\!>I(T_{12};X_{1},Y_{1}|X_{2},Y_{2},S_{12},S_{21})}\label{dec12}
\\&{\!r'_{21,s}\!>I(T_{21};X_{2},Y_{2}|X_{1},Y_{1},S_{12},S_{21})}\label{dec13}
\\&{\!r'_{13,s}\!>I(T_{13};X_{1},Y_{1}|Y_{3},S_{13},S_{23},T_{23})}\label{dec14}
\\&{\!r'_{23,s}\!>I(T_{23};X_{2},Y_{2}|Y_{3},S_{13},S_{23},T_{13})}\label{dec15}
\\&{\!r'_{13,s}\!+\!r'_{23,s}>I(T_{13},T_{23};X_{1},Y_{1},X_{2},Y_{2}|Y_{3},S_{13},S_{23})}\label{dec16}
\end{align}
The necessary condition (\ref{necessary cond}) in definition of the functions $f_{12},f_{13},f_{21}$ and $f_{23}$ holds according to the rate constraints (\ref{constraints}) in Theorem \ref{th3} and equations (\ref{dec1})-(\ref{dec5}) and (\ref{dec12})-(\ref{dec16}).
\end{rem}

\begin{rem}\label{rem:5}
In the code construction of the primary keys, we implicitly assumed that $I(S_{12} ;X_{2},Y_{2})\geq I(S_{12};Y_{3},S_{13},S_{23},T_{13},T_{23})$. In the case where $I(S_{12} ;X_{2},Y_{2})< I(S_{12};Y_{3},S_{13},S_{23},T_{13},T_{23})$, User 1 randomly maps $k'_{12,s}$ into a space with $2^{n(I(S_{12} ;X_{2},Y_{2})-\delta'(\epsilon))}$ elements and no primary key is chosen by User 1 to be shared with User 2 and the bound on the rate of the primary key between Users 1 and 2 is equal to $[I(S_{21} ;X_{1},Y_{1})-I(S_{21} ;Y_{3} ,S_{13},S_{23},T_{13},T_{23})]^{+}$ in (\ref{rate-define2}). Similarly, we implicitly assumed that $I(S_{13} ;Y_{3}|S_{23})\geq I(S_{13} ;X_{2},Y_{2},S_{12},T_{12}|S_{23})$. In the case where $I(S_{13} ;Y_{3}|S_{23})< I(S_{13} ;X_{2},Y_{2},S_{12},T_{12}|S_{23})$, User 1 randomly maps $k'_{13,s}$ into a space with $2^{n(I(S_{13} ;Y_{3}|S_{23})-\delta''(\epsilon))}$ elements and no primary key is chosen by User 1 to be shared with User 3 and the bound on the sum rate of $r_{13,p}+r_{23,p}$ is equal to $[I(S_{23} ;Y_{3}|S_{13})-I(S_{23} ;X_{1},Y_{1},S_{21},T_{21}|S_{13})]^{+}$ in (\ref{rate-define2}). The same is true for User 2's codebook.
\end{rem}

\begin{rem}\label{rem:6}
In the code construction of the secondary keys, we assumed that $I(T_{12} ;X_{2},Y_{2}|S_{12},S_{21})-I(T_{12};Y_{3},S_{13},S_{23},T_{13},T_{23}|S_{12},S_{21})\geq 0$, respectively $I(T_{21} ;X_{1},Y_{1}|S_{12},S_{21})-I(T_{21};Y_{3},S_{13},S_{23},t_{13},t_{23}|S_{12},S_{21})\geq 0$. Otherwise, we set $T_{12}=\phi$, respectively $T_{21}=\phi$. The same is true in deriving $r_{13,s} +r_{23,s}$.
\end{rem}


\subsection*{Security Analysis}
We now analyze the security condition of Definition 1.
Performing the described encoding and decoding procedures in $B$ blocks, the first secrecy constraint in (\ref{eq3}) specializes as
\begin{equation}\label{seccon}
I((K_{12,p},K_{21,p},K_{12,s},K_{21,s})^{1:B};\textbf{Y}_{3}^{1:B}|(K''_{12,s},K''_{21,s})^{1:B})<\epsilon
\end{equation}
where $(K_{12,s},K_{21,s})^{B}=\emptyset$. We have:
\small{
\begin{align}
&{I((K_{12,p},K_{21,p},K_{12,s},K_{21,s})^{1:B};\textbf{Y}_{3}^{1:B}|(K''_{12,s},K''_{21,s})^{1:B})}\no
\\&{\leq I((K_{12,p},K_{21,p},K_{12,s},K_{21,s},K''_{12,s},K''_{21,s})^{1:B};\textbf{Y}_{3}^{1:B})}\no
\\&{=\sum_{i=1}^{B}I((K_{12,p},K_{21,p},K_{12,s},K_{21,s},K''_{12,s},K''_{21,s})^{i};\textbf{Y}_{3}^{1:B}|(K_{12,p},K_{21,p},K_{12,s},K_{21,s},K''_{12,s},K''_{21,s})^{1:i-1})}\no
\\&{\leq\sum_{i=1}^{B}I((K_{12,p},K_{21,p},K_{12,s},K_{21,s},K''_{12,s},K''_{21,s})^{i};\textbf{Y}_{3}^{1:B},(K_{12,p},K_{21,p},K_{12,s},K_{21,s},K''_{12,s},K''_{21,s})^{1:i-1})}\label{independence}
\\&{=\sum_{i=1}^{B}[I((K_{12,p},K_{21,p})^{i};\textbf{Y}_{3}^{1:B},(K_{12,p},K_{21,p},K_{12,s},K_{21,s},K''_{12,s},K''_{21,s})^{1:i-1})}\no
\\&{+I((K_{12,s},K_{21,s},K''_{12,s},K''_{21,s})^{i};\textbf{Y}_{3}^{1:B},(K_{12,p},K_{21,p},K_{12,s},K_{21,s},K''_{12,s},K''_{21,s})^{1:i-1}|(K_{12,p},K_{21,p})^{i})]}\no
\\&{\leq\sum_{i=1}^{B}[\overbrace{I((K_{12,p},K_{21,p})^{i};\textbf{Y}_{3}^{1:B},(K_{12,p},K_{21,p},K_{12,s},K_{21,s},K''_{12,s},K''_{21,s})^{1:i-1})}^{A_{i}}}\no
\\&{+\underbrace{I((K_{12,s},K_{21,s},K''_{12,s},K''_{21,s})^{i};\textbf{Y}_{3}^{1:B},(K_{12,p},K_{21,p},K_{12,s},K_{21,s},K''_{12,s},K''_{21,s})^{1:i-1},(K_{12,p},K_{21,p})^{i})}_{B_{i}}]}\no
\end{align}}
\normalsize{
We analyze each of the above terms separately.}
Some Markov chains useful in the security analysis are given in (\ref{mrf1})-(\ref{mrf5}). These Markov chains arise from the coding scheme.
\small{
\begin{align} &{(K_{12,p},K_{21,p})^{i}\!-\!(K'_{12,s},K'_{21,s},K'_{13,s},K'_{23,s})^{i-1}\!-\!(K_{12,p},K_{21,p},K_{12,s},K_{21,s},K''_{12,s},K''_{21,s},\textbf{Y}_{3})^{1:i-1}}\label{mrf1}
\\&{(K_{12,p},K_{21,p})^{i}\!-\!(K'_{12,s},K'_{21,s},K'_{13,s},K'_{23,s})^{i}\!-\!(\textbf{Y}_{3})^{i+1:B}}\label{mrf2}
\\&{(K_{12,p},K_{21,p})^{i}\!-\!(\textbf{Y}_{3},\textbf{S}_{13},\textbf{S}_{23})^{i}\!-\!(K'_{12,s},K'_{21,s},K'_{13,s},K'_{23,s})^{i-1}}\label{mrf3}
\\&{(K_{12,s},K_{21,s},K''_{12,s},K''_{21,s})^{i}\!-\!(\textbf{S}_{12},\textbf{S}_{21},\textbf{S}_{13},\textbf{S}_{23})^{i}\!-\!(K_{12,p},K_{21,p},K_{12,s},K_{21,s},K''_{12,s},K''_{21,s},\textbf{Y}_{3})^{1:i-1}}\label{mrf4}
\\&{(K_{12,s},K_{21,s},K''_{12,s},K''_{21,s})^{i}\!-\!(K'_{12,s},K'_{21,s},K'_{13,s},K'_{23,s})^{i}\!-\!(\textbf{Y}_{3})^{i+1:B}}\label{mrf5}
\end{align}}
%
%
\normalsize{
For term $A_{i}$, we have:}
\vspace{.3cm}

\small{
$\begin{array}{l}
{A_{i}=I((K_{12,p},K_{21,p})^{i};\textbf{Y}_{3}^{1:B},(K_{12,p},K_{21,p},K_{12,s},K_{21,s},K''_{12,s},K''_{21,s})^{1:i-1})}
\\{\leq I((K_{12,p},K_{21,p})^{i};\textbf{Y}_{3}^{1:B},(K_{12,p},K_{21,p},K_{12,s},K_{21,s},K''_{12,s},K''_{21,s})^{1:i-1},(K'_{12,s},K'_{21,s},K'_{13,s},K'_{23,s})^{i-1:i})}
\\{\mathop{=}\limits^{(a)}I((K_{12,p},K_{21,p})^{i};\textbf{Y}_{3}^{i:B},(K'_{12,s},K'_{21,s},K'_{13,s},K'_{23,s})^{i-1:i})}
\\{\mathop{=}\limits^{(b)}I((K_{12,p},K_{21,p})^{i};\textbf{Y}_{3}^{i},(K'_{12,s},K'_{21,s},K'_{13,s},K'_{23,s})^{i-1:i})}
\\{\leq I((K_{12,p},K_{21,p})^{i};\textbf{Y}_{3}^{i},\textbf{S}_{13}^{i},\textbf{S}_{23}^{i},(K'_{12,s},K'_{21,s},K'_{13,s},K'_{23,s})^{i-1:i})}
\\{\mathop{=}\limits^{(c)}I((K_{12,p},K_{21,p})^{i};\textbf{Y}_{3}^{i},\textbf{S}_{13}^{i},\textbf{S}_{23}^{i},(K'_{12,s},K'_{21,s},K'_{13,s},K'_{23,s})^{i})}
\\{\mathop{\le}\limits^{(d)}I((K_{12,p},K_{21,p})^{i};\textbf{Y}_{3}^{i},\textbf{S}_{13}^{i},\textbf{S}_{23}^{i},\textbf{T}_{13}^{i},\textbf{T}_{23}^{i},(K'_{12,s},K'_{21,s})^{i})}
\\{= I((K_{12,p},K_{21,p})^{i};\textbf{Y}_{3}^{i},\textbf{S}_{13}^{i},\textbf{S}_{23}^{i},\textbf{T}_{13}^{i},\textbf{T}_{23}^{i})+I((K_{12,p},K_{21,p})^{i};(K'_{12,s},K'_{21,s})^{i}|\textbf{Y}_{3}^{i},\textbf{S}_{13}^{i},\textbf{S}_{23}^{i},\textbf{T}_{13}^{i},\textbf{T}_{23}^{i})}
\\{\mathop{\le}\limits^{(e)}\overbrace{I((K_{12,p},K_{21,p})^{i};\textbf{Y}_{3}^{i},\textbf{S}_{13}^{i},\textbf{S}_{23}^{i},\textbf{T}_{13}^{i},\textbf{T}_{23}^{i})}^{A_{1i}}+\overbrace{I((K'_{12,s},K'_{21,s})^{i};\textbf{Y}_{3}^{i},\textbf{S}_{12}^{i},\textbf{S}_{21}^{i},\textbf{S}_{13}^{i},\textbf{S}_{23}^{i},\textbf{T}_{13}^{i},\textbf{T}_{23}^{i})}^{A_{2i}}}
\end{array}$}
\vspace{.3cm}

\normalsize{
\noindent where in the above equations, (a), (b) and (c) are, respectively, due to Markov chains (\ref{mrf1}), (\ref{mrf2}) and (\ref{mrf3}). (d) holds since $k'_{13,s}$ and $k'_{23,s}$ are indices of sequences $\textbf{t}_{13}$ and $\textbf{t}_{23}$, respectively. (e) is true due to the fact that $k_{12,p}$ and $k_{21,p}$ are indices of sequences $\textbf{s}_{12}$ and $\textbf{s}_{21}$, respectively.

For term $B_{i}$, we have:}
\vspace{.3cm}

\small{
$\begin{array}{l}{B_{i}=I((K_{12,s},K_{21,s},K''_{12,s},K''_{21,s})^{i};\textbf{Y}_{3}^{1:B},(K_{12,p},K_{21,p},K_{12,s},K_{21,s},K''_{12,s},K''_{21,s})^{1:i-1},(K_{12,p},K_{21,p})^{i})}
\\{\mathop{\le}\limits^{(a)}  I((K_{12,s},K_{21,s},K''_{12,s},K''_{21,s})^{i};\textbf{Y}_{3}^{1:B},(K_{12,p},K_{21,p},K_{12,s},K_{21,s},K''_{12,s},K''_{21,s})^{1:i-1},(\textbf{S}_{12},\textbf{S}_{21})^{i})}
\\{\leq I((K_{12,s},K_{21,s},K''_{12,s},K''_{21,s})^{i};\textbf{Y}_{3}^{1:B},(K_{12,p},K_{21,p},K_{12,s},K_{21,s},K''_{12,s},K''_{21,s})^{1:i-1},(\textbf{S}_{12},\textbf{S}_{21},\textbf{S}_{13},\textbf{S}_{23})^{i})}
\\{\mathop{=}\limits^{(b)} I((K_{12,s},K_{21,s},K''_{12,s},K''_{21,s})^{i};\textbf{Y}_{3}^{i:B},(\textbf{S}_{12},\textbf{S}_{21},\textbf{S}_{13},\textbf{S}_{23})^{i})}
\\{\leq I((K_{12,s},K_{21,s},K''_{12,s},K''_{21,s})^{i};\textbf{Y}_{3}^{i:B},(\textbf{S}_{12},\textbf{S}_{21},\textbf{S}_{13},\textbf{S}_{23})^{i},(K'_{12,s},K'_{21,s},K'_{13,s},K'_{23,s})^{i})}
\\{\mathop{=}\limits^{(c)} I((K_{12,s},K_{21,s},K''_{12,s},K''_{21,s})^{i};\textbf{Y}_{3}^{i},(\textbf{S}_{12},\textbf{S}_{21},\textbf{S}_{13},\textbf{S}_{23})^{i},(K'_{12,s},K'_{21,s},K'_{13,s},K'_{23,s})^{i})}
\\{\leq I((K_{12,s},K_{21,s},K''_{12,s},K''_{21,s})^{i};\textbf{Y}_{3}^{i},(\textbf{S}_{12},\textbf{S}_{21},\textbf{S}_{13},\textbf{S}_{23},\textbf{T}_{13},\textbf{T}_{23})^{i},(K'_{12,s},K'_{21,s})^{i})}
\\{\mathop{=}\limits^{(d)} \overbrace{I((K_{12,s},K_{21,s},K''_{12,s},K''_{21,s})^{i};\textbf{Y}_{3}^{i},(\textbf{S}_{12},\textbf{S}_{21},\textbf{S}_{13},\textbf{S}_{23},\textbf{T}_{13},\textbf{T}_{23})^{i}|(K'_{12,s},K'_{21,s})^{i})}^{B_{1i}}}
\end{array}$}
\vspace{.3cm}

\normalsize{
\noindent where in the above equations, (a) is due to the fact that $k_{12,p}$ and $k_{21,p}$ are indices of sequences $\textbf{s}_{12}$ and $\textbf{s}_{21}$, respectively. (b) and (c) are deduced from Markov chains (\ref{mrf4}) and (\ref{mrf5}), respectively. (d) holds since the indices are independent of each other.

We continue by combining three terms $A_{1i},A_{2i}$ and $B_{1i}$. Since in all the three terms, only block index $i$ appears, we drop it in the following. Security condition (\ref{seccon}) appears as:
\small{
\begin{align}
&I((K_{12,p},K_{21,p},K_{12,s},K_{21,s})^{1:B};\textbf{Y}_{3}^{1:B}|(K''_{12,s},K''_{21,s})^{1:B})\leq B (A_{1}+A_{2}+B_{1})=\no
\\& B(\!I(\!K_{12,p},\!K_{21,p};\!\textbf{Y}_{3},\!\textbf{S}_{13},\!\textbf{S}_{23},\!\textbf{T}_{13},\!\textbf{T}_{23}\!)\!+\!I(\!K_{12,s},\!K_{21,s},K'_{12,s},\!K'_{21,s},\!K''_{12,s},\!K''_{21,s};\!\textbf{Y}_{3},\!\textbf{S}_{12},\!\textbf{S}_{21},\!\textbf{S}_{13},\!\textbf{S}_{23},\!\textbf{T}_{13},\!\textbf{T}_{23}\!)\!)\label{totsec}
\end{align}}

\normalsize{
We analyze the two terms in (\ref{totsec}) separately. Specifically, we use Lemma \ref{lm:4} and Lemma \ref{lm:5} as follows.}
\begin{lem}
  \label{lm:4}
If
\begin{align}
&{r'_{12,p}\!+r'_{21,p}>\!I(S_{12},S_{21};Y_{3},S_{13},S_{23},T_{13},T_{23})+2\epsilon'}\label{r12+r21,p}
\\&{r'_{12,p}\!>\!I(S_{12};Y_{3},S_{13},S_{23},T_{13},T_{23})+2\epsilon'}\label{r12,p}
\\&{r'_{21,p}\!>\!I(S_{21};Y_{3},S_{13},S_{23},T_{13},T_{23})+2\epsilon'}\label{r21,p}
\\&{r'_{13,p}\!>\!I(S_{13};X_{2},Y_{2},S_{12},T_{12})\!+2\epsilon'}\label{r13,p}
\\&{r'_{23,p}\!>\!I(S_{23};X_{1},Y_{1},S_{21},T_{21})\!+2\epsilon'}\label{r23,p}
\end{align}
\noindent then we have
\begin{align}
&I(K_{12,p},K_{21,p};\textbf{Y}_{3},\textbf{S}_{13},\textbf{S}_{23},\textbf{T}_{13},\textbf{T}_{23})<\frac{\epsilon}{2B}\label{prisec1}
\\&I(K_{13,p};\textbf{Y}_{2},\textbf{X}_{2},\textbf{S}_{12},\textbf{T}_{12})<\frac{\epsilon}{2B}\label{prisec2}
\\&I(K_{23,p};\textbf{Y}_{1},\textbf{X}_{1},\textbf{S}_{21},\textbf{T}_{21})<\frac{\epsilon}{2B}\label{prisec3}
\end{align}
\end{lem}
The proof of Lemma \ref{lm:4} is given in Appendix~\ref{App2}.

\begin{lem}
  \label{lm:5}
If
\noindent \begin{align}
&{r_{12,s}+r'_{12,s}+r''_{12,s}\!+r_{21,s}+r'_{21,s}+r''_{21,s}<\!H(T_{12},T_{21}|Y_{3},S_{12},S_{21},S_{13},S_{23},T_{13},T_{23})\!-2\epsilon''}\label{r12+r21,s}
\\&{r_{12,s}+r'_{12,s}+r''_{12,s}\!<\!H(T_{12}|Y_{3},S_{12},S_{21},S_{13},S_{23},T_{13},T_{23})-2\epsilon''}\label{r12,s}
\\&{r_{21,s}+r'_{21,s}+r''_{21,s}\!<\!H(T_{21}|Y_{3},S_{12},S_{21},S_{13},S_{23},T_{13},T_{23})-2\epsilon''}\label{r21,s}
\\&{r_{13,s}+r'_{13,s}+r''_{13,s}\!<\!H(T_{13}|X_{2},Y_{2},S_{12},S_{13},T_{12})\!-2\epsilon''}\label{r13,s}
\\&{r_{23,s}+r'_{23,s}+r''_{23,s}\!<\!H(T_{23}|X_{1},Y_{1},S_{21},S_{23},T_{21})\!-2\epsilon''}\label{r23,s}
\end{align}
\noindent then we have
\begin{align}
&I(K_{12,s},K_{21,s},K'_{12,s},K'_{21,s},K''_{12,s},K''_{21,s};\textbf{Y}_{3},\textbf{S}_{12},\textbf{S}_{21},\textbf{S}_{13},\textbf{S}_{23},\textbf{T}_{13},\textbf{T}_{23})<\frac{\epsilon}{2B}\label{secsec1}
\\&I(K_{13,s},K'_{13,s},K''_{13,s};\textbf{Y}_{2},\textbf{X}_{2},\textbf{S}_{12},\textbf{S}_{13},\textbf{T}_{12})<\frac{\epsilon}{2B}\label{secsec2}
\\&I(K_{23,s},K'_{23,s},K''_{23,s};\textbf{Y}_{1},\textbf{X}_{1},\textbf{S}_{21},\textbf{S}_{23},\textbf{T}_{21})<\frac{\epsilon}{2B}\label{secsec3}
\end{align}
\end{lem}
The proof of Lemma \ref{lm:5} is given in Appendix~\ref{App4}.

Combining (\ref{totsec}), (\ref{prisec1}) and (\ref{secsec1}), the strong secrecy condition of the key between Users 1 and 2 is deduced as:
\begin{equation}
I((K_{12,p},K_{21,p},K_{12,s},K_{21,s})^{1:B};\textbf{Y}_{3}^{1:B}|(K''_{12,s},K''_{21,s})^{1:B})\leq \epsilon
\end{equation}

Using similar Markov chains as in (\ref{mrf1})-(\ref{mrf5}) for keys $K_{13,p},K_{23,p},K_{13,s},K_{23,s}$ and exploiting Lemma \ref{lm:4} and \ref{lm:5}, the other strong secrecy conditions in (\ref{eq3}) are deduced as:

\begin{align}
&I((K_{13,p},K_{13,s})^{1:B};(\textbf{X}_{2},\textbf{Y}_{2})^{1:B}|(K''_{13,s})^{1:B})\leq \epsilon\\
&I((K_{23,p},K_{23,s})^{1:B};(\textbf{X}_{1},\textbf{Y}_{1})^{1:B}|(K''_{23,s})^{1:B})\leq \epsilon
\end{align}

\noindent Replacing equations (\ref{r12+r21,p})-(\ref{r23,p}) in (\ref{dec1})-(\ref{dec5}), we obtain:
\begin{align}
&{r_{12,p}< I(S_{12};X_{2},Y_{2})-I(S_{12};Y_{3},S_{13},S_{23},T_{13},T_{23})\triangleq \textbf{r}_{\textbf{12,p}},}\no
\\&{r_{21,p} < I(S_{21};X_{1},Y_{1})-I(S_{21};Y_{3},S_{13},S_{23},T_{13},T_{23})\triangleq \textbf{r}_{\textbf{21,p}},}\no
\\&{r_{12,p} +r_{21,p} < I(S_{12};X_{2},Y_{2})+I(S_{21};X_{1},Y_{1})-I(S_{12},S_{21};Y_{3},S_{13},S_{23},T_{13},T_{23})}\no
\\&{= \textbf{r}_{\textbf{12,p}}+\textbf{r}_{\textbf{21,p}}-\textbf{I}_{\textbf{12,p}},}\no
\\&{r_{13,p}< I(S_{13};Y_{3}|S_{23})-I(S_{13};Y_{2},X_{2},S_{12},T_{12}|S_{23})\triangleq \textbf{r}_{\textbf{13,p}},}\no
\\&{r_{23,p}< I(S_{23};Y_{3}|S_{13})-I(S_{23};Y_{1},X_{1},S_{21},T_{21}|S_{13})\triangleq \textbf{r}_{\textbf{23,p}},}\no
\\&{r_{13,p}+r_{23,p}< I(S_{13},S_{23};Y_{3})-I(S_{13};Y_{2},X_{2},S_{12},T_{12}|S_{23})-I(S_{23};Y_{1},X_{1},S_{21},T_{21}|S_{13})}\no
\\&{=\textbf{r}_{\textbf{13,p}}+\textbf{r}_{\textbf{23,p}}-\textbf{I}_{\textbf{3,p}}}\label{primary}
\end{align}
By setting $\overline{r}_{12,p}=r_{12,p}+r_{21,p},\overline{r}_{13,p}=r_{13,p},\overline{r}_{23,p}=r_{23,p}$ and applying Fourier-Motzkin elimination \cite{fouri-motz} to the above region, the primary keys rates of Theorem \ref{th3} (and also Theorem 1) are derived.

\noindent Replacing equations \eqref{r12+r21,s}-\eqref{r23,s} with \eqref{dec6}-\eqref{dec11}, the following rates are achievable for the secondary keys:
\begin{align}
&{r_{12,s}<I(T_{12} ;X_{2},Y_{2}|S_{12},S_{21})-I(T_{12};Y_{3},S_{13},S_{23},T_{13},T_{23}|S_{12},S_{21})\triangleq \textbf{r}_{\textbf{12,s}},}\no
\\&{r_{21,s}<I(T_{21} ;X_{1},Y_{1}|S_{12},S_{21})-I(T_{21};Y_{3},S_{13},S_{23},T_{13},T_{23}|S_{12},S_{21})\triangleq \textbf{r}_{\textbf{21,s}},}\no
\\&{r_{12,s}+r_{21,s}<I(T_{12} ;X_{2},Y_{2}|S_{12},S_{21})+I(T_{21} ;X_{1},Y_{1}|S_{12},S_{21})-I(T_{12};T_{21}|S_{12},S_{21})}\no
\\&{-I(T_{12},T_{21}  ;Y_{3},S_{13},S_{23},T_{13},T_{23}|S_{12},S_{21})=\textbf{r}_{\textbf{12,s}}+\textbf{r}_{\textbf{21,s}}-\textbf{I}_{\textbf{12,s}},}\no
\\&{r_{13,s}<I(T_{13} ;Y_{3}|S_{13},S_{23},T_{23})-I(T_{13} ;X_{2},Y_{2},S_{12},T_{12}|S_{13},S_{23},T_{23})\triangleq \textbf{r}_{\textbf{13,s}},}\no
\\&{r_{23,s}<I(T_{23} ;Y_{3}|S_{13},T_{13},S_{23})-I(T_{23} ;X_{1},Y_{1},S_{21},T_{21}|S_{13},T_{13},S_{23})\triangleq \textbf{r}_{\textbf{23,s}},}\no
\\&{r_{13,s}+r_{23,s}\!<\!I(T_{13},\!T_{23} ;\!Y_{3}|S_{13},\!S_{23})\!-\!I(T_{13};\!T_{23}|S_{13},\!S_{23})\!-\!I(T_{13} ;\!X_{2},\!Y_{2},\!S_{12},\!T_{12}|S_{13},\!S_{23},\!T_{23})-}\no
\\&{I(T_{23} ;X_{1},Y_{1},S_{21},T_{21}|S_{13},T_{13},S_{23})=\textbf{r}_{\textbf{13,s}}+ \textbf{r}_{\textbf{23,s}}-\textbf{I}_{\textbf{3,s}}}\label{secondary}
\end{align}
By setting $\overline{r}_{12,s}=r_{12,s}+r_{21,s},\overline{r}_{13,s}=r_{13,s},\overline{r}_{23,s}=r_{23,s}$ and applying Fourier-Motzkin elimination \cite{fouri-motz} to the above region, the secondary keys rates of Theorem \ref{th3} are derived.

To show that the total rate of the secret key between Users 1 and 2 is the sum of the rates $\overline{r}_{12,p}$ and $\overline{r}_{12,s}$, we should prove the independence of the primary and the secondary keys. Since $K_{12,p}$ and $K_{21,p}$ are indices of $\textbf{S}_{12}$ and $\textbf{S}_{21}$, respectively, (\ref{secsec1}) implies that:

$$I(K_{12,s},K_{21,s};K_{12,p},K_{21,p})\leq \epsilon,$$

\noindent and hence:
\vspace{.4cm}

\noindent
$\begin{array}{l} {H(K_{12,s},K_{21,s},K_{12,p},K_{21,p})\ge H(K_{12,s},K_{21,s})+H(K_{12,p},K_{21,p})-\epsilon.}
\end{array}$
\vspace{.4cm}

Furthermore, we need to prove the independence of the primary and the secondary keys of different blocks. Referring to (\ref{independence}), we have:
$$\sum_{i=1}^{B}I((K_{12,p},K_{21,p},K_{12,s},K_{21,s})^{i};\textbf{Y}_{3}^{1:B},(K_{12,p},K_{21,p},K_{12,s},K_{21,s})^{1:i-1})<\epsilon$$
and hence:
$$\sum_{i=1}^{B}I((K_{12,p},K_{21,p},K_{12,s},K_{21,s})^{i};(K_{12,p},K_{21,p},K_{12,s},K_{21,s})^{1:i-1})<\epsilon$$
which proves the independence of the keys of different blocks.

Similar arguments as above hold for the key between Users 1 and 3, and also the key between Users 2 and 3.

\indent Finally the following rate is achievable between Users $j\in\{1,2\}$ and $l\in\{1,2,3\}$ where $j\neq l$:
$$R_{jl}=\frac{nB\bar{r}_{jl,p}+n(B-1)\bar{r}_{jl,s}}{nB}$$
which is approximately equal to $\bar{r}_{jl,p}+\bar{r}_{jl,s}$ if $B$ is large enough. Hence, the achievability of the secret-key rate region in Theorem \ref{th3} is deduced according to (\ref{primary}) and (\ref{secondary}).

\noindent This completes achievability of the key rate region in Theorem \ref{th3} in strong sense.

\section{Proof of strong secrecy of the primary keys in Theorem \ref{th3}}
\label{App2}

\noindent We note that in the pre-generated keys scheme, the generalized nature of the GDMMAC as feedback is not exploited, and the observations are merely used as side information. Consequently, the mechanisms exploited for secrecy reduce to secure communication over a wiretap channel. In this scheme, strong secrecy can be obtained as a byproduct of channel resolvability. Specifically, given a message $M$ to be secured against an observation $Z^n$, notice that:
\begin{align}\label{channel res.}
  I(M;Z^n)=D(p_{MZ^n}\parallel p_M p_{Z^n})\leq E_M(D(p_{Z^n|M}\parallel{q_{Z^n}}))
\end{align}
for any $q_{Z^n}\in\Delta(\mathcal{Z}^n)$.

We start by establishing a general result that we use in the sequel to prove the strong secrecy of the pre-generated keys in Lemma 2 of Appendix A. We consider the general model illustrated in Fig.~\ref{fig:model_channel_resolvability} in which each user is active and generates keys to share with the other two users. In this model, $k_{ij}$ is the key generated by User $i$ to be shared with User $j$ which should be kept secret from the remaining user as the potential eavesdropper where $i,j\in \{1,2,3\}, i\neq j$ and $S_{ij}$ is the respective auxiliary random variable to $k_{ij}$. The channel output $Y^{n}$ represents the output received by the potential eavesdropper which can be any of the three users. Since the model is symmetric, it is sufficient to prove strong secrecy of the keys shared between each pair of the users and then, the result can be extended to the other pairs' secret keys. In continue, we consider the case where User 3 is the eavesdropper and we intend to prove strong secrecy of the keys between Users 1 and 2, i.e., $(k_{12},k_{21})$. We assume that User 3 has already decoded his intended codewords $s_{13}^{n}(k_{13},k'_{13})$ and $s_{23}^{n}(k_{23},k'_{23})$ from Users 1 and 2, respectively. Hence, User 3's observation is $z^{n}=(y^{n},x_{3}^{n},k_{13},k'_{13},k_{23},k'_{23})$. In Proposition \ref{prop:channel_resolvability}, strong secrecy of the keys between Users 1 and 2 against User 3's observation is given by using the inequality in (\ref{channel res.}). In continue, the notations are borrowed from Appendix A.

\begin{figure}
\centering
\includegraphics[width=13cm]{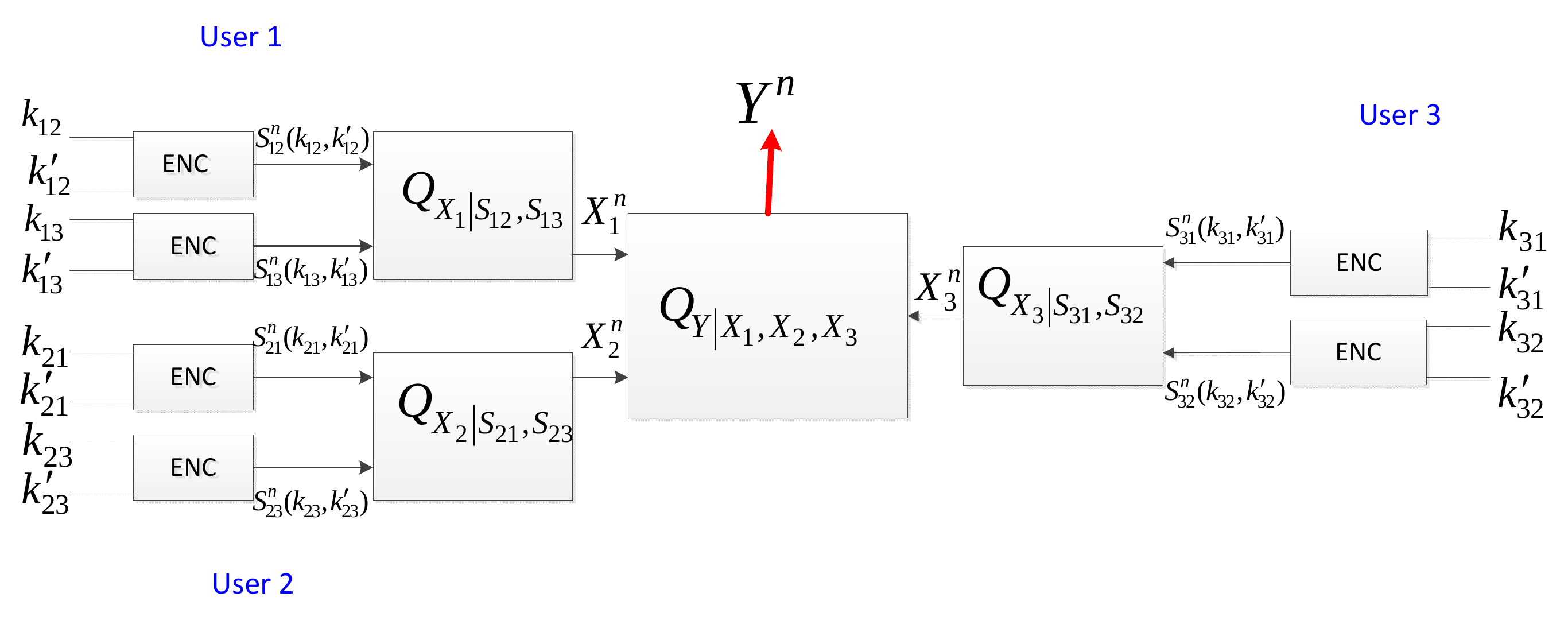}
\vspace{-.3cm}
\caption{\footnotesize{Model for Proposition~\ref{prop:channel_resolvability}}}
  \label{fig:model_channel_resolvability}
\vspace{-.5cm}
\end{figure}
\begin{pro}
  \label{prop:channel_resolvability}
  \begin{align}
    E(D(P^n_{Y^{n}X_{3}^{n}K_{13}K'_{13}K_{23}K'_{23}|K_{12}K_{21}}\parallel{\widehat{P}^{n}_{Y^{n}X_{3}^{n}K_{13}K'_{13}K_{23}K'_{23}}}))\rightarrow 0
  \end{align}
for sufficiently large $n$.
\end{pro}
\begin{IEEEproof}

Let $Q\in\Delta(\mathcal{S}_{12}\times \mathcal{S}_{13}\times \mathcal{S}_{21}\times\mathcal{S}_{23}\times\mathcal{S}_{31}\times\mathcal{S}_{32}\times\mathcal{X}_1\times\mathcal{X}_2\times\mathcal{X}_3\times\mathcal{Y})$ denote the PMF defined by the channel and the random coding argument. The independent and identically distributed (iid) product distribution on $\Delta(\mathcal{S}_{12}^n\times \mathcal{S}_{13}^n\times \mathcal{S}_{21}^n\times\mathcal{S}_{23}^n\times\mathcal{S}_{31}^n\times\mathcal{S}_{32}^n\times\mathcal{X}_1^n\times\mathcal{X}_2^n\times\mathcal{X}_3^n\times\mathcal{Y}^n)$ is denoted by $Q^{\otimes n}$. In contrast let $P^n\in \Delta(\mathcal{S}_{12}^n\times \mathcal{S}_{13}^n\times \mathcal{S}_{21}^n\times\mathcal{S}_{23}^n\times\mathcal{S}_{31}^n\times\mathcal{S}_{32}^n\times\mathcal{X}_1^n\times\mathcal{X}_2^n\times\mathcal{X}_3^n\times\mathcal{Y}^n)$ denote the distribution induced by the coding scheme. By construction:
\begin{multline}
  P^n\left(k_{12},k'_{12},k_{13},k'_{13},k_{21},k'_{21},k_{23},k'_{23},k_{31},k'_{31},k_{32},k'_{32},\mathbf{x}_{1},\mathbf{x}_{2},\mathbf{x}_{3},\mathbf{y}\right)\triangleq\\
Q^{\otimes n}\left(\mathbf{y}|\mathbf{x}_{1},\mathbf{x}_{2},\mathbf{x}_{3}\right)Q^{\otimes n}(\mathbf{x}_1|\mathbf{s}_{12}(k_{12},k'_{12}) \mathbf{s}_{13}(k_{13},k'_{13})) Q^{\otimes n}(\mathbf{x}_2|\mathbf{s}_{21}(k_{21},k'_{21}) \mathbf{s}_{23}(k_{23},k'_{23}))\\Q^{\otimes n}(\mathbf{x}_3|\mathbf{s}_{31}(k_{31},k'_{31}) \mathbf{s}_{32}(k_{32},k'_{32}))
\frac{1}{|\mathcal{K}_{12}|\!|\mathcal{K}'_{12}|\!|\mathcal{K}_{13}|\!|\mathcal{K}'_{13}|\!|\mathcal{K}_{21}|\!|\mathcal{K}'_{21}|\!|\mathcal{K}_{23}|\!|\mathcal{K}'_{23}|\!|\mathcal{K}_{31}|\!|\mathcal{K}'_{31}|\!|\mathcal{K}_{32}|\!|\mathcal{K}'_{32}|}.
\end{multline}
where $|\mathcal{K}_{ij}|$ is the cardinality of key set $\mathcal{K}_{ij}$.

Define:
\begin{multline}
  \widehat{P}^n_{Y^nX_3^nK_{13}K'_{13}K_{23}K'_{23}}(\mathbf{y},\mathbf{x}_3,k_{13},k'_{13},k_{23},k'_{23})\triangleq Q^{\otimes n}_{Y^n|X_3^nS_{13}^nS_{23}^n}(\mathbf{y}|\mathbf{x}_3,\mathbf{s}_{13}(k_{13},k'_{13}),\mathbf{s}_{23}(k_{23},k'_{23}))\\
P^n_{X_3^nK_{13}K'_{13}K_{23}K'_{23}}(\mathbf{x}_3,k_{13},k'_{13}k_{23},k'_{23}).
\end{multline}
 We analyze $E(D(P^n_{Y^nX_3^nK_{13}K'_{13}K_{23}K'_{23}|K_{12}=k_{12}K_{21}=k_{21}}\parallel \widehat{P}^{n}_{Y^nX_3^nK_{13}K'_{13}K_{23}K'_{23}}))$, where the average is over the randomly generated code. Note that:
 \begin{multline}
D(P^n_{Y^nX_3^nK_{13}K'_{13}K_{23}K'_{23}|K_{12}=k_{12}K_{21}=k_{21}}\parallel \widehat{P}^{n}_{Y^nX_3^nK_{13}K'_{13}K_{23}K'_{23}})\\
   \begin{split} &=\sum_{\mathbf{y}\mathbf{x}_3k_{13}k'_{13}k_{23}k'_{23}}P^n(\mathbf{y},\mathbf{x}_3,k_{13},k'_{13},k_{23},k'_{23}|k_{12},k_{21})\log_2\frac{P^n(\mathbf{y},\mathbf{x}_3,k_{13},k'_{13},k_{23},k'_{23}|k_{12},k_{21})}{\widehat{P}^{n}(\mathbf{y},\mathbf{x}_3,k_{13},k'_{13},k_{23},k'_{23})}\\
&=\sum_{\mathbf{x}_3k_{13}k'_{13}k_{23}k'_{23}}P^n(\mathbf{x}_3,k_{13},k'_{13},k_{23},k'_{23})\sum_{\mathbf{y}}P^n(\mathbf{y}|\mathbf{x}_3,k_{13},k'_{13},k_{23},k'_{23},k_{12},k_{21})
   \end{split}\\
\log_2\frac{P^n(\mathbf{y}|\mathbf{x}_3,k_{13},k'_{13},k_{23},k'_{23},k_{12},k_{21})}{Q^{\otimes n}_{Y^{n}|X_3^nS_{13}^nS_{23}^n}(\mathbf{y}|\mathbf{x}_3\mathbf{s}_{13}(k_{13},k'_{13})\mathbf{s}_{23}(k_{23},k'_{23}))}
 \end{multline}
where by construction:
 \begin{align}
   P^n\!(\!\mathbf{x}_3,\!k_{13},\!k'_{13},\!k_{23},\!k'_{23}\!) &\!\!=\!\!\!\!\!\!\!\!\!\!\!\! \sum_{k_{31}k'_{31}k_{32}k'_{32}}\!\!\!\!\!\! \!\!\!\!\!\!Q^{\otimes n}(\!\mathbf{x}_3|\mathbf{s}_{31}(\!k_{31},\!k'_{31}\!) \mathbf{s}_{32}\!(\!k_{32},\!k'_{32}\!)\!)\frac{1}{|\mathcal{K}_{13}|\!|\mathcal{K}'_{13}|\!|\mathcal{K}_{23}|\!|\mathcal{K}'_{23}|\!|\mathcal{K}_{31}|\!|\mathcal{K}'_{31}|\!|\mathcal{K}_{32}|\!|\mathcal{K}'_{32}|}
 \end{align}
and
 \begin{multline}
P^n(\mathbf{y}|\mathbf{x}_3,k_{13},k'_{13},k_{23},k'_{23},k_{12},k_{21})=\\
\sum_{\mathbf{x}_1\mathbf{x}_2k'_{12}k'_{21}}\!\!\!\!\!\!\!\!Q^{\otimes n}(\!\mathbf{y}\!|\mathbf{x}_1 \mathbf{x}_2 \mathbf{x}_3\!)Q^{\otimes n}(\!\mathbf{x}_1\!|\mathbf{s}_{12}(k_{12},\!k'_{12}\!) \mathbf{s}_{13}(k_{13},\!k'_{13})\!) Q^{\otimes n}(\!\mathbf{x}_2\!|\mathbf{s}_{21}(\!k_{21},\!k'_{21}\!) \mathbf{s}_{23}(\!k_{23},\!k'_{23})\!)\frac{1}{|\mathcal{K}'_{12}||\mathcal{K}'_{21}|}
\end{multline}
so that

\small{
\begin{multline}
  D({P^n_{Y^nX_3^nK_{13}K'_{13}K_{23}K'_{23}|K_{12}=k_{12}K_{21}=k_{21}}}\parallel{\widehat{P}^{n}_{Y^nX_3^nK_{13}K'_{13}K_{23}K'_{23}}})=\\
  \frac{1}{|\mathcal{K}'_{12}||\mathcal{K}'_{21}| |\mathcal{K}_{13}|\!|\mathcal{K}'_{13}|\!|\mathcal{K}_{23}|\!|\mathcal{K}'_{23}|\!|\mathcal{K}_{31}|\!|\mathcal{K}'_{31}|\!|\mathcal{K}_{32}|\!|\mathcal{K}'_{32}|}\\
\sum_{\mathbf{y}\mathbf{x}_1\mathbf{x}_2 \mathbf{x}_3k'_{12}k'_{21}k_{13}k'_{13}k_{23}k'_{23} k_{31}k'_{31}k_{32}k'_{32}}Q^{\otimes n}(\mathbf{y}|\mathbf{x}_1 \mathbf{x}_2 \mathbf{x}_3)Q^{\otimes n}(\mathbf{x}_1 |\mathbf{s}_{12}(k_{12},k'_{12}) \mathbf{s}_{13}(k_{13},k'_{13})) \\
Q^{\otimes n}(\mathbf{x}_2 |\mathbf{s}_{21}(k_{21},k'_{21}) \mathbf{s}_{23}(k_{23},k'_{23})) Q^{\otimes n}(\mathbf{x}_3|\mathbf{s}_{31}(k_{31},k'_{31}) \mathbf{s}_{32}(k_{32},k'_{32}))\\
\log_2\!\!\frac{\sum_{\tilde{\mathbf{x}}_1\tilde{\mathbf{x}}_2\tilde{k}'_{12}\tilde{k}'_{21}}Q^{\otimes n}\!(\!\mathbf{y}|\tilde{\mathbf{x}}_1 \tilde{\mathbf{x}}_2 \mathbf{x}_3\!)Q^{\otimes n}\!(\tilde{\mathbf{x}}_1 |\mathbf{s}_{12}(\!k_{12},\!\tilde{k}'_{12}\!) \mathbf{s}_{13}(\!k_{13},\!k'_{13})\!) Q^{\otimes n}\!(\tilde{\mathbf{x}}_2 |\mathbf{s}_{21}(k_{21},\!\tilde{k}'_{21}) \mathbf{s}_{23}(k_{23},\!k'_{23}))\frac{1}{|\mathcal{K}'_{12}||\mathcal{K}'_{21}|}}{Q^{\otimes n}_{Y^n|X_3^nS_{13}^nS_{23}^n}(\!\mathbf{y}|\mathbf{x}_3\mathbf{s}_{13}(k_{13},k'_{13})\mathbf{s}_{23}(k_{23},k'_{23})\!)}
\end{multline}}
\normalsize{
Let us focus on the average of the log term, taking the average over all codewords $\mathbf{s}_{12}$ and $\mathbf{s}_{21}$  \emph{except} than $\mathbf{s}_{12}(k_{12},k'_{12})$ and $\mathbf{s}_{21}(k_{21},k'_{21})$. Then,}
\small{
\begin{align}
  &\no E\!({\log_2\frac{\sum_{\tilde{\mathbf{x}}_1\tilde{\mathbf{x}}_2\tilde{k}'_{12}\tilde{k}'_{21}}Q^{\otimes n}\!(\!\mathbf{y}\!|\!\tilde{\mathbf{x}}_1 \tilde{\mathbf{x}}_2 \mathbf{x}_3)Q^{\otimes n}(\tilde{\mathbf{x}}_1\!|\!\mathbf{s}_{12}\!(\!k_{12},\!\tilde{k}'_{12}\!) \mathbf{s}_{13}(k_{13},\!k'_{13}\!)\!) Q^{\otimes n}(\!\tilde{\mathbf{x}}_2 \!|\!\mathbf{s}_{21}\!(\!k_{21},\tilde{k}'_{21}\!) \mathbf{s}_{23}(k_{23},\!k'_{23}\!)\!)\frac{1}{|\mathcal{K}'_{12}||\mathcal{K}'_{21}|}}{Q^{\otimes n}_{Y^n|X_3^nS_{13}^nS_{23}^n}(\mathbf{y}\!|\!\mathbf{x}_3\mathbf{s}_{13}(\!k_{13},\!k'_{13}\!)\mathbf{s}_{23}(k_{23},k'_{23}))}})\\ \no
  &\mathop{\leq}\limits^{{\rm (a)}} \log_2\frac{\!\!\!\!E({\sum_{\tilde{\mathbf{x}}_1\tilde{\mathbf{x}}_2\tilde{k}'_{12}\tilde{k}'_{21}}Q^{\otimes n}(\!\mathbf{y}\!|\!\tilde{\mathbf{x}}_1 \tilde{\mathbf{x}}_2 \mathbf{x}_3\!)Q^{\otimes n}(\!\tilde{\mathbf{x}}_1\!|\!\mathbf{s}_{12}(\!k_{12},\!\tilde{k}'_{12}\!) \mathbf{s}_{13}(\!k_{13},\!k'_{13}\!)\!) Q^{\otimes n}(\!\tilde{\mathbf{x}}_2\!|\!\mathbf{s}_{21}(\!k_{21},\!\tilde{k}'_{21}\!) \mathbf{s}_{23}(\!k_{23},\!k'_{23}\!)\!)\frac{1}{|\mathcal{K}'_{12}||\mathcal{K}'_{21}|}})}{Q^{\otimes n}_{Y^n|X_3^nS_{13}^nS_{23}^n}(\!\mathbf{y}\!|\!\mathbf{x}_3\mathbf{s}_{13}(\!k_{13},\!k'_{13}\!)\mathbf{s}_{23}(\!k_{23},\!k'_{23}\!))}\\\no
  &=\log_2\left(\frac{1}{|\mathcal{K}'_{12}||\mathcal{K}'_{21}|}\frac{\sum_{\tilde{\mathbf{x}}_1\tilde{\mathbf{x}}_2}Q^{\otimes n}(\!\mathbf{y}\!|\!\tilde{\mathbf{x}}_1 \tilde{\mathbf{x}}_2 \mathbf{x}_3\!)Q^{\otimes n}(\!\tilde{\mathbf{x}}_1\!|\!\mathbf{s}_{12}(\!k_{12},\!{k}'_{12}\!)\mathbf{s}_{13}(\!k_{13},\!k'_{13}\!)\!) Q^{\otimes n}(\tilde{\mathbf{x}}_2 |\mathbf{s}_{21}(k_{21},{k}'_{21}) \mathbf{s}_{23}(k_{23},k'_{23}))}{Q^{\otimes n}_{Y^n|X_3^nS_{13}^nS_{23}^n}(\mathbf{y}|\mathbf{x}_3\mathbf{s}_{13}(k_{13},k'_{13})\mathbf{s}_{23}(k_{23},k'_{23}))}\right.\\\no
  &\phantom{===}+\frac{|\mathcal{K}'_{12}|-1}{|\mathcal{K}'_{12}||\mathcal{K}'_{21}|}\frac{\sum_{\tilde{\mathbf{x}}_1\tilde{\mathbf{x}}_2}Q^{\otimes n}(\mathbf{y}|\tilde{\mathbf{x}}_1 \tilde{\mathbf{x}}_2 \mathbf{x}_3)Q^{\otimes n}(\tilde{\mathbf{x}}_1 | \mathbf{s}_{13}(k_{13},k'_{13})) Q^{\otimes n}(\tilde{\mathbf{x}}_2 |\mathbf{s}_{21}(k_{21},{k}'_{21}) \mathbf{s}_{23}(k_{23},k'_{23}))}{Q^{\otimes n}_{Y^n|X_3^nS_{13}^nS_{23}^n}(\mathbf{y}|\mathbf{x}_3\mathbf{s}_{13}(k_{13},k'_{13})\mathbf{s}_{23}(k_{23},k'_{23}))}\\\no
  &\phantom{===}+\frac{|\mathcal{K}'_{21}|-1}{|\mathcal{K}'_{12}||\mathcal{K}'_{21}|}\frac{\sum_{\tilde{\mathbf{x}}_1\tilde{\mathbf{x}}_2}Q^{\otimes n}(\mathbf{y}|\tilde{\mathbf{x}}_1 \tilde{\mathbf{x}}_2 \mathbf{x}_3)Q^{\otimes n}(\tilde{\mathbf{x}}_1 | \mathbf{s}_{12}(k_{12},k'_{12})\mathbf{s}_{13}(k_{13},k'_{13})) Q^{\otimes n}(\tilde{\mathbf{x}}_2 |\mathbf{s}_{23}(k_{23},k'_{23}))}{Q^{\otimes n}_{Y^n|X_3^nS_{13}^nS_{23}^n}(\mathbf{y}|\mathbf{x}_3\mathbf{s}_{13}(k_{13},k'_{13})\mathbf{s}_{23}(k_{23},k'_{23}))}\\\no
  &\left.\phantom{===}+\frac{(|\mathcal{K}'_{12}|-1) (|\mathcal{K}'_{21}|-1)}{|\mathcal{K}'_{12}||\mathcal{K}'_{21}|}\frac{\sum_{\tilde{\mathbf{x}}_1\tilde{\mathbf{x}}_2}Q^{\otimes n}(\mathbf{y}|\tilde{\mathbf{x}}_1 \tilde{\mathbf{x}}_2 \mathbf{x}_3)Q^{\otimes n}(\tilde{\mathbf{x}}_1 | \mathbf{s}_{13}(k_{13},k'_{13})) Q^{\otimes n}(\tilde{\mathbf{x}}_2 |\mathbf{s}_{23}(k_{23},k'_{23}))}{Q^{\otimes n}_{Y^n|X_3^nS_{13}^nS_{23}^n}(\mathbf{y}|\mathbf{x}_3\mathbf{s}_{13}(k_{13},k'_{13})\mathbf{s}_{23}(k_{23},k'_{23}))}\right)\\\no
&\leq \log_2\left(\frac{1}{|\mathcal{K}'_{12}||\mathcal{K}'_{21}|}\frac{Q^{\otimes n}(\mathbf{y}|\mathbf{x}_3\mathbf{s}_{12}(k_{12},{k}'_{12}) \mathbf{s}_{13}(k_{13},k'_{13})\mathbf{s}_{21}(k_{21},{k}'_{21}) \mathbf{s}_{23}(k_{23},k'_{23}))}{Q^{\otimes n}_{Y^n|X_3^nS_{13}^nS_{23}^n}(\mathbf{y}|\mathbf{x}_3\mathbf{s}_{13}(k_{13},k'_{13})\mathbf{s}_{23}(k_{23},k'_{23}))}\right.\\\no
  &\phantom{===}+\frac{1}{|\mathcal{K}'_{21}|}\frac{Q^{\otimes n}(\mathbf{y}|\mathbf{x}_3\mathbf{s}_{13}(k_{13},k'_{13})\mathbf{s}_{21}(k_{21},{k}'_{21}) \mathbf{s}_{23}(k_{23},k'_{23}))}{Q^{\otimes n}_{Y^n|X_3^nS_{13}^nS_{23}^n}(\mathbf{y}|\mathbf{x}_3\mathbf{s}_{13}(k_{13},k'_{13})\mathbf{s}_{23}(k_{23},k'_{23}))}\\\no
  &\phantom{===}+\frac{1}{|\mathcal{K}'_{12}|}\frac{Q^{\otimes n}(\mathbf{y}|\mathbf{x}_3 \mathbf{s}_{12}(k_{12},{k}'_{12}) \mathbf{s}_{13}(k_{13},k'_{13})\mathbf{s}_{23}(k_{23},k'_{23}))}{Q^{\otimes n}_{Y^n|X_3^nS_{13}^nS_{23}^n}(\mathbf{y}|\mathbf{x}_3\mathbf{s}_{13}(k_{13},k'_{13})\mathbf{s}_{23}(k_{23},k'_{23}))}\\\no
  &\left.\phantom{===}+\frac{Q^{\otimes n}(\mathbf{y}|\mathbf{x}_3 \mathbf{s}_{13}(k_{13},k'_{13})\mathbf{s}_{23}(k_{23},k'_{23}))}{Q^{\otimes n}_{Y^n|X_3^nS_{13}^nS_{23}^n}(\mathbf{y}|\mathbf{x}_3\mathbf{s}_{13}(k_{13},k'_{13})\mathbf{s}_{23}(k_{23},k'_{23}))}\right)\label{ineq1}\\
\end{align}}
\normalsize{
where (a) is deduced from Jensen's inequality.}

Substituting the upper bound of (\ref{ineq1}) back and taking the average over all other codewords, we obtain,
\begin{multline}
  E(D({P^n_{Y^nX_3^nK_{13}K'_{13}K_{23}K'_{23}|K_{12}K_{21}}}\parallel{\widehat{P}^{n}_{Y^nX_3^nK_{13}K'_{13}K_{23}K'_{23}}}))\\
  \leq\sum_{\mathbf{y}\mathbf{x_{3}}\mathbf{s_{12}}\mathbf{s_{13}}\mathbf{s_{21}}\mathbf{s_{23}}}Q^{\otimes n}(\mathbf{y}\mathbf{x_{3}}\mathbf{s}_{12} \mathbf{s}_{13}\mathbf{s}_{21}\mathbf{s}_{23}) \log_2\left(\frac{1}{|\mathcal{K}'_{12}||\mathcal{K}'_{21}|}\frac{Q^{\otimes n}(\mathbf{y}|\mathbf{x}_3\mathbf{s}_{12} \mathbf{s}_{13}\mathbf{s}_{21}\mathbf{s}_{23})}{Q^{\otimes n}_{Y^n|X_3^nS_{13}^nS_{23}^n}(\mathbf{y}|\mathbf{x}_3,\mathbf{s}_{13},\mathbf{s}_{23})}\right.\\
\left.  \phantom{===}+\frac{1}{|\mathcal{K}'_{21}|}\frac{Q^{\otimes n}(\mathbf{y}|\mathbf{x}_3\mathbf{s}_{13}\mathbf{s}_{21} \mathbf{s}_{23})}{Q^{\otimes n}_{Y^n|X_3^nS_{13}^nS_{23}^n}(\mathbf{y}|\mathbf{x}_3,\mathbf{s}_{13},\mathbf{s}_{23})}+\frac{1}{|\mathcal{K}'_{12}|}\frac{Q^{\otimes n}(\mathbf{y}|\mathbf{x}_3 \mathbf{s}_{12}\mathbf{s}_{13}\mathbf{s}_{23}))}{Q^{\otimes n}_{Y^n|X_3^nS_{13}^nS_{23}^n}(\mathbf{y}|\mathbf{x}_3,\mathbf{s}_{13},\mathbf{s}_{23})}+1\right)\label{sum}
\end{multline}
  We split the sum between the sequences that are jointly typical, and those that are not.
  \begin{itemize}
  \item If $(\mathbf{y},\mathbf{x}_3,\mathbf{s}_{12},\mathbf{s}_{13},\mathbf{s}_{21},\mathbf{s}_{23})\in \mathcal{T}_{\epsilon'}^n(Q_{YX_3S_{12}S_{13}S_{21}S_{23}})$, the log argument is upper bounded by
\begin{align}
\hspace{-1cm}\frac{2^{n(I(S_{12},S_{21};Y|X_3,S_{13},S_{23})+2\epsilon')}}{|\mathcal{K}'_{12}||\mathcal{K}'_{21}|}\!+\!\frac{2^{n(I(S_{21};Y|X_3,S_{13},S_{23})+2\epsilon')}}{|\mathcal{K}'_{21}|} \!+\!\frac{2^{n(I(S_{12};Y|X_3,S_{13},S_{23})+2\epsilon')}\!}{|\mathcal{K}'_{12}|}+\!1
\end{align}
If we have
\begin{align}
      r'_{12}+r'_{21} &> I(S_{12},S_{21};Y|X_3,S_{13},S_{23})+2\epsilon'.\\
      r'_{21}&>I(S_{21};Y|X_3,S_{13},S_{23})+2\epsilon'\\
      r'_{12}&>I(S_{12};Y|X_3,S_{13},S_{23})+2\epsilon'.
\end{align}

then for sufficiently large $n$, the log argument can be bounded by $3\epsilon'+1$.

  \item If $(\mathbf{y},\mathbf{x}_3,\mathbf{s}_{12},\mathbf{s}_{13},\mathbf{s}_{21},\mathbf{s}_{23})\notin \mathcal{T}_{\epsilon'}^n(Q_{YX_3S_{12}S_{13}S_{21}S_{23}})$, the log term is upper bounded by $\log_2(3\mu^{-n}+1)$ where
    \begin{align}
      \mu\triangleq \min_{Q(y|x_3s_{13}s_{23})>0}Q(y|x_3s_{13}s_{23}).
    \end{align}
    Hence, the sum over the non typical sequences is bounded by $\epsilon'\log_2(3\mu^{-n}+1)$.
  \end{itemize}

\indent Finally, the sum in (\ref{sum}) is bounded as
 \begin{align}
  E(D({P^n_{Y^nX_3^nK_{13}K'_{13}K_{23}K'_{23}|K_{12}K_{21}}}\parallel{\widehat{P}^{n}_{Y^nX_3^nK_{13}K'_{13}K_{23}K'_{23}}}))\leq \epsilon'\log_2(3\mu^{-n}+1)+\log(3\epsilon'+1)
\end{align}
which vanishes as $n$ goes to infinity and $\epsilon'$ goes to zero.
\end{IEEEproof}

Now, we follow secure communication between each pair of the users in the pre-generated keys scheme using the general result in Proposition \ref{prop:channel_resolvability}. Since the model in Fig.~\ref{fig:model_channel_resolvability} is symmetric, we can appropriately substitute the random variables to deduce secrecy from resolvability condition against the intended user. In particular:

\begin{itemize}
  \item by substituting $Y=Y_{3},X_{3}=S_{31}=S_{32}=K_{31}=K_{32}=K'_{31}=K'_{32}=\phi$ in Fig.~\ref{fig:model_channel_resolvability} and using Proposition \ref{prop:channel_resolvability}, secrecy from resolvability condition against User 3 is obtained.
  \item by substituting $Y=Y_{2},X_{3}=S_{31}=S_{32}=K_{31}=K_{32}=K'_{31}=K'_{32}=\phi$ in Fig.~\ref{fig:model_channel_resolvability} and using Proposition \ref{prop:channel_resolvability}, secrecy from resolvability condition against User 2 is obtained.
  \item by substituting $Y=Y_{1},X_{3}=S_{31}=S_{32}=K_{31}=K_{32}=K'_{31}=K'_{32}=\phi$ in Fig.~\ref{fig:model_channel_resolvability} and using Proposition \ref{prop:channel_resolvability}, secrecy from resolvability condition against User 1 is obtained.
\end{itemize}

\begin{lem}[Secrecy from channel resolvability conditions]
  \label{lm:1}
If \eqref{r12+r21,p}-\eqref{r23,p} hold, then there exist $\alpha>0, \beta>0$ and $\gamma>0$ such that
  \begin{align}
    E(D({P^n_{Y_{3}^nK_{13}K'_{13}K_{23}K'_{23}|K_{12}K_{21}}}\parallel{\widehat{P}^{n}_{Y_{3}^nK_{13}K'_{13}K_{23}K'_{23}}})) \leq 2^{-\alpha n}\label{n.21}\\
    E(D({P^n_{Y_{2}^nX_2^nK_{12}K'_{12}|K_{13}}}\parallel{\widehat{P}^{n}_{Y_{2}^nX_2^nK_{12}K'_{12}}})) \leq 2^{-\beta n}\label{n.22}\\
    E(D({P^n_{Y_{1}^nX_1^nK_{21}K'_{21}|K_{23}}}\parallel{\widehat{P}^{n}_{Y_{1}^nX_1^nK_{21}K'_{21}}})) \leq 2^{-\gamma n}\label{n.23}
  \end{align}
\end{lem}

To deduce (\ref{prisec1})-(\ref{prisec3}) in Lemma \ref{lm:4} in Appendix A, we substitute $Y_3=(Y_{3},T_{13},T_{23}),Y_{2}=(Y_{2},T_{12})$ and $Y_{1}=(Y_{1},T_{21})$ in \eqref{n.21}-\eqref{n.23} and use the independence of auxiliary random variables $S_{12},S_{13},S_{21}$ and $S_{23}$.

\section{Proof of strong secrecy of the secondary keys in Theorem \ref{th3}}
\label{App4}
In the generalized scheme, both wiretap coding and secret key generation are used where the latter uses the channel outputs correlation as induced sources.  For the wiretap codebook, the channel resolvability was used in Appendix B to prove strong secrecy of the primary keys. For the secondary keys, the outputs of the GDMMAC are exploited as source observations to share secret keys and the transmitted information by Users 1 and 2 needs to satisfy \eqref{dec6}-\eqref{dec11}. We consider the secondary keys generation between Users 1 and 2 and prove strong secrecy for the secondary key pair $(k_{12,s},k_{21,s})$ where the notations are borrowed from Appendix A. Here, we drop the index $s$ since we just deal with the secondary keys. The secondary keys generation between Users 1 and 2 against User 3 is shown in Fig.~\ref{fig:model_source_resolvability}.

\begin{figure}
\centering
\includegraphics[width=13cm]{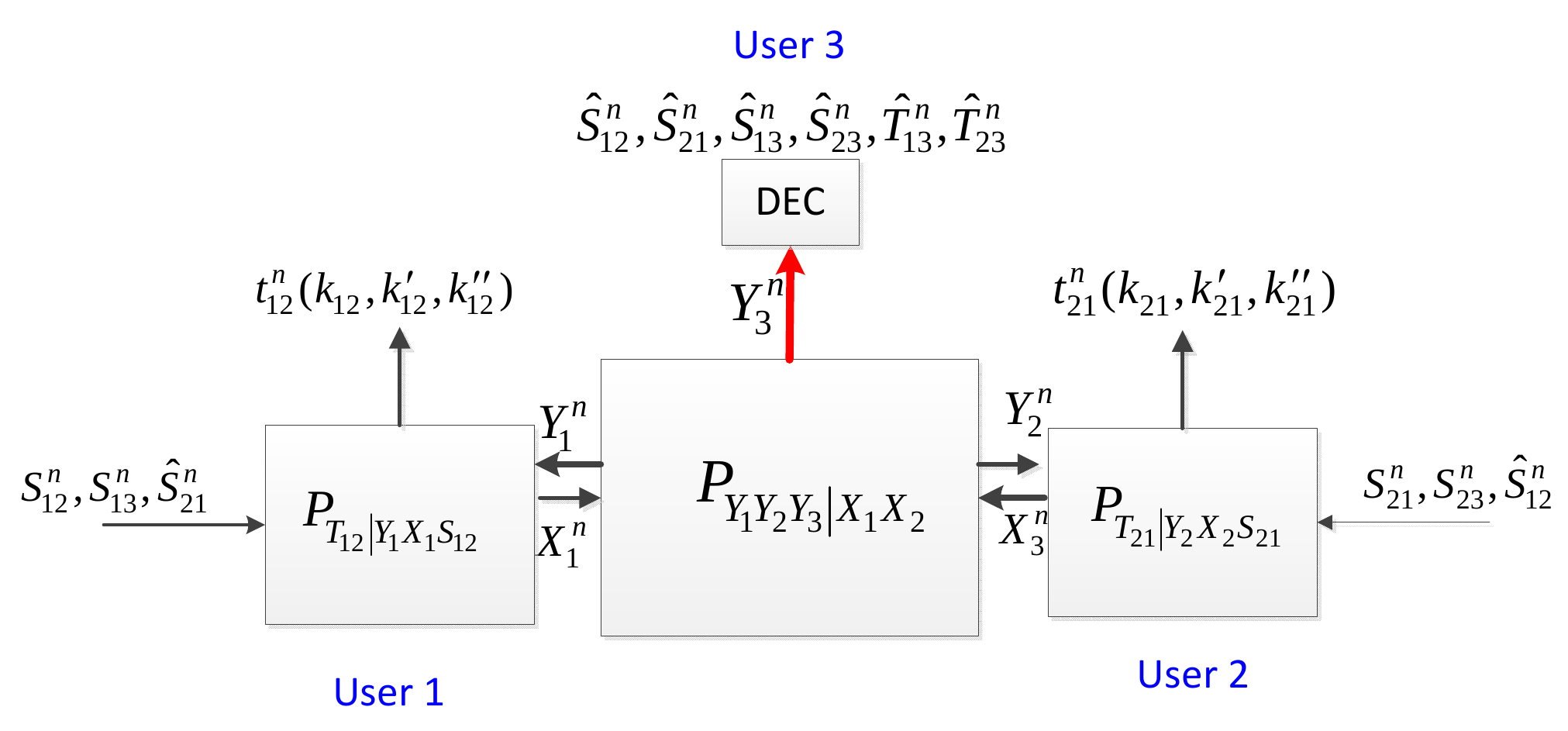}
\vspace{-.3cm}
\caption{\footnotesize{Model for Proposition~\ref{pro2}}}
  \label{fig:model_source_resolvability}
\vspace{-.5cm}
\end{figure}

The objective is to extract uniformly distributed keys $k_{12}\in\{1,\!...,\!2^{nr_{12}}\}$ and $k_{21}\in\{1,\!...,\!2^{nr_{21}}\}$ at User 2 and User 1, respectively, where the keys are independent of each other and of User 3's observation, i.e.,
\begin{align}
\mathbf{z}=z^{n}=(y_{3}^{n},t_{13}^{n},t_{23}^{n},s_{13}^{n},s_{23}^{n},s_{12}^{n},s_{21}^{n})\label{z}
\end{align}

We assume that User 3 has already decoded his intended codewords $t_{13}^{n}$ and $t_{23}^{n}$ of the secondary keys from Users 1 and 2, respectively, and all the codewords of the primary keys, i.e., $(s_{13}^{n},s_{23}^{n},s_{12}^{n},s_{21}^{n})$. To prove Lemma \ref{lm:5} in Appendix A, we show that:

\begin{pro}\label{pro2}
  \begin{align}
     &{I(K_{12},K_{21},K'_{12},K'_{21},K''_{12},K''_{21};Z^n)\leq}\no
     \\& {\ \ \ \ \ \ \ \ \ \ \ \ \ \ \ E(D(P_{K_{12},K_{21},K'_{12},K'_{21},K''_{12},K''_{21},Z^{n}}\parallel{q_{K_{12}}q_{K_{21}}q_{K'_{12}}q_{K'_{21}}q_{K''_{12}}q_{K''_{21}}P_{Z^{n}}}))\rightarrow 0 } \label{strong-source}
  \end{align}
for sufficiently large $n$, where $q_{K_{ij}}$, $q_{K'_{ij}}$ and $q_{K''_{ij}}$ have uniform distributions over $\{1,\!...,\!2^{nr_{ij}}\},\{1,\!...,\!2^{nr'_{ij}}\}$ and $\{1,\!...,\!2^{nr''_{ij}}\}$, respectively, for $i=1,2$, $j=1,2$ and $i\neq j$.
\end{pro}

\begin{IEEEproof}
To prove Proposition \ref{pro2}, we use the random binning mappings $\phi_{ij},\psi_{ij}$ and $\theta_{ij}$ as defined in Appendix A.

Using the law of total probability, the joint distribution $p_{k_{12},k_{21},k'_{12},k'_{21},k''_{12},k''_{21},z^{n}}$  can be expressed as follows:
\begin{align}
  &{p(k_{12},k_{21},k'_{12},k'_{21},k''_{12},k''_{21},z^{n})=\sum_{\mathbf{t}_{12}\mathbf{t}_{21}} p(\mathbf{t}_{12},\mathbf{t}_{21},\mathbf{z})\Pr\{\phi_{12}(\mathbf{t}_{12})=k_{12}\}\Pr\{\psi_{12}(\mathbf{t}_{12})=k'_{12}\}}\no
  \\&{\Pr\{\theta_{12}(\mathbf{t}_{12})=k''_{12}\}\Pr\{\phi_{21}(\mathbf{t}_{21})=k_{21}\}\Pr\{\psi_{21}(\mathbf{t}_{21})=k'_{21}\}\Pr\{\theta_{21}(\mathbf{t}_{21})=k''_{21}\}}
\end{align}
To simplify the notation, we define:
\begin{align}
  {P(\phi_{ij},\psi_{ij},\theta_{ij},\mathbf{t}_{ij},k_{ij},k'_{ij},k''_{ij})\!\triangleq\! \Pr\{\phi_{ij}(\mathbf{t}_{ij})\!=\!k_{ij}\}\!\Pr\{\psi_{ij}(\mathbf{t}_{ij})\!=\!k'_{ij}\}\!\Pr\{\theta_{ij}(\mathbf{t}_{ij})\!=\!k''_{ij}\}}
\end{align}

Now, we evaluate \small{$E_{\Phi_{12}\Psi_{12}\Theta{12}\Phi_{21}\Psi_{21}\Theta{21}}(\!D(\!P_{K_{12},K_{21},K'_{12},K'_{21},K''_{12},K''_{21},Z^{n}}\!\!\parallel\!\!{q_{K_{12}}q_{K_{21}}q_{K'_{12}}q_{K'_{21}}q_{K''_{12}}q_{K''_{21}}P_{Z^{n}}}))$}.

\indent \normalsize{By definition, we have:}
\small{\begin{multline}\label{ddef1}
D(P_{K_{12},K_{21},K'_{12},K'_{21},K''_{12},K''_{21},Z^{n}}\parallel{q_{K_{12}}q_{K_{21}}q_{K'_{12}}q_{K'_{21}}q_{K''_{12}}q_{K''_{21}}P_{Z^{n}}})\\
=\sum_{k_{12}k_{21}k'_{12}k'_{21}k''_{12}k''_{21}\mathbf{z}}p(k_{12},k_{21},k'_{12},k'_{21},k''_{12},k''_{21},\mathbf{z})\log_2\frac{p(k_{12},k_{21},k'_{12},k'_{21},k''_{12},k''_{21},\mathbf{z})}{q_{k_{12}}q_{k_{21}}q_{k'_{12}}q_{k'_{21}}q_{k''_{12}}q_{k''_{21}}p_{\mathbf{z}}}\\
=\sum_{k_{12}k_{21}k'_{12}k'_{21}k''_{12}k''_{21}\mathbf{t}_{12}\mathbf{t}_{21}\mathbf{z}} \!\!\!\!\!\!\!\!\!\!\!\!\!\!\!\!\!\!p(\mathbf{t}_{12},\mathbf{t}_{21},\mathbf{z})P(\phi_{12},\psi_{12},\theta_{12},\mathbf{t}_{12},k_{12},k'_{12},k''_{12})P(\phi_{21},\psi_{21},\theta_{21},\mathbf{t}_{21},k_{21},k'_{21},k''_{21})\times
\\ \log_2\frac{\sum_{\tilde{\mathbf{t}}_{12}\tilde{\mathbf{t}}_{21}} p(\tilde{\mathbf{t}}_{12},\tilde{\mathbf{t}}_{21},\mathbf{z})P(\phi_{12},\psi_{12},\theta_{12},\tilde{\mathbf{t}}_{12},k_{12},k'_{12},k''_{12})P(\phi_{21},\psi_{21},\theta_{21},\tilde{\mathbf{t}}_{21},k_{21},k'_{21},k''_{21})}{q_{k_{12}}q_{k_{21}}q_{k'_{12}}q_{k'_{21}}q_{k''_{12}}q_{k''_{21}}p_{\mathbf{z}}}
 \end{multline}}
 \normalsize{
Then, we focus on the log term in (\ref{ddef1}) and take the average over possible values of $\Phi_{ij}(\tilde{t}_{ij}^{n})\Psi_{ij}(\tilde{t}_{ij}^{n})$ and $\Theta_{ij}(\tilde{t}_{ij}^{n})$ for all $\tilde{t}_{12}^{n}\neq t_{12}^{n}$ and $\tilde{t}_{21}^{n}\neq t_{21}^{n}$ which is denoted by $\tilde{E}$. We have:}
\begin{multline}\label{ddef2}
\tilde{E}(\log_2\frac{\sum_{\tilde{\mathbf{t}}_{12}\tilde{\mathbf{t}}_{21}} p(\tilde{\mathbf{t}}_{12},\!\tilde{\mathbf{t}}_{21},\!\mathbf{z})P(\phi_{12},\!\psi_{12},\!\theta_{12},\!\tilde{\mathbf{t}}_{12},\!k_{12},\!k'_{12},\!k''_{12}\!)P(\phi_{21},\!\psi_{21},\!\theta_{21},\tilde{\mathbf{t}}_{21},\!k_{21},\!k'_{21},\!k''_{21})}{q_{k_{12}}q_{k_{21}}q_{k'_{12}}q_{k'_{21}}q_{k''_{12}}q_{k''_{21}}p_{\mathbf{z}}})
\\\mathop{\leq}\limits^{{\rm (a)}}\log_2\frac{\tilde{E}(\sum_{\tilde{\mathbf{t}}_{12}\tilde{\mathbf{t}}_{21}} p(\tilde{\mathbf{t}}_{12},\!\tilde{\mathbf{t}}_{21},\!\mathbf{z}\!)P(\!\phi_{12},\!\psi_{12},\!\theta_{12},\!\tilde{\mathbf{t}}_{12},\!k_{12},\!k'_{12},\!k''_{12}\!)P(\!\phi_{21},\!\psi_{21},\!\theta_{21},\!\tilde{\mathbf{t}}_{21},\!k_{21},\!k'_{21},\!k''_{21}))}{q_{k_{12}}q_{k_{21}}q_{k'_{12}}q_{k'_{21}}p_{\mathbf{z}}}
 \end{multline}
where (a) is deduced from Jensen's inequality.

We extend various terms inside $\tilde{E}$ in (\ref{ddef2}) as follow:
\begin{multline}\label{ddef3}
\tilde{E}(\sum_{\tilde{\mathbf{t}}_{12}\tilde{\mathbf{t}}_{21}} p(\tilde{\mathbf{t}}_{12},\tilde{\mathbf{t}}_{21},\mathbf{z})P(\phi_{12},\psi_{12},\theta_{12},\tilde{\mathbf{t}}_{12},k_{12},k'_{12},k''_{12})P(\phi_{21},\psi_{21},\theta_{21},\tilde{\mathbf{t}}_{21},k_{21},k'_{21},k''_{21}))
\\= p(\mathbf{t}_{12},\mathbf{t}_{21},\mathbf{z})P(\phi_{12},\psi_{12},\theta_{12},\mathbf{t}_{12},k_{12},k'_{12},k''_{12})P(\phi_{21},\psi_{21},\theta_{21},\mathbf{t}_{21},k_{21},k'_{21},k''_{21})+
\\\sum_{\tilde{\mathbf{t}}_{21}\neq\mathbf{t}_{21}} p(\mathbf{t}_{12},\tilde{\mathbf{t}}_{21},\mathbf{z})P(\phi_{12},\psi_{12},\theta_{12},\mathbf{t}_{12},k_{12},k'_{12},k''_{12})\frac{1}{2^{n(r_{21}+r'_{21}+r''_{21})}}+
\\\sum_{\tilde{\mathbf{t}}_{12}\neq\mathbf{t}_{12}} p(\tilde{\mathbf{t}}_{12},\mathbf{t}_{21},\mathbf{z})P(\phi_{21},\psi_{21},\theta_{21},\mathbf{t}_{21},k_{21},k'_{21},k''_{21})\frac{1}{2^{n(r_{12}+r'_{12}+r''_{12})}}+
\\\sum_{\tilde{\mathbf{t}}_{12}\neq\mathbf{t}_{12}\tilde{\mathbf{t}}_{21}\neq\mathbf{t}_{21}} p(\tilde{\mathbf{t}}_{12},\tilde{\mathbf{t}}_{21},\mathbf{z})\frac{1}{2^{n(r_{12}+r'_{12}+r''_{12}+r_{21}+r'_{21}+r''_{21})}}
\\\leq p(\mathbf{t}_{12},\mathbf{t}_{21},\mathbf{z})P(\phi_{12},\psi_{12},\theta_{12},\mathbf{t}_{12},k_{12},k'_{12},k''_{12})P(\phi_{21},\psi_{21},\theta_{21},\mathbf{t}_{21},k_{21},k'_{21},k''_{21})+
\\p(\mathbf{t}_{12},\mathbf{z})P(\phi_{12},\psi_{12},\theta_{12},\mathbf{t}_{12},k_{12},k'_{12},k''_{12})\frac{1}{2^{n(r_{21}+r'_{21}+r''_{21})}}+
\\p(\mathbf{t}_{21},\mathbf{z})P(\phi_{21},\psi_{21},\theta_{21},\mathbf{t}_{21},k_{21},k'_{21},k''_{21})\frac{1}{2^{n(r_{12}+r'_{12}+r''_{12})}}+ p(\mathbf{z})\frac{1}{2^{n(r_{12}+r'_{12}+r''_{12}+r_{21}+r'_{21}+r''_{21})}}
\end{multline}
Since $q_{k_{ij}}=\frac{1}{2^{nr_{ij}}},,q_{k'_{ij}}=\frac{1}{2^{nr'_{ij}}}$ and $q_{k''_{ij}}=\frac{1}{2^{nr''_{ij}}}$, the log term in (\ref{ddef2}) is obtained as:
\small{
\begin{multline}\label{ddef4}
\log_2\frac{\tilde{E}(\sum_{\tilde{\mathbf{t}}_{12}\tilde{\mathbf{t}}_{21}} p(\tilde{\mathbf{t}}_{12},\tilde{\mathbf{t}}_{21},\mathbf{z})P(\phi_{12},\psi_{12},\theta_{12},\tilde{\mathbf{t}}_{12},k_{12},k'_{12},k''_{12})P(\phi_{21},\psi_{21},\theta_{21},\tilde{\mathbf{t}_{21}},k_{21},k'_{21},k''_{21}))}{q_{k_{12}}q_{k_{21}}q_{k'_{12}}q_{k'_{21}}p_{\mathbf{z}}}
\\\leq\log_2\!\!\Big(\! p(\mathbf{t}_{12},\!\mathbf{t}_{21}\!|\mathbf{z}\!)P(\!\phi_{12},\!\psi_{12},\!\theta_{12},\!\mathbf{t}_{12},\!k_{12},\!k'_{12},\!k''_{12}\!)P(\!\phi_{21},\!\psi_{21},\!\theta_{21},\!\mathbf{t}_{21},\!k_{21},\!k'_{21},\!k''_{21}\!)2^{n(r_{12}\!+\!r_{21}\!+\!r'_{12}\!+\!r'_{21}\!+\!r''_{12}\!+\!r''_{21}\!)}\!+
\\p(\mathbf{t}_{12}|\mathbf{z})P(\phi_{12},\psi_{12},\theta_{12},\mathbf{t}_{12},k_{12},k'_{12},k''_{12})2^{n(r_{12}+r'_{12}+r''_{12})}+
\\p(\mathbf{t}_{21}|\mathbf{z})P(\phi_{21},\psi_{21},\theta_{21},\mathbf{t}_{21},k_{21},k'_{21},k''_{21})2^{n(r_{21}+r'_{21}+r''_{21})}+1\Big)
\end{multline}}
\normalsize{By substituting (\ref{ddef4}) in (\ref{ddef1}) and taking the average over all random binning mappings, we have:}
\small{
\begin{multline}\label{ddef5}
E_{\Phi_{12}\Psi_{12}\Theta_{12}\Phi_{21}\Psi_{21}\Theta_{21}}(D(P_{K_{12},K_{21},K'_{12},K'_{21},K''_{12},K''_{21},Z^{n}}\parallel{q_{K_{12}}q_{K_{21}}q_{K'_{12}}q_{K'_{21}}q_{K''_{12}}q_{K''_{21}}P_{Z^{n}}}))\leq
\\\sum_{k_{12}k_{21}k'_{12}k'_{21}k''_{12}k''_{21}\mathbf{t}_{12}\mathbf{t}_{21}\mathbf{z}}\!\!\!\!\!\!\!\!\!\!\!\!\!\!\! p(\mathbf{t}_{12},\mathbf{t}_{21},\mathbf{z})\!\!\!\!\!\!\!\!\!\!\!\!\!\sum_{\phi_{12}(\!\mathbf{t}_{12})\psi_{12}(\mathbf{t}_{12}\!)\theta_{12}(\mathbf{t}_{12})}\!\!\!\!\!\!\!\!\!\!\!\!\!\!\!\!\!\!\!\!\!\!\!\frac{P(\!\phi_{12},\!\psi_{12},\!\theta_{12},\!\mathbf{t}_{12},\!k_{12},\!k'_{12},\!k''_{12})\!}{2^{n(r_{12}\!+\!r'_{12}\!+r''_{12})}}\!\!\!\!\!\!\!\!\!\!\!\!\!\!\!\!\!\!\sum_{\phi_{21}(\mathbf{t}_{21})\psi_{21}(\mathbf{t}_{21})\theta_{21}(\mathbf{t}_{21})}\!\!\!\!\!\!\!\!\!\!\!\!\!\!\!\!\!\!\!\frac{P(\!\phi_{21},\!\psi_{21},\!\theta_{21},\!\mathbf{t}_{21},\!k_{21},\!k'_{21},\!k''_{21}\!)}{2^{n(r_{21}\!+\!r'_{21}\!+\!r''_{21})}}
\\\times \log_2\!\!\Big( p(\!\mathbf{t}_{12},\mathbf{t}_{21}|\mathbf{z}\!)P(\!\phi_{12},\psi_{12},\theta_{12},\mathbf{t}_{12},k_{12},k'_{12},k''_{12}\!)P(\!\phi_{21},\psi_{21},\theta_{21},\mathbf{t}_{21},k_{21},k'_{21},k''_{21}\!)2^{n(r_{12}+r_{21}+r'_{12}+r'_{21}+r''_{12}+r''_{21})}+
\\p(\mathbf{t}_{12}|\mathbf{z})P(\phi_{12},\psi_{12},\theta_{12},\mathbf{t}_{12},k_{12},k'_{12},k''_{12})2^{n(r_{12}+r'_{12}+r''_{12})}+
\\p(\mathbf{t}_{21}|\mathbf{z})P(\phi_{21},\psi_{21},\theta_{21},\mathbf{t}_{21},k_{21},k'_{21},k''_{21})2^{n(r_{21}+r'_{21}+r''_{21})}+1\Big)
\\\leq\!\!\sum_{k_{12}k_{21}k'_{12}k'_{21}k''_{12}k''_{21}\mathbf{t}_{12}\mathbf{t}_{21}\mathbf{z}}\!\!\!\!\!\!\!\!\!\!\!\!\!\!\! p(\mathbf{t}_{12},\mathbf{t}_{21},\mathbf{z})\!\!\!\!\!\!\!\!\!\!\!\!\sum_{\phi_{12}(\!\mathbf{t}_{12})\psi_{12}(\mathbf{t}_{12}\!)\theta_{12}(\mathbf{t}_{12})}\!\!\!\!\!\!\!\!\!\!\!\!\!\!\!\!\!\frac{P(\!\phi_{12},\!\psi_{12},\!\theta_{12},\!\mathbf{t}_{12},\!k_{12},\!k'_{12},\!k''_{12})\!}{2^{n(r_{12}\!+\!r'_{12}\!+\!r''_{12})}}\!\!\!\!\!\!\!\!\!\!\!\!\!\!\!\!\!\!\sum_{\phi_{21}(\mathbf{t}_{21})\psi_{21}(\mathbf{t}_{21})\theta_{21}(\mathbf{t}_{21})}\!\!\!\!\!\!\!\!\!\!\!\!\!\!\!\!\!\!\!\frac{P(\!\phi_{21},\!\psi_{21},\!\theta_{21},\!\mathbf{t}_{21},\!k_{21},\!k'_{21},\!k''_{21}\!)}{2^{n(r_{21}\!+\!r'_{21}\!+\!r''_{21})}}
\\\times \log_2\bigg( p(\mathbf{t}_{12},\mathbf{t}_{21}|\mathbf{z})2^{n(r_{12}+r'_{12}+r''_{12}+r_{21}+r'_{21}+r''_{21})}+p(\mathbf{t}_{12}|\mathbf{z})2^{n(r_{12}+r'_{12})}+p(\mathbf{t}_{21}|\mathbf{z})2^{n(r_{21}+r'_{21})}+1\bigg)
\\=\sum_{\mathbf{t}_{12}\mathbf{t}_{21}\mathbf{z}}\!\!p(\!\mathbf{t}_{12},\!\mathbf{t}_{21},\!\mathbf{z}\!)\log_2 \bigg(\! p(\!\mathbf{t}_{12},\!\mathbf{t}_{21}\!|\mathbf{z}\!)2^{n(\!r_{12}\!+\!r'_{12}\!+\!r''_{12}\!+\!r'_{21}\!+\!r_{21}\!+\!r''_{21}\!)}\!+\!p(\!\mathbf{t}_{12}\!|\mathbf{z}\!)2^{n(\!r_{12}\!+\!r'_{12}\!+\!r''_{12}\!)}\!+\! p(\!\mathbf{t}_{21}\!|\mathbf{z}\!)2^{n(\!r_{21}\!+\!r'_{21}\!+\!r''_{21}\!)}\!+1\!\bigg)
\end{multline}}
\normalsize{To analyze (\ref{ddef5}), we split the sum between the following sequences:}
 \begin{itemize}
  \item If $(\mathbf{t}_{12},\mathbf{t}_{21},\mathbf{z})\in \mathcal{T}_{\epsilon''}^n(P_{T_{12}T_{21}Z})$, then the log term in (\ref{ddef5}) is upper bounded by
 \begin{align}
      &{\log_2 \bigg( 2^{-n(H(T_{12},T_{21}|Z)-2\epsilon'')}2^{n(r_{12}+r'_{12}+r''_{12}+r_{21}+r'_{21}+r''_{21})}+2^{-n(H(T_{12}|Z)-2\epsilon'')}2^{n(r_{12}+r'_{12}+r''_{12})}+}\no \\&{\ \ \ \ \ \ \ \ \ \ \ \ \ \ \ \ 2^{-n(H(T_{21}|Z)-2\epsilon'')}2^{n(r_{21}+r'_{21}+r''_{21})}+1\bigg)}\label{ddef6}
     \end{align}

\normalsize{By substituting $Z=(Y_{3},T_{13},T_{23},S_{13},S_{23},S_{12},S_{21})$ in \eqref{ddef6} and using the conditions \eqref{r12+r21,s}-\eqref{r21,s} in Lemma \ref{lm:5}, the log term in \eqref{ddef6} can be bounded by $\log(3\epsilon''+1)$ for sufficiently large $n$.}
  \item If $(\mathbf{t}_{12},\mathbf{t}_{21},\mathbf{z})\notin \mathcal{T}_{\epsilon''}^n(P_{T_{12}T_{21}Z})$, then the log term in (\ref{ddef5}) is upper bounded by
    \begin{align}
      &{\log_2\! \bigg(\!p(\!\mathbf{t}_{12},\!\mathbf{t}_{21}\!|\mathbf{z}\!)2^{n(\!r_{12}\!+\!r'_{12}\!+\!r''_{12}\!+\!r'_{21}\!+\!r_{21}\!+\!r''_{21}\!)}\!+\!p(\!\mathbf{t}_{12}|\mathbf{z})2^{n(r_{12}\!+\!r'_{12}\!+\!r''_{12}\!)}\!+\! p(\!\mathbf{t}_{21}\!|\mathbf{z})2^{n(\!r_{21}\!+\!r'_{21}\!+\!r''_{21}\!)}\!+\!1\bigg)}\no
     \\ &{\leq\log_2 \bigg( \mu^{-n} 2^{n(r_{12}+r'_{12}+r''_{12}+r_{21}+r'_{21}+r''_{21})}+\mu^{-n} 2^{n(r_{12}+r'_{12}+r''_{12})}+ \mu^{-n} 2^{n(r_{21}+r'_{21}+r''_{21})}+1\bigg)}\no
    \\ &{\leq \log_2( 3\mu^{-n} 2^{n(r_{12}+r'_{12}+r''_{12}+r_{21}+r'_{21}+r''_{21})}+1)}\no
     \end{align}
    where
    \begin{align}
      \mu\triangleq \min_{p(z)>0}p(z).
    \end{align}
    Hence, the sum over the non-typical sequences is bounded by $$\epsilon''\log_2(3\mu^{-n} 2^{n(\!r_{12}\!+\!r'_{12}\!+\!r''_{12}\!+\!r_{21}\!+\!r'_{21}\!+\!r''_{21}\!)}\!+\!1\!).$$
  \end{itemize}
\normalsize{
\indent Finally, the sum in (\ref{ddef5}) is bounded as}
 \begin{align}
 &{E(D(P_{K_{12},K_{21},K'_{12},K'_{21},K''_{12},K''_{21},Y_{3}^{n},T_{13}^{n},T_{23}^{n},S_{13}^{n},S_{23}^{n},S_{12}^{n},S_{21}^{n}}\parallel{q_{K_{12}}q_{K_{21}}q_{K'_{12}}q_{K'_{21}}q_{K''_{12}}q_{K''_{21}}P_{Z^{n}}}))\leq}\no \\&{\ \ \ \ \ \ \ \ \ \ \ \epsilon''\log_2(3\mu^{-n} 2^{n(\!r_{12}\!+\!r'_{12}\!+\!r''_{12}\!+\!r_{21}\!+\!r'_{21}\!+\!r''_{21}\!)}\!+\!1\!)\!+\!\log(\!3\epsilon''\!+\!1\!)}\label{n.31}
\end{align}
which vanishes as $n$ goes to infinity and $\epsilon''$ goes to zero. To obtain \eqref{n.31} we substituted $z$ as in \eqref{z}.
\end{IEEEproof}
The same approach as in Proposition \ref{pro2} can be applied to the secondary keys between Users 1 and 3 and, between Users 2 and 3. In particular, by using \eqref{r13,s} and \eqref{r23,s} in Lemma \ref{lm:5}, we have:
\begin{align}
  &{E_(D(P_{K_{13},K'_{13},K''_{13},Y_{2}^{n},X_{2}^{n},S_{12}^{n},T_{12}^{n},S_{13}^{n},S_{23}^{n},T_{23}^{n}}\parallel{q_{K_{13}}q_{K'_{13}}q_{K''_{13}}P_{Y_{2}^{n},X_{2}^{n},S_{12}^{n},T_{12}^{n},S_{13}^{n},S_{23}^{n},T_{23}^{n}}}))\leq}\no \\&{\ \ \ \ \ \ \ \ \ \ \ \epsilon''\log_2(\mu'^{-n} 2^{n(r_{13}+r'_{13}+r''_{13})}+1)+\log(\epsilon''+1)\label{str5}}
  \\&{E_(D(P_{K_{23},K'_{23},K''_{23},Y_{1}^{n},X_{1}^{n},S_{21}^{n},T_{21}^{n},S_{13}^{n},S_{23}^{n},T_{13}^{n}}\parallel{q_{K_{23}}q_{K'_{23}}q_{K''_{23}}P_{Y_{1}^{n},X_{1}^{n},S_{21}^{n},T_{21}^{n},S_{13}^{n},S_{23}^{n},T_{13}^{n}}}))\leq}\no \\&{\ \ \ \ \ \ \ \ \ \ \ \epsilon''\log_2(\mu''^{-n} 2^{n(r_{23}+r'_{23}+r''_{23})}+1)+\log(\epsilon''+1)\label{str6}}
\end{align}

Now, we follow the result in Proposition \ref{pro2} and also (\ref{str5})-(\ref{str6}) to show strong secrecy of the secondary keys between the users.

\begin{lem}[Strong secrecy of the secondary keys]
  \label{lm:2}
If \eqref{r12+r21,s}-\eqref{r23,s} hold, then there exist $\alpha>0, \beta>0$ and $\gamma>0$ such that
  \begin{align}
    &{E_{\Phi\Psi}(D(P_{K_{12},K_{21},K'_{12},K'_{21},K''_{12},K''_{21},Y_{3}^{n},T_{13}^{n},T_{23}^{n},S_{13}^{n},S_{23}^{n},S_{12}^{n},S_{21}^{n}}\!\parallel\!{q_{K_{12}}q_{K_{21}}P_{Y_{3}^{n},T_{13}^{n},T_{23}^{n},S_{13}^{n},S_{23}^{n},S_{12}^{n},S_{21}^{n}}})) \!\!\leq\!\! 2^{-\alpha n}}\\
    &{E_(D(P_{K_{13},K'_{13},K''_{13},Y_{2}^{n},X_{2}^{n},S_{12}^{n},T_{12}^{n},S_{13}^{n},S_{23}^{n},T_{23}^{n}}\parallel{q_{K_{13}}P_{Y_{2}^{n},X_{2}^{n},S_{12}^{n},T_{12}^{n},S_{13}^{n},S_{23}^{n},T_{23}^{n}}})) \leq 2^{-\beta n}}\\
    &{E_(D(P_{K_{23},K'_{23},K''_{23},Y_{1}^{n},X_{1}^{n},S_{21}^{n},T_{21}^{n},S_{13}^{n},S_{23}^{n},T_{13}^{n}}\parallel{q_{K_{23}}P_{Y_{1}^{n},X_{1}^{n},S_{21}^{n},T_{21}^{n},S_{13}^{n},S_{23}^{n},T_{13}^{n}}}))\leq 2^{-\gamma n}}
  \end{align}
\end{lem}

It should be note that we performed the security analysis based on the simpler distribution $P$ as in \eqref{dist.contruc1} in Appendix A. To show that the same analysis is valid for the induced distribution $\tilde{P}$ from the encoding scheme as in \eqref{dist.contruc3}, we use the fact that according to \eqref{dif.ent1}, the distance between distributions $P$ and $\tilde{P}$ is arbitrarily small and hence the same security analysis holds for distribution $\tilde{P}$.

Furthermore, applying the same approach as in the decoding error probability analysis in Appendix A, we can show:
\small{
\begin{align}
    &{E_{K''_{12}K''_{21}}(D(\!P_{K_{12},\!K_{21},\!K'_{12},\!K'_{21},\!Y_{3}^{n},\!T_{13}^{n},\!T_{23}^{n},\!S_{13}^{n},\!S_{23}^{n},\!S_{12}^{n},\!S_{21}^{n}\!|K''_{12},\!K''_{21}}\!\parallel\!{q_{K_{12}}q_{K_{21}}q_{K'_{12}}q_{K'_{21}}P_{Y_{3}^{n},\!T_{13}^{n},\!T_{23}^{n},\!S_{13}^{n},\!S_{23}^{n},\!S_{12}^{n},\!S_{21}^{n}}})) \!\leq\! 2^{-\alpha n}}\\
    &{E_{K''_{13}}(D(P_{K_{13},K'_{13},Y_{2}^{n},X_{2}^{n},S_{12}^{n},T_{12}^{n},S_{13}^{n},S_{23}^{n},T_{23}^{n}|K''_{13}}\parallel{q_{K_{13}}q_{K'_{13}}P_{Y_{2}^{n},X_{2}^{n},S_{12}^{n},T_{12}^{n},S_{13}^{n},S_{23}^{n},T_{23}^{n}}})) \leq 2^{-\beta n}}\\
    &{E_{K''_{23}}(D(P_{K_{23},K'_{23},Y_{1}^{n},X_{1}^{n},S_{21}^{n},T_{21}^{n},S_{13}^{n},S_{23}^{n},T_{13}^{n}|K''_{23}}\parallel{q_{K_{23}}q_{K'_{23}}P_{Y_{1}^{n},X_{1}^{n},S_{21}^{n},T_{21}^{n},S_{13}^{n},S_{23}^{n},T_{13}^{n}}}))\leq 2^{-\gamma n}}
  \end{align}}
\normalsize{and similarly, for the chosen $k_{12}^{''*},k_{21}^{''*},k_{13}^{''*}$ and $k_{23}^{''*}$ in Appendix A, we have:}
\small{
\begin{align}
    &{D(P_{K_{12},K_{21},K'_{12},K'_{21},Y_{3}^{n},T_{13}^{n},T_{23}^{n},S_{13}^{n},S_{23}^{n},S_{12}^{n},S_{21}^{n}|k_{12}^{''*},k_{21}^{''*}}\parallel{q_{K_{12}}q_{K_{21}}q_{K'_{12}}q_{K'_{21}}P_{Y_{3}^{n},T_{13}^{n},T_{23}^{n},S_{13}^{n},S_{23}^{n},S_{12}^{n},S_{21}^{n}}})) \leq 3.2^{-\alpha n}}\\
    &{D(P_{K_{13},K'_{13},Y_{2}^{n},X_{2}^{n},S_{12}^{n},T_{12}^{n},S_{13}^{n},S_{23}^{n},T_{23}^{n}|k_{13}^{''*}}\parallel{q_{K_{13}}q_{K'_{13}}P_{Y_{2}^{n},X_{2}^{n},S_{12}^{n},T_{12}^{n},S_{13}^{n},S_{23}^{n},T_{23}^{n}}})) \leq 3.2^{-\beta n}}\\
    &{D(P_{K_{23},K'_{23},Y_{1}^{n},X_{1}^{n},S_{21}^{n},T_{21}^{n},S_{13}^{n},S_{23}^{n},T_{13}^{n}|k_{23}^{''*}}\parallel{q_{K_{23}}q_{K'_{23}}P_{Y_{1}^{n},X_{1}^{n},S_{21}^{n},T_{21}^{n},S_{13}^{n},S_{23}^{n},T_{13}^{n}}}))\leq 3.2^{-\gamma n}}
  \end{align}}
\normalsize{Hence, Lemma \ref{lm:5} in Appendix A is deduced.}

\section{Proof of the outer bound in Theorem \ref{th2}}
\label{App3}
Since the pre-generated keys scheme is a special case of the generalized scheme, the outer bound on the secret-key capacity region of the generalized scheme applies to the pre-generated keys scheme, as well. Since, the upper bounds on $R_{12}$ are the same in Theorem \ref{th2} and Theorem \ref{th4}, we refer the reader to the bound on $R_{12}$ derived as part of the proof of Theorem \ref{th4} in Appendix E. In the following, we only prove upper bounds on $R_{13}$ and $R_{23}$.

In the pre-generated keys scheme, Users 1 and 2 independently generate the uniformly distributed keys $K_{13}$ and $K_{23}$, respectively, to share with User 3. After sending the channel inputs by Users 1 and 2, $Y_{i}^{n}$ is received by User $i$ for $i\in\{1,2,3\}$. Applying Fano's inequality on $K_{13}$ and $K_{23}$, for an arbitrary small $\epsilon>0$, we obtain:
\begin{equation}{H(K_{13},K_{23}|Y_{3}^{n})\le n\left(\frac{h(\epsilon)}{n}+\epsilon \log(|{\rm {\mathcal K}}_{13}||{\rm {\mathcal K}}_{23}|-1)\right)\triangleq n\epsilon_{1}}\label{fano} \end{equation}
where $|\mathcal{K}_{13}|$ is the cardinality of key set $\mathcal{K}_{13}$, and $\epsilon_{1}\to0$ if $\epsilon\to0$. Furthermore, the security conditions should be satisfied as:
\begin{align}
&I(K_{13};X_{2}^{n},Y_{2}^{n})<n\epsilon\label{sec.cond1}
\\&I(K_{23};X_{1}^{n},Y_{1}^{n})<n\epsilon\label{sec.cond2}
\end{align}
Next, we show that for secret key $K_{13}$ satisfying the reliability and security conditions, the upper bound on $R_{13}$ in Theorem \ref{th2} is satisfied. The upper bound on $R_{23}$ may be proved in an identical way.
\begin{align*}
  nR_{13}\le H(K_{13})+n\epsilon&\mathop{\le}\limits^{{\rm(a)}}H(K_{13}|X_{2}^{n},Y_{2}^{n})+2n\epsilon 
  \\ & {\mathop{\le}\limits^{{\rm (b)}} H(K_{13}|X_{2}^{n},Y_{2}^{n})-H(K_{13}|Y_{3}^{n})+2n\epsilon+n\epsilon_{1}}
  \\ & {\le H(K_{13}|X_{2}^{n},Y_{2}^{n})-H(K_{13}|X_{2}^{n},Y_{2}^{n},Y_{3}^{n})+2n\epsilon+n\epsilon_{1} }
  \\ & {= I(K_{13};Y_{3}^{n}|X_{2}^{n},Y_{2}^{n})+2n\epsilon+n\epsilon_{1} }
  \\ & {\mathop{\le}\limits^{{\rm (c)}} I(X_{13}^{n};Y_{3}^{n}|X_{2}^{n},Y_{2}^{n})+2n\epsilon+n\epsilon_{1} }
  \\ & {\mathop{\le}\limits^{{\rm (d)}} nI(X_{13};Y_{3}|X_{2},Y_{2})+2n\epsilon+n\epsilon_{1} }  
\end{align*}
where (a) results from the security condition (\ref{sec.cond1}), (b) from Fano's inequality in (\ref{fano}), (c) from the Markov chain $K_{13}-X_{1}^{n}-(X_{2}^{n},Y_{1}^{n},Y_{2}^{n},Y_{3}^{n})$ and (d) from the memoryless property of the channel.


\section{Proof of the outer bound in Theorem \ref{th4}}
\label{App5}
In the generalized scheme, after receiving the channel outputs, the users generate the corresponding keys as stochastic functions of the information available at them. In particular, User 1 generates $K_{12}$ and $K_{13}$ for sharing with Users 2 and 3, respectively, as stochastic functions of $(X_{1}^{n},Y_{1}^{n})$. Similarly, $\hat{K}_{12}$ and $K_{23}$ are generated by User 2 for sharing with Users 1 and 3, respectively, as functions of $(X_{2}^{n},Y_{2}^{n})$. User 3 estimates $\hat{K}_{13}$ and $\hat{K}_{23}$ as stochastic functions of $Y_{3}^{n}$.

For an arbitrary $\epsilon>0$,  applying Fano's inequality to the keys yields
\begin{align}
H(K_{12}|\hat{K}_{12})&\le n\left(\frac{h(\epsilon)}{n}+\epsilon \log(|{\rm {\mathcal K}}_{12}|-1)\right)\triangleq n\epsilon_{1}\label{fano1}\\
H(K_{13}|\hat{K}_{13})&\le n\left(\frac{h(\epsilon)}{n}+\epsilon \log(|{\rm {\mathcal K}}_{13}|-1) \right)\triangleq n\epsilon_{2}\label{fano2}\\
H(K_{23}|\hat{K}_{23})&\le n\left(\frac{h(\epsilon)}{n}+\epsilon \log(|{\rm {\mathcal K}}_{23}|-1) \right)\triangleq n\epsilon_{3}\label{fano3}
 \end{align}
where $|{\rm{\mathcal K}}_{12}|$ is the cardinality of key set ${\rm{\mathcal K}}_{12}$, and for $i=1,2,3$, $\epsilon_{i}\to0$ if $\epsilon\to0$. Furthermore, the secrecy conditions should be satisfied, i.e.,
\begin{align}
I(K_{12} ;Y_{3}^{n})&<n\epsilon\label{seccon1}\\
I(K_{13} ;X_{2}^{n},Y_{2}^{n})&<n\epsilon\label{seccon2}\\
I(K_{23} ;X_{1}^{n},Y_{1}^{n})&<n\epsilon\label{seccon3}
 \end{align}

We show next that for secret keys $K_{12}$ and $K_{13}$ satisfying the reliability and security conditions, $R_{12}$ and $R_{13}$  must satisfy the upper bounds in Theorem \ref{th4}. The upper bound on $R_{23}$ may be proved in similar way to  $R_{13}$. To upper bound on $R_{12}$, first note that
\begin{align*}
nR_{12}&\le H(K_{12})+n\epsilon\\
       &\mathop{\le}\limits^{{\rm(a)}}H(K_{12}|Y_{3}^{n})+2n\epsilon
\\ &\mathop{\le }\limits^{{\rm (b)}} H(K_{12}|Y_{3}^{n})-H(K_{12}|\hat{K}_{12})+2n\epsilon+n\epsilon_{1}
\\ &\le H(K_{12}|Y_{3}^{n})-H(K_{12}|\hat{K}_{12},X_{2}^{n},Y_{2}^{n})+2n\epsilon+n\epsilon_{1}
\\ &\mathop{= }\limits^{{\rm (c)}} H(K_{12}|Y_{3}^{n})-H(K_{12}|X_{2}^{n},Y_{2}^{n})+2n\epsilon+n\epsilon_{1}
\\ &\le I(K_{12};X_{2}^{n},Y_{2}^{n}|Y_{3}^{n})+2n\epsilon+n\epsilon_{1}
\\ &\le I(K_{12},X_{1}^{n},Y_{1}^{n};X_{2}^{n},Y_{2}^{n}|Y_{3}^{n})+2n\epsilon+n\epsilon_{1}
\\ &\mathop{= }\limits^{{\rm (d)}} I(X_{1}^{n},Y_{1}^{n};X_{2}^{n},Y_{2}^{n}|Y_{3}^{n})+2n\epsilon+n\epsilon_{1}
\\ &= I(X_{1}^{n};Y_{2}^{n}|X_{2}^{n},Y_{3}^{n})+I(X_{2}^{n};Y_{1}^{n}|X_{1}^{n},Y_{3}^{n})+I(Y_{1}^{n};Y_{2}^{n}|X_{1}^{n},X_{2}^{n},Y_{3}^{n})+I(X_{1}^{n};X_{2}^{n}|Y_{3}^{n})+n\epsilon_{1}+2n\epsilon
\\ &\mathop{\le}\limits^{{\rm (e)}} \sum _{i=1}^{n}[I(X_{1,i};\!Y_{2,i}|X_{2,i},\!Y_{3,i})\!+I(X_{2,i};\!Y_{1,i}|X_{1,i},\!Y_{3,i})+\!I(Y_{1,i};\!Y_{2,i}|X_{1,i},\!X_{2,i},\!Y_{3,i})]\!+\!I(X_{1}^{n};\!X_{2}^{n}|Y_{3}^{n})\!+\!n\epsilon_{1}\!+\!2n\epsilon
\end{align*}
where (a) results from the security condition (\ref{seccon1}), (b) from Fano's inequality in (\ref{fano1}), (c) and (d) from the fact that $\hat{K}_{12}$ and $K_{12}$ are stochastic functions of $(X_{2}^{n},Y_{2}^{n})$ and $(X_{1}^{n},Y_{1}^{n})$, respectively, so that $\hat{K}_{12}-(X_{2}^{n},Y_{2}^{n})-(X_{1}^{n},Y_{1}^{n})-K_{12}$ forms a  Markov chain. Inequality (e) follows since the first three terms can be single-letterized using the memoryless property of the channel as, e.g.,
\begin{align*}
I(X_{1}^{n};Y_{2}^{n}|X_{2}^{n},Y_{3}^{n})&\le \sum _{i=1}^{n}H(Y_{2,i}|X_{2,i},Y_{3,i})-H(Y_{2}^{n}|X_{1}^{n},X_{2}^{n},Y_{3}^{n})\\
&= \sum _{i=1}^{n}[H(Y_{2,i}|X_{2,i},Y_{3,i})-H(Y_{2,i}|X_{1,i},X_{2,i},Y_{3,i})].
\end{align*}
To single-letterize the fourth term, note that
\begin{align*}
I(X_{1}^{n};X_{2}^{n}|Y_{3}^{n})&\mathop{=}\limits^{{\rm(a)}}I(X_{1}^{n};Y_{3}^{n}|X_{2}^{n})-I(X_{1}^{n};Y_{3}^{n})
\\ &= \sum _{i=1}^{n}I(X_{1}^{n};Y_{3,i}|X_{2}^{n},Y_{3}^{i-1})-I(X_{1}^{n};Y_{3}^{n})\displaybreak[0]
\\ &\mathop{\le}\limits^{{\rm(b)}} \sum _{i=1}^{n}I(X_{1,i};Y_{3,i}|X_{2,i},Y_{3}^{i-1})-I(X_{1}^{n};Y_{3}^{n}) \displaybreak[0]
\\ &=\sum _{i=1}^{n}I(X_{1,i};Y_{3,i}|X_{2,i},Y_{3}^{i-1})-\sum _{i=1}^{n}I(X_{1}^{n};Y_{3,i}|Y_{3}^{i-1}) \displaybreak[0]
\\ &\le \sum _{i=1}^{n}I(X_{1,i};Y_{3,i}|X_{2,i},Y_{3}^{i-1})-\sum _{i=1}^{n}I(X_{1,i};Y_{3,i}|Y_{3}^{i-1}) \displaybreak[0]
\\ &\mathop{=}\limits^{{\rm(c)}} \sum _{i=1}^{n}[I(X_{1,i};Y_{3,i}|X_{2,i},U_{i})-I(X_{1,i};Y_{3,i}|U_{i})]
\end{align*}
where (a) results from the independence of $X_{1}^{n}$ and $X_{2}^{n}$, (b) from the memoryless property of the channel and (c) from the definition of auxiliary random variable $U_{i}\triangleq Y_{3}^{i-1}$. Combining all the above, we obtain
\begin{multline*}
  nR_{12}\le \sum _{i=1}^{n}[I(X_{1,i};\!Y_{2,i}|X_{2,i},\!Y_{3,i})\!+\!I(X_{2,i};\!Y_{1,i}|X_{1,i},\!Y_{3,i})\!+\!I(Y_{1,i};\!Y_{2,i}|X_{1,i},X_{2,i},\!Y_{3,i})\\+  I(X_{1,i};\!Y_{3,i}|X_{2,i},U_{i})-\!I(X_{1,i};\!Y_{3,i}|U_{i})]\!+\!n\epsilon_{1}\!+\!2n\epsilon.
\end{multline*}

Finally, introducing random variable $Q$ which is uniformly distributed on $\{1,2,...,n\}$ and independent of all the other variables, and defining $X_{1}=X_{1,Q},X_{2}=X_{2,Q},Y_{1}=Y_{1,Q},Y_{2}=Y_{2,Q},Y_{3}=(Y_{3,Q},Q),U=(U_{Q},Q)$, we have
\begin{equation}\no
R_{12}\!\le\! I(\!X_{1};\!Y_{2}|X_{2},\!Y_{3}\!)\!+\!I(\!X_{2};\!Y_{1}|X_{1},\!Y_{3}\!)\!+\!I(\!Y_{1};\!Y_{2}|X_{1},\!X_{2},\!Y_{3}\!)\!+\!I(\!X_{1};\!Y_{3}|X_{2},\!U\!)\!-\!I(\!X_{1};\!Y_{3}|U\!)\!+\!\epsilon_{1}\!+\!2\epsilon \end{equation}

To upper bound on $R_{13}$, note that
\begin{align*}
  nR_{13}\le H(K_{13})+n\epsilon&\mathop{\le}\limits^{{\rm(a)}}H(K_{13}|X_{2}^{n},Y_{2}^{n})+2n\epsilon
\\ &\mathop{\le }\limits^{{\rm (b)}} H(K_{13}|X_{2}^{n},Y_{2}^{n})-H(K_{13}|\hat{K}_{13})+2n\epsilon+n\epsilon_{2}
\\ &\le H(K_{13}|X_{2}^{n},Y_{2}^{n})-H(K_{13}|\hat{K}_{13},Y_{3}^{n})+2n\epsilon+n\epsilon_{2}
\\ &\mathop{=}\limits^{{\rm (c)}} H(K_{13}|X_{2}^{n},Y_{2}^{n})-H(K_{13}|Y_{3}^{n})+2n\epsilon+n\epsilon_{2}
\\ &\le I(K_{13};Y_{3}^{n}|X_{2}^{n},Y_{2}^{n})+2n\epsilon+n\epsilon_{2}
\\ &\le I(K_{13},X_{1}^{n},Y_{1}^{n};Y_{3}^{n}|X_{2}^{n},Y_{2}^{n})+2n\epsilon+n\epsilon_{2}
\\ &\mathop{=}\limits^{{\rm (d)}}I(X_{1}^{n},Y_{1}^{n};Y_{3}^{n}|X_{2}^{n},Y_{2}^{n})+2n\epsilon+n\epsilon_{2}
\\ &\mathop{\le}\limits^{{\rm (e)}}  nI(Y_{3};X_{1},Y_{1}|X_{2},Y_{2})+n\epsilon_{2}+2n\epsilon
\end{align*}
where (a) results from the security condition (\ref{seccon2}), (b) from Fano's inequality in (\ref{fano2}), (c) and (d) from the fact that $\hat{K}_{13}$ and $K_{13}$ are stochastic functions of $Y_{3}^{n}$ and $(X_{1}^{n},Y_{1}^{n})$, respectively, so that $\hat{K}_{13}-Y_{3}^{n}-(X_{1}^{n},Y_{1}^{n})-K_{13}$ forms a Markov chain. Inequality (e) follows from the memoryless property of the channel.


\bibliographystyle{IEEEtran}
 

\end{document}